%% file: bggs_new.tex
\documentclass[preprint,notoc]{JHEP3} 
\usepackage{epsfig,multicol}

\usepackage{axodraw}

\usepackage{amsmath}

\newcommand{\gsim}{\stackrel{\scriptstyle >}{{ }_{\sim}}}
\newcommand{\lsim}{\stackrel{\scriptstyle <}{{ }_{\sim}}}

\newcommand{\CtbH}{\ensuremath{H^-t\bar{b}}}
\newcommand{\tbH}{\ensuremath{H^+\bar{t}b}}
\newcommand{\gb}{\ensuremath{\bar{b}g}}

\newcommand{\mH}{\ensuremath{M_{H^+}}}
\newcommand{\mb}{\ensuremath{m_b}}

\newcommand{\mt}{\ensuremath{m_t}}

\newcommand{\bsg}{\ensuremath{b \to s\gamma}}
\newcommand{\pptbH}{\ensuremath{p\bar{p}(pp)\to\tbH +X}}

\newcommand{\Dmb}[1][]{\ensuremath{\Delta\mb^{#1}}}
\newcommand{\aS}[1][]{\ensuremath{\alpha_S^{#1}}}

\newcommand{\Hpm}{\ensuremath{H^\pm}}
\newcommand{\tb}[1][]{\ensuremath{\tan^{#1}\!\beta}}
\newcommand{\MSUSY}{\ensuremath{M_\mathrm{SUSY}}}
\newcommand{\mg}{\ensuremath{m_{\tilde{g}}}}
\newcommand{\msb}[1]{\ensuremath{m_{\tilde{b}_{#1}}}}
\newcommand{\mst}[1]{\ensuremath{m_{\tilde{t}_{#1}}}}
\newcommand{\be}{\begin{equation}}
\newcommand{\ee}{\end{equation}}
\newcommand{\eps}{\epsilon}
\newcommand{\fb}{\ensuremath{~{\rm fb}}}
\newcommand{\pb}{\ensuremath{~{\rm pb}}}
\newcommand{\GeV}{\ensuremath{\,{\rm GeV}}}
\newcommand{\TeV}{\ensuremath{\,{\rm TeV}}}

\title{Prospects for heavy supersymmetric charged Higgs boson
        searches  at hadron colliders}
\author{Alexander Belyaev,\thanks{On leave of absence from Nuclear Physics 
Institute, Moscow State University.}\\ 
Physics Department, Florida State University
Tallahassee, FL 32306-4350, USA}
\author{David Garcia,\\
Theory Division, CERN, CH--1211 Geneva 23, Switzerland}
\author{Jaume Guasch,\thanks{Present address: Theory 
  Group LTP, Paul Scherrer Institut, CH-5232 Villigen PSI,
  Switzerland.} \\
Institut f\"{u}r Theoretische Physik, Universit\"at Karlsruhe, 
D--76128 Karlsruhe, Germany }
\author{Joan Sol\`{a}\\
Departament d'Estructura i Constituents de la
Mat\`eria, Universitat de Barcelona,\\
E-08028, Diagonal 647, Barcelona, Catalonia, Spain, and\\
Institut de F\'{\i}sica d'Altes Energies, Universitat Aut\`onoma de Barcelona,\\
E-08193, Bellaterra, Barcelona, Catalonia, Spain}

\preprint{CERN-TH/2001-051, 
KA-TP-7-2001, 
UB-ECM-PF-02/02, 
FSU-HEP-020301, 
\hepph{0203031}}

\abstract{We
investigate the production of a heavy charged Higgs boson at
hadron colliders within the context of the MSSM. A detailed study
is performed for all important production modes and basic
background processes for the $t\bar{t}b\bar{b}$ signature. In
our analysis we include effects of initial and final state
showering, hadronization, and 
principal detector effects.
For the signal
production rate we include the leading SUSY quantum effects at
high $\tan \beta\gtrsim m_{t}/m_{b}$. Based on the obtained
efficiencies for the signal and background we 
estimate the
discovery and exclusion mass limits of the charged Higgs boson at
high values of $\tan\beta$.
At the upgraded Tevatron
the discovery
of a heavy charged Higgs boson ($M_{H^{\pm}}\gtrsim
200\GeV$) is  impossible 
for the tree-level cross-section values.
 However, if QCD and SUSY effects happen
to reinforce mutually, there are indeed regions of the MSSM
parameter space which could provide 
$3\,\sigma$ evidence
and, at best, 
$5\,\sigma$  charged Higgs boson discovery 
at the Tevatron for  masses $M_{H^{\pm}}\lesssim 300\GeV$
and $M_{H^{\pm}}\lesssim 250\GeV$, respectively, even assuming
squark and gluino masses in the $ (500-1000)\GeV$ range. On the
other hand, at the LHC one can 
discover a $H^{\pm}$
as heavy as $1\TeV$ at the canonical confidence level of
$5\,\sigma$; or else exclude its existence at $95\,\%$ C.L. up to
masses $\sim 1.5\TeV$. Again the presence of SUSY quantum effects
can be very important here as they may shift the LHC limits by a
few hundred $\mathrm{GeV}$.
}


\begin{document} 

\section{Introduction}

The full experimental confirmation of the Standard Model (SM) is still
waiting for the finding of the Higgs boson. Last LEP results, suggesting a
light Higgs boson of about $115\GeV$~\cite{LEP115}, are encouraging, but
we will 
have to wait 
 new data from the upgraded Tevatron or the advent of the
LHC to see this result
either confirmed or dismissed~{%
\cite{HiggsRunII,tev3-t-hb,CMS2001,Ellisnew}}. {Moreover, at the
LHC a large amount of top-quark pairs will be produced, allowing
for high-precision measurements of all its properties, providing
strong checks of the SM, or new physics signals~\cite{LHCtop}.}
Even if a neutral Higgs boson is discovered, the principal
question will still be present at the forefront of Elementary
Particle research, namely: is the SM realized in nature or does a
model beyond the SM take place with extended Higgs sectors of
various kinds (extra doublets, singlets, even triplets)? In most
of these extensions, the physical spectrum contains charged Higgs
bosons and this introduces a distinctive feature. For example, in
the general two-Higgs-doublet model (2HDM)~\cite{Hunter} one just
adds up another doublet of scalars and then the spectrum of the
model contains three neutral Higgs bosons and two charged ones,
the latter being commonly denoted by $H^{\pm}$. This is also the
case of the Higgs sector of the Minimal Supersymmetric Standard
Model (MSSM) \cite{MSSMreps}, which is a prominent Type II 2HDM
of a very restricted kind.

While the detection of a charged Higgs boson would still leave a
lot of questions unanswered, it would immediately offer (in
contrast to the detection of a neutral one) indisputable evidence
of physics beyond the SM. Then the next step in charged Higgs
boson physics would be the precise measurement of its mass and
couplings to fermions, to give us the understanding of whether
those couplings and masses are compatible with the ones in the
MSSM or belong to a more general 2HDM. Of course this task cannot
be accomplished without including the information provided by
radiative corrections. These are not only potentially large in
the computation of the MSSM Higgs boson masses
themselves~\cite{CHHHWW} but also in the interaction vertices and
self-energies. The latter features have been studied in great
detail for the decay of the top quark into a charged
Higgs boson~\cite{SUSYtbH} and also for the
hadronic decays of the 
Higgs bosons of the MSSM~{\cite{CJS1,SUSYHtoTB}}.

Focusing on the charged Higgs boson, the information we have nowadays is
rather limited. First, there is a direct LEP limit, $M_{H^{\pm}}>79.9\GeV$,
for a 2HDM $H^{\pm}$ decaying exclusively into $\tau\nu_{\tau}$ \cite
{LEPHpm}.
Second, Tevatron direct and indirect searches for a $H^{\pm}$
place constraints on the $M_{H^{+}}$--$BR(t\rightarrow bH^{+})$
plane~{\cite{TeVHpm,CDF}}, which are usually translated to the
$M_{H^{+}}$--$\tb$ plane once the relevant MSSM parameters are
fixed~\cite{CDF,DPR1,tbmh}; here {\tb\ stands for
the ratio between the two
  Higgs doublets vacuum expectation values, $\tb\equiv v_2/v_1$}~\cite{Hunter}. 
Such searches 
have 
been done for $\mH<\mt-\mb$ so far. Finally, the bound coming from
the rare process \bsg, for a pure Type~II 2HDM, places a tight constraint in
the \mH--\tb\ plane, although in the MSSM lighter $H^{\pm}$ masses can be
traded for a constraint in the $A_{t}$--$\mu$ plane when the $R$-parity is
conserved \cite{BSG-HPM} 
or broken~\cite{BSG-HPM-R}. {This constraint has
been found to be robust under higher order effects~\cite{Bsg2l}.} 
Unlike the direct searches at LEP and the Tevatron, which rely on the
electromagnetic coupling and on a few, quite universal, decay modes, the
weak point of the 
indirect 
limits is their model dependence.

Several other processes have been proposed in the literature that would
allow detection of a charged Higgs boson in various kinematic ranges. Among
others, there are: pair production at $e^{+}e^{-}$~\cite{eeHH} and
hadron~\cite{ppHH} machines; associated production with a $W$ boson in
$e^{+}e^{-}$~\cite{eeHW} and hadron~\cite{KZ,ppHW} colliders. Indirect
searches in
B-physics observables (apart from \bsg) are also promising, specially in $%
b\rightarrow s\,l^{+}l^{-}$ and $b\rightarrow c\,\tau \,\nu _{\tau }$~\cite
{BH}. {Very high precision measurements on $\tau $ and $K$ physics can give
also interesting indirect limits~\cite{TauK}.}
Further production channels in hadron colliders specific to SUSY models
include the possibility of the charged Higgs boson being produced in
cascade decays of strong interacting SUSY particles, as well as
associated production with squarks~\cite{Dattaetal}. These
channels are mostly useful in intermediate ranges of $\tb$, and are
therefore complementary to the channel studied in the present work.

The aim of this paper is to extend the  analysis of
the associated $H^{\pm}$ production with top (and bottom)
{quarks} at hadron colliders \pptbH\ to both the Tevatron and the
LHC. 
In this work we will be focusing exclusively on the
MSSM case. The general 2HDM case will be addressed
elsewhere~\cite {prep}; following~\cite{CGHS} we also expect here
sizeable quantum effects, though of very different origin and
distinguishable from the MSSM ones.

{From our point of view, the relevance of the process \pptbH\ is
manifold. A positive signal would be instant
evidence of new physics and at the same time it would strongly
indicate a large value of $\tan\beta$. In fact, as we shall see,
the great virtue of this process is its ability to test the
charged Higgs boson coupling to the third generation of quarks and
leptons}, which in general 2HDM's (and in particular in the MSSM)
can be greatly augmented (or suppressed) not only at the
tree-level but also due to model-dependent quantum effects.
Such an enhancement could be
critical for heavy $H^{\pm}$ production, especially for the
Tevatron. Furthermore, if one would 
be able to correlate the quantum effects  on the charged Higgs boson
Yukawa coupling  with the corresponding effect in the measured
value of the neutral Higgs boson Yukawa couplings, it would mean a
strong hint for a MSSM Higgs
sector, even if no supersymmetric particle is detected at all~%
{\cite{CJS1,SUSYHtoTB,GHP,Carenanew}}. {That this
kind of scenario is possible is corroborated by the fact that the
process \pptbH\ is affected by exceptional SUSY quantum effects,
namely effects that do not necessarily vanish in the limit of
large \MSUSY\ masses{~\cite {Dmb}}}.  This
remarkable property has been greatly emphasized from the
phenomenological
point of view for applications in both high- and low-energy processes in~%
{\cite{SUSYtbH,CJS1,SUSYHtoTB,GHP,Carenanew,Haberetal,eff}}
and~\cite
{tau} respectively.\footnote{%
For a detailed analysis of the neutral MSSM Higgs bosons production at the
Tevatron, with the inclusion of leading SUSY effects, see~\cite{CMW}.}
Certainly, here is another place where it might play a fundamental role to
help uncovering the SUSY nature of the charged Higgs boson in high-energy
experiments. Indeed, we will show that these
Higgs boson production processes could receive large radiative corrections even
if the direct cascade SUSY channels mentioned above are kinematically closed.
In other words, the Higgs production modes addressed in the present 
study are not only generally competitive with the direct SUSY channels, but 
they could also be highly relevant by themselves.

At this point, a review of the previous works on \pptbH\ is in order. To the
best of our knowledge, the first analyses on this process in the literature
are the works of Refs.~\cite{Gunion,Barger}. Recently, there have been new
contributions retaking this issue in more detail~\cite
{treepptbH,Borzumati,morepptbH,DP-4b,Guchait}. 
Nevertheless all of these works stick to a tree-level computation, in spite
of the fact that some of them explicitly admit that the sort of charged
Higgs boson they are dealing with is of the MSSM type. Therefore, they are
unavoidably affected by some drawbacks. On the one hand, the tree-level
approach can be quite inaccurate (e.g. gluino radiative corrections are
potentially of order 1 for large \tb\ and have, a priori, no definite sign)
and, on the other, it is unable to distinguish between a generic $H^{\pm }$
and one with the particular couplings of the MSSM.

In Ref.~\cite{Coarasa}, a first treatment of the SUSY radiative corrections
was presented although it was not sufficiently complete and moreover no
background analysis was attempted.\footnote{{A close comparison of the
results from Refs.~\cite{Borzumati,Coarasa} can be found in~\cite{HiggsRunII}.}%
} Therefore, we believe that considerably more discussion is needed before
any definite predictions on the discovery limits for the charged Higgs boson
as a function of the MSSM parameter space can be made.

The paper we are presenting here gives a significant further step in this
direction. It adds, with respect to previous studies, a simultaneous
treatment of the leading SUSY radiative corrections (both strong and
electroweak) including an analysis of the off-shell effects. Also, in
contrast with previous works we perform a beyond-the-parton-level simulation
of events, which includes the toy-detector simulation, jet fragmentation,
initial and final radiation effects. We show that some of those effects,
such as final radiation, do play a crucial role in the charged Higgs boson
studies. Another new aspect of the present paper is the proper kinematical
analysis of $gg\rightarrow H^{+}\bar{t}b$ and $g\bar{b}\rightarrow H^{+}\bar{%
t}$ processes after their correct combination.\footnote{{Ref.~\cite
{Borzumati} makes also the proper combination of the $gg\rightarrow H^{+}%
\bar{t}b $ and $g\bar{b}\rightarrow H^{+}\bar{t} $ channels for
the total effective cross-section, however no analysis of the
differential cross-section is performed.}} {First results of the
present analysis were presented in~\cite{BGGS1}.}

{Last but not least, it will also be important to discuss the
possible effect from the conventional QCD corrections, which
 unfortunately are not available for \pptbH\ at
present. However, two independent
calculations for the related process in the Standard Model, }$p\bar{p}%
(pp)\rightarrow H\,t\bar{t}+X${,  have recently appeared, namely the
  associated SM Higgs 
boson production at the Tevatron and the LHC, which have been
carried out at the next-to-leading order (NLO) in QCD \thinspace
\cite{SpiraNLO,DRNLO}. The result is that at the Tevatron the NLO
effects for SM Higgs boson production off
top quarks are negative and can be approximately described by a
$K$-factor ranging between }$0.8-1$ {whereas at the LHC they are
positive and the $K$-factor lies between }$1.2-1.4$ {depending on the
renormalization and factorization scales.
Obviously
this process bears relation to ours, $ p\bar{%
p}(pp)\rightarrow H^{+}\,\bar{t}\,b+X$, and so a few comments on
the expected NLO QCD corrections on the latter are in order (see
Section~\ref{sec:Xseccomp}). }

The rest of the paper is organized as follows: in Section~\ref{sec:Xseccomp}
we discuss the general procedure for the computation of the cross-section.
In Section~\ref{sect:svb} we perform a signal and background study based on
the PYTHIA6.1 simulations, and work out kinematical cuts and a suitable
strategy to suppress the background and extract the signal in the most
efficient way. Then we present signal and background efficiencies and signal
rates that could be viable at the Tevatron and LHC. In Section~\ref
{sec:SUSYcorr} we discuss in detail the role of the SUSY corrections both
analytically and numerically, {and then evaluate their impact on the
discovery limit or exclusion region of the charged Higgs boson}. The last
section, Section~\ref{sec:conclu}, is devoted to the summary and conclusions.

\section{Cross-section computation}

\label{sec:Xseccomp} The relevant charged Higgs boson production processes under
study are the following:

\begin{equation}
\pptbH\hspace{1cm}\mathrm{Tevatron\;(LHC)}  \label{tbh}
\end{equation}
At the parton level, the reaction (\ref{tbh}) proceeds through three
channels: i) $q\bar{q}$-annihilation for light quarks\footnote{%
We shall omit the charge-conjugate process, $p\bar{p}(pp)\rightarrow
\CtbH+X$, for the sake of brevity. Including this process just amounts
to multiplying our cross-section by a factor of 2. }
\begin{equation}
q\bar{q}\rightarrow \tbH,  \label{qq-tbh}
\end{equation}
where $q=u,d$ (the $s$ contribution can be safely neglected), a channel only
relevant for the Tevatron~\cite{Borzumati,Coarasa}; ii) $gg$-fusion
\begin{equation}
gg\rightarrow \tbH\,,  \label{gg-tbh}
\end{equation}
which is dominant at the LHC, but it can also be important at the Tevatron
for increasing $H^{+}$ masses \cite{Borzumati,Coarasa}; and finally there
is the iii) bottom-gluon 2-body channel
\begin{equation}
\bar{b}g\rightarrow H^{+}\bar{t}\,\,.  \label{gb-th}
\end{equation}

\FIGURE[t]{
\par
\begin{tabular}[t]{cc}
\multicolumn{2}{c}{
\begin{picture}(95,79)(0,0) \Text(15.0,70.0)[r]{$g$}
\DashLine(16.0,70.0)(37.0,60.0){3.0} \Text(15.0,50.0)[r]{$g$}
\DashLine(16.0,50.0)(37.0,60.0){3.0} \Text(47.0,61.0)[b]{$g$}
\DashLine(37.0,60.0)(58.0,60.0){3.0} \Text(80.0,70.0)[l]{$\bar{t}$}
\ArrowLine(79.0,70.0)(58.0,60.0) \Text(54.0,50.0)[r]{$t$}
\ArrowLine(58.0,60.0)(58.0,40.0) \Text(80.0,50.0)[l]{$H^+$}
\DashArrowLine(58.0,40.0)(79.0,50.0){1.0} \Text(80.0,30.0)[l]{$b$}
\ArrowLine(58.0,40.0)(79.0,30.0) \end{picture}
}
\begin{picture}(95,79)(0,0) \Text(15.0,70.0)[r]{$g$}
\DashLine(16.0,70.0)(37.0,60.0){3.0} \Text(15.0,50.0)[r]{$g$}
\DashLine(16.0,50.0)(37.0,60.0){3.0} \Text(47.0,61.0)[b]{$g$}
\DashLine(37.0,60.0)(58.0,60.0){3.0} \Text(80.0,70.0)[l]{$b$}
\ArrowLine(58.0,60.0)(79.0,70.0) \Text(54.0,50.0)[r]{$b$}
\ArrowLine(58.0,40.0)(58.0,60.0) \Text(80.0,50.0)[l]{$H^+$}
\DashArrowLine(58.0,40.0)(79.0,50.0){1.0} \Text(80.0,30.0)[l]{$\bar{t}$}
\ArrowLine(79.0,30.0)(58.0,40.0) \end{picture} {} \ 
\begin{picture}(95,79)(0,0) \Text(15.0,60.0)[r]{$g$}
\DashLine(16.0,60.0)(37.0,60.0){3.0} \Line(37.0,60.0)(58.0,60.0)
\Text(80.0,70.0)[l]{$\bar{t}$} \ArrowLine(79.0,70.0)(58.0,60.0)
\Text(33.0,50.0)[r]{$t$} \ArrowLine(37.0,60.0)(37.0,40.0)
\Text(15.0,40.0)[r]{$g$} \DashLine(16.0,40.0)(37.0,40.0){3.0}
\Text(47.0,44.0)[b]{$t$} \ArrowLine(37.0,40.0)(58.0,40.0)
\Text(80.0,50.0)[l]{$H^+$} \DashArrowLine(58.0,40.0)(79.0,50.0){1.0}
\Text(80.0,30.0)[l]{$b$} \ArrowLine(58.0,40.0)(79.0,30.0) \end{picture} {} \ 
\begin{picture}(95,79)(0,0) \Text(15.0,70.0)[r]{$g$}
\DashLine(16.0,70.0)(58.0,70.0){3.0} \Text(80.0,70.0)[l]{$\bar{t}$}
\ArrowLine(79.0,70.0)(58.0,70.0) \Text(54.0,60.0)[r]{$t$}
\ArrowLine(58.0,70.0)(58.0,50.0) \Text(80.0,50.0)[l]{$H^+$}
\DashArrowLine(58.0,50.0)(79.0,50.0){1.0} \Text(54.0,40.0)[r]{$b$}
\ArrowLine(58.0,50.0)(58.0,30.0) \Text(15.0,30.0)[r]{$g$}
\DashLine(16.0,30.0)(58.0,30.0){3.0} \Text(80.0,30.0)[l]{$b$}
\ArrowLine(58.0,30.0)(79.0,30.0) \end{picture} {} \  
 \\
\multicolumn{2}{c}{\begin{picture}(95,79)(0,0) \Text(15.0,60.0)[r]{$g$}
\DashLine(16.0,60.0)(37.0,60.0){3.0} \Text(47.0,64.0)[b]{$t$}
\ArrowLine(37.0,60.0)(58.0,60.0) \Text(80.0,70.0)[l]{$H^+$}
\DashArrowLine(58.0,60.0)(79.0,70.0){1.0} \Text(80.0,50.0)[l]{$b$}
\ArrowLine(58.0,60.0)(79.0,50.0) \Text(33.0,50.0)[r]{$t$}
\ArrowLine(37.0,40.0)(37.0,60.0) \Text(15.0,40.0)[r]{$g$}
\DashLine(16.0,40.0)(37.0,40.0){3.0} \Line(37.0,40.0)(58.0,40.0)
\Text(80.0,30.0)[l]{$\bar{t}$} \ArrowLine(79.0,30.0)(58.0,40.0)
\end{picture} {} \ \begin{picture}(95,79)(0,0) \Text(15.0,60.0)[r]{$g$}
\DashLine(16.0,60.0)(37.0,60.0){3.0} \Text(47.0,64.0)[b]{$b$}
\ArrowLine(58.0,60.0)(37.0,60.0) \Text(80.0,70.0)[l]{$H^+$}
\DashArrowLine(58.0,60.0)(79.0,70.0){1.0} \Text(80.0,50.0)[l]{$\bar{t}$}
\ArrowLine(79.0,50.0)(58.0,60.0) \Text(33.0,50.0)[r]{$b$}
\ArrowLine(37.0,60.0)(37.0,40.0) \Text(15.0,40.0)[r]{$g$}
\DashLine(16.0,40.0)(37.0,40.0){3.0} \Line(37.0,40.0)(58.0,40.0)
\Text(80.0,30.0)[l]{$b$} \ArrowLine(58.0,40.0)(79.0,30.0) \end{picture} {} \ %
\begin{picture}(95,79)(0,0) \Text(15.0,60.0)[r]{$g$}
\DashLine(16.0,60.0)(37.0,60.0){3.0} \Line(37.0,60.0)(58.0,60.0)
\Text(80.0,70.0)[l]{$b$} \ArrowLine(58.0,60.0)(79.0,70.0)
\Text(33.0,50.0)[r]{$b$} \ArrowLine(37.0,40.0)(37.0,60.0)
\Text(15.0,40.0)[r]{$g$} \DashLine(16.0,40.0)(37.0,40.0){3.0}
\Text(47.0,44.0)[b]{$b$} \ArrowLine(58.0,40.0)(37.0,40.0)
\Text(80.0,50.0)[l]{$H^+$} \DashArrowLine(58.0,40.0)(79.0,50.0){1.0}
\Text(80.0,30.0)[l]{$\bar{t}$} \ArrowLine(79.0,30.0)(58.0,40.0)
\end{picture} {} \ \begin{picture}(95,79)(0,0) \Text(15.0,70.0)[r]{$g$}
\DashLine(16.0,70.0)(58.0,70.0){3.0} \Text(80.0,70.0)[l]{$b$}
\ArrowLine(58.0,70.0)(79.0,70.0) \Text(54.0,60.0)[r]{$b$}
\ArrowLine(58.0,50.0)(58.0,70.0) \Text(80.0,50.0)[l]{$H^+$}
\DashArrowLine(58.0,50.0)(79.0,50.0){1.0} \Text(54.0,40.0)[r]{$t$}
\ArrowLine(58.0,30.0)(58.0,50.0) \Text(15.0,30.0)[r]{$g$}
\DashLine(16.0,30.0)(58.0,30.0){3.0} \Text(80.0,30.0)[l]{$\bar{t}$}
\ArrowLine(79.0,30.0)(58.0,30.0) \end{picture} {} \ } \\[-1cm]
\multicolumn{2}{c}{\large (a)} \\[12pt]
\begin{picture}(95,79)(0,0) \Text(15.0,70.0)[r]{$q$}
\ArrowLine(16.0,70.0)(37.0,60.0) \Text(15.0,50.0)[r]{$\bar{q}$}
\ArrowLine(37.0,60.0)(16.0,50.0) \Text(47.0,61.0)[b]{$g$}
\DashLine(37.0,60.0)(58.0,60.0){3.0} \Text(80.0,70.0)[l]{$\bar{t}$}
\ArrowLine(79.0,70.0)(58.0,60.0) \Text(54.0,50.0)[r]{$t$}
\ArrowLine(58.0,60.0)(58.0,40.0) \Text(80.0,50.0)[l]{$H^+$}
\DashArrowLine(58.0,40.0)(79.0,50.0){1.0} \Text(80.0,30.0)[l]{$b$}
\ArrowLine(58.0,40.0)(79.0,30.0) \end{picture} {} \ %
\begin{picture}(95,79)(0,0) \Text(15.0,70.0)[r]{$q$}
\ArrowLine(16.0,70.0)(37.0,60.0) \Text(15.0,50.0)[r]{$\bar{q}$}
\ArrowLine(37.0,60.0)(16.0,50.0) \Text(47.0,61.0)[b]{$g$}
\DashLine(37.0,60.0)(58.0,60.0){3.0} \Text(80.0,70.0)[l]{$b$}
\ArrowLine(58.0,60.0)(79.0,70.0) \Text(54.0,50.0)[r]{$b$}
\ArrowLine(58.0,40.0)(58.0,60.0) \Text(80.0,50.0)[l]{$H^+$}
\DashArrowLine(58.0,40.0)(79.0,50.0){1.0} \Text(80.0,30.0)[l]{$\bar{t}$}
\ArrowLine(79.0,30.0)(58.0,40.0) \end{picture} {} \  & %
\begin{picture}(95,79)(0,0) \Text(15.0,70.0)[r]{$\bar{b}$}
\ArrowLine(37.0,60.0)(16.0,70.0) \Text(15.0,50.0)[r]{$g$}
\DashLine(16.0,50.0)(37.0,60.0){3.0} \Text(47.0,64.0)[b]{$b$}
\ArrowLine(58.0,60.0)(37.0,60.0) \Text(80.0,70.0)[l]{$H^+$}
\DashArrowLine(58.0,60.0)(79.0,70.0){1.0} \Text(80.0,50.0)[l]{$\bar{t}$}
\ArrowLine(79.0,50.0)(58.0,60.0) \end{picture} {} \ %
\begin{picture}(95,79)(0,0) \Text(15.0,70.0)[r]{$\bar{b}$}
\ArrowLine(58.0,70.0)(16.0,70.0) \Text(80.0,70.0)[l]{$H^+$}
\DashArrowLine(58.0,70.0)(79.0,70.0){1.0} \Text(54.0,60.0)[r]{$t$}
\ArrowLine(58.0,50.0)(58.0,70.0) \Text(15.0,50.0)[r]{$g$}
\DashLine(16.0,50.0)(58.0,50.0){3.0} \Text(80.0,50.0)[l]{$\bar{t}$}
\ArrowLine(79.0,50.0)(58.0,50.0) \end{picture} {} \  \\[-1cm]
{\large (b)} & {\large (c)}
\end{tabular}
\par
\caption{\label{fig:diagr}
Diagrams for the tree-level processes (\ref{gg-tbh}) --
\textbf{a)}, (\ref{qq-tbh}) -- \textbf{b)} and (\ref{gb-th}) -- \textbf{c)}.
}
}

We remark that the latter 
can also be significant because the bottom quark mass, $m_{b}$,
is small with respect to the energy of the process, and therefore
parton distribution functions (PDFs) for $b$ and
$\overline{b}$-quarks (i.e. $b$-densities) have to be introduced,
allowing for the resummation of collinear logarithms~\cite
{Olness}. This provides an extra channel contributing to the
cross-section which must be appropriately combined with the
$gg$-fusion channel {as we will comment further below}.

We will compute the cross-section for the charged Higgs boson production process (%
{\ref{tbh}}) at the leading order (LO) in QCD, namely at
  $\mathcal{O}(\alpha
_{S}^{2})$. Feynman diagrams for the partonic subprocess (\ref{qq-tbh}), (%
\ref{gg-tbh}) and (\ref{gb-th}) are presented at the tree-level in Fig.~\ref
{fig:diagr}. However, the QCD corrections at
the next-to-leading order (NLO) or $\mathcal{O}(\alpha _{S}^{3})$ could be
important. We have already mentioned in the introduction that the full set
of QCD corrections to the process of SM Higgs boson radiation off top quarks in
hadron colliders,
\begin{equation}
p\bar{p}(pp)\rightarrow H\,t\bar{t}+X\,\,,  \label{SMHtt}
\end{equation}
have been computed at the NLO by two independent
groups~\cite{SpiraNLO,DRNLO}. The result is that, at the Tevatron, the
NLO effects from
QCD are negative, hence diminishing the signal cross-section, whereas at the
LHC they are positive and so enhancing the signal. To be more specific, if
one defines the scale  $\mu _{0}=m_{t}+M_{H}/2$, then the NLO effects can be
approximately encoded in a $K$-factor as follows \cite{SpiraNLO,DRNLO}.
Assuming equal factorization and renormalization scales $Q=\mu _{R}\equiv
\mu $, the $K$-factor  for the Tevatron ranges between $K=0.8$ (for the
central scale value $\mu =\mu _{0}$) and $K=1$ (for the threshold energy
value $\mu =2\mu _{0}=2\,m_{t}+M_{H}$), whereas at the LHC it varies between
$K=1.2$ (for $\mu =\mu _{0}$) and $K=1.4$ (for $\mu =2\mu _{0}$).  These
results are in agreement with the expectations from the fragmentation model
of Ref.\cite{Dawson} where an approximate calculation of the NLO cross-section
can be performed in the limit of small Higgs boson masses. In particular, the
negative sign of the QCD effects at the Tevatron can be understood from the
dominance of the $q\overline{q}$ $\rightarrow H\,\overline{t}t$ partonic
mode at the Tevatron energies (or, equivalently, for heavy quark masses such
as the top quark mass) which is subject to (moderate) negative QCD effects
both near and above the threshold. Clearly, the value of the $K$-factor can
be important, and even critical, especially at the Tevatron where the size
of the signal is not too conspicuous to allow for a comfortable separation
of it from the QCD background. Remarkably, in the MSSM framework the neutral
Higgs bosons production processes in association with bottom quarks can be, in
contrast to the SM case, even more important than the associated production
with top quarks. It can proceed through
\begin{equation}
p\bar{p}(pp)\rightarrow h\,b\overline{b}+X\,\,,  \label{hbb}
\end{equation}
with $h=h^{0},H^{0},A^{0}$ any of the neutral Higgs bosons of the
MSSM \cite {Hunter}.  These processes
are certainly related, in fact they are complementary, to our
charged Higgs boson case ({\ref{tbh}}), and different aspects of
them have already been analyzed in the SUSY context in several
places of the 
literature~\cite{HiggsRunII,CMW,barnett}.  
As in the charged Higgs boson case, the QCD
NLO corrections for the
processes (\ref{hbb}) have not been computed yet\footnote{%
Notice that in the MSSM the process (\ref{SMHtt}), with $H$
replaced with any of $h=h^{0},H^{0},A^{0}$, is not favored at high
$\tan \beta $, and moreover for $\tan \beta =\mathcal{O}(1)$ the
QCD effects should not be essentially different from those in the
SM case\,\cite{SpiraNLO,DRNLO}. }. Nevertheless we wish to remark
that in all of these MSSM Higgs boson production processes in
hadron colliders, with at least one $b$-quark in the final state,
we expect important QCD effects at the NLO (and even at higher
orders)\,\cite{Nason}. In fact, in contradistinction to the SM
case (\ref{SMHtt}), the QCD corrections for these processes should
entail a large and positive $K$-factor for both the Tevatron and
the LHC~\cite{spira_privat}. This circumstance could be highly
favorable to enhance the MSSM Higgs boson production and perhaps
make one of these particles visible already at the Tevatron.

In the absence of a fully-fledged
calculation, we can still have some hints on the approximate
value of the QCD corrections as follows. First, the result of the
NLO QCD corrections on the SM process~(\ref{SMHtt}) from
Refs.~\cite{SpiraNLO,DRNLO} does not directly apply to the
charged Higgs boson case under study (\ref{tbh}), since the
presence of a (nearly) massless quark in the final and
intermediate states will introduce additional logarithms of the
light quark mass. Second, part of these logarithms is taken into
account by the proper combination (see below) of the
processes~(\ref{gg-tbh}) and~(\ref{gb-th}), which should absorb
most of the collinear logarithms due to the light bottom quark
mass into the PDF of the bottom quark.
 In this respect we warn the reader
that the convenience of the use of bottom-quark PDFs to
appropriately describe Higgs boson production is not entirely
settled down yet~\cite{spira_privat,SpiraHouches}. This is
specially so in the case of neutral Higgs boson
production~(\ref{hbb}), where the result of adding up the $2\to2$
with the $2\to3$ process yields a cross-section an order of magnitude larger
than the leading order $2\to3$ alone~\cite{HiggsRunII,SpiraHouches}, thus giving a strong hint
that the bottom-quark PDF description does not approximate well
the full result. Third, for the charged Higgs boson production
cross-section~(\ref{tbh}) the situation is different, since the
result of the combination of processes~(\ref{gg-tbh}) and
(\ref{gb-th}) gives a cross-section of the same order of
magnitude than that of process~(\ref{gg-tbh}) alone -- see e.g.
Tables~\ref{tab:signal_tev} and~\ref{tab:signal_lhc} below. So it
is fair to think that the combination of subprocesses~(\ref{gg-tbh}) and
(\ref{gb-th}) gives a better approximation to
the charged Higgs boson production process than the computation
based on using~(\ref{gg-tbh}) only. Still, one
should keep in mind the situation of the neutral Higgs bosons,
whose resolution might  lead to a new and
interesting description of these processes \cite{spira_privat}. Aside
from the collinear logarithms, other
contributions to the NLO corrections exist. For example, in
Ref.~\cite{shouhua} the standard NLO QCD corrections to the
subprocess~(\ref{gb-th}) at the LHC are computed, obtaining a large
K-factor between $\sim1.6$ and $\sim 1.8$ for $\tb\gsim 20$.
Taking into account these considerations, a large $K$-factor for
the full process (\ref{tbh}) coming from the standard QCD corrections
 (e.g. $K^{\rm QCD}\simeq 1.5$) is not ruled out.

{On account of the previous
discussion, we will 
make use of pole quark masses
throughout our study. The use of the running quark mass would be
justified only if we would have good control on the value of the
remaining QCD corrections. It is well known that in decay
processes, like for instance }$t\rightarrow bH^{+}${, the use of
the bottom quark running mass }$\overline{m}_{b}(Q)$
{actually accounts for most of the  QCD virtual effects (at }$Q=%
\mt ${)~\cite{QCDtbH}. But in production processes the QCD
corrections cannot be parametrized in this way. Indeed, a clear
hint of the inappropriateness of this description ensues from the
fact that one expects a large 
$K$-factor
increasing the  cross-section, whereas the use of the running
masses would imply a  negative correction on the
cross-section. For this reason we prefer to use the pole masses
in our calculation, and fully parameterize our ignorance of the
QCD corrections by means of a }$K^{\rm QCD}${-factor }
 whose precise value will be easily incorporated
once it will be known in the future.

 After some digression let us come back to the
general description of the cross-section computation for the
process (\ref{tbh}).
 Once a PDF for $b$-quarks is used, there is some amount of overlap between $%
\gb$- and $gg$-initiated amplitudes, which has to be removed~\cite
{Olness}. The overlap arises because the $b$-density in the $\gb$
amplitude receives contributions from gluon splitting which was already
counted in the $gg$ amplitude (see e.g. the last diagram in
Fig.~\ref{fig:diagr}a), so 
we have to avoid double counting
by  the subtracting of the gluon splitting term. 
The net partonic cross-section from the $\gb$- and $gg$-initiated
subprocesses is
\begin{align}
\sigma (\gb+gg\rightarrow H^{+}\bar{t}+X)_{net}=& \sigma (g\bar{b}%
\rightarrow H^{+}\bar{t})+\sigma (gg\rightarrow H^{+}\bar{t}b)  \notag \\
& -\sigma (g\rightarrow b\bar{b}\otimes g\bar{b}\rightarrow H^{+}\bar{t}).
\label{netgb}
\end{align}
Recalling that the $b$-density evolution equation is given by the standard
formula \cite{AP}
\begin{equation}
\frac{d}{d\ln Q^{2}}f_{b/h}(x,Q^{2})=\frac{\alpha _{S}(Q^{2})}{2\pi }%
\,\int_{x}^{1}\frac{dz}{z}\,\left\{ P_{b\leftarrow b}(z)\,f_{b/h}(\frac{x}{z}%
,Q^{2})+P_{b\leftarrow g}(z)\,f_{g/h}(\frac{x}{z},Q^{2})\right\}
\label{AP/b}
\end{equation}
{where $f_{i/h}(x,Q^{2})$ is the PDF for the parton $i$ carrying
  a momentum  fraction $x$ at scale $Q^{2}$ to be found in the hadron $h$}.
The subtraction term can be approximated by integrating the gluon-splitting
part $g\rightarrow $ $b$ of this equation at the leading order in $\alpha
_{S}(Q^{2})$. Therefore the modified $b$-density is 
  found  to be
\begin{equation}
{\tilde{f}_{b/h}}(x,Q^{2})=\frac{\alpha _{S}(Q^{2})}{2\,\pi }\ln \left(
\frac{Q^{2}}{m_{b}^{2}}\right) \int_{x}^{1}\frac{dz}{z}\left[ \frac{%
z^{2}+(1-z)^{2}}{2}\right] \,f_{g/h}\left( \frac{x}{z},Q^{2}\right) \,\,.
\label{modifiedPDF}
\end{equation}
where $P_{b\leftarrow g}(z)=(1/2)$ $\left[ z^{2}+(1-z)^{2}\right] $ is the
gluon splitting function. Thus the {}modified $b$-density{} contains the
collinear logarithm and splitting function $P_{b\leftarrow g}$ convoluted
with the $g$-density. Accordingly the expression for the subtraction term in
({\ref{netgb}}) is
\begin{align}
\mbox{Subt.term}(AB& \rightarrow H^{+}\bar{t}b)=  \notag \\
& \int dx_{1}dx_{2}\left\{
\begin{array}{c}
{\tilde{f}_{b/A}}(x_{1},Q^{2})f_{g/B}(x_{2},Q^{2}){\sigma }(\bar{b}%
g\rightarrow \bar{t}H^{+}) \\
+{f_{g/A}}(x_{1},Q^{2}){\tilde{f}_{b/B}}(x_{2},Q^{2}){\sigma }(g\bar{b}%
\rightarrow \bar{t}H^{+})
\end{array}
\right\}  \label{subtract}
\end{align}
The subtraction
term $g\rightarrow b\bar{b}\otimes \gb\rightarrow \bar{t}H^{+}$ is not
negligible at all as it involves a leading log. We shall see explicitly its
numerical significance in the next section.

The hadronic  cross-section (\ref{tbh}) at LO in QCD is obtained by
convoluting in the usual way the tree-level partonic cross-sections with the
partonic densities of the two colliding hadrons: specifically $\sigma (q%
\overline{q}\rightarrow H^{+}\overline{t}b)$ must be functionally convoluted
with the light quark densities, also the first two terms on the 
right-hand-side (RHS) 
of (%
{\ref{netgb}}) are convoluted with the $g$ and $b$~-densities, 
and
finally one sums these results and subtracts the term
(\ref{subtract}). It is seen that all
these contributions to the cross-section are effectively of leading order $%
\alpha _{S}^{2}$. Moreover, from the previous considerations we
will apply in the end an overall QCD \mbox{$K$-factor} to account for the
NLO (or higher) effects. For definiteness we will take
 $K^{\rm QCD}\simeq 1.5$ whenever a QCD $K$-factor
is invoked. Once the exact value of $K$ will be known, it will be
easy to  rescale our plots to take into account the effect of the ({%
process-dependent)} gluon loops.

Also, for the study of the various differential distributions 
one has to
properly combine the $2\rightarrow2$ and $2\rightarrow3$ processes in order
to reproduce not only the total cross-section but also the \textit{correct
event kinematics.} The point is that we know the \textit{total amount} of
double counting but not a priori which part of this value should be
subtracted from the $H^{+}\bar{t}$ process and which part from the $H^{+}%
\bar{t}b$ one. We apply here the method proposed in~\cite{TW} for the
analogous process of the single top quark production. According to this
method we use the cut on the transverse momenta of the $b$-quark associated
with charged Higgs boson production to separate and recombine $gg$- and $\gb$%
-initiated processes. We will discuss this in detail in the next section. As
a result we obtain the correct combination of kinematical distributions. To
the best of our knowledge this has not yet been done in charged Higgs boson
study searches in the literature.

Let us next briefly comment on the relevant MSSM quantum effects for the
process~(\ref{tbh}). A more detailed discussion is given in
Section~\ref{sec:SUSYcorr}. 
Amplitudes that are proportional to the bottom quark Yukawa coupling receive
supersymmetric quantum corrections that can be very sizeable for large \tb\
values~\cite{Dmb}. This feature has been exploited for phenomenological
applications to the physics of the on-shell $tbH^{+}$ vertex in Ref.~\cite
{SUSYtbH}. However, it can also be important for the corresponding off-shell
vertex involved in the production process (\ref{tbh}), as first pointed out
in~\cite{Coarasa} where a first estimation was attempted. For $\tb\gsim\sqrt{%
\mt/\mb}$ process (\ref{tbh}) is dominated by the piece proportional to the
bottom quark Yukawa coupling. Moreover, corrections of order 
$(\alpha/4\pi )^{n}\tb[n]$, where $\alpha =\alpha _{S},\alpha _{W}$,
appear at $n$-th order of 
perturbation theory~\cite{eff}. They can all be summed up in the definition
of the renormalized bottom Yukawa coupling,
\begin{equation}
h_{b}=\frac{\mb}{v_{1}}\frac{1}{1+\Dmb}\rightarrow \frac{\mb}{v}\frac{1}{1+%
\Dmb}\,\tan \beta  \label{hb}
\end{equation}
where $v_{1}=v\,\cos \beta $ and so the last expression is valid only for $%
\tan \beta \gg 1$, showing the prominence of the large $\tan \beta $ region
for our process. The quantity \Dmb\ in~eq.(\ref{hb}) is driven by both
strong (SUSY-QCD) and electroweak (SUSY-EW) supersymmetric effects
(depending on whether $\alpha =\alpha _{S}$ or $\alpha =\alpha _{W}$) and it
increases linearly with $\tan \beta $ (Cf. Section~\ref{sec:SUSYcorr}), therefore reaching
values that can be of order $1$. As a result any realistic analysis of the
reach of process (\ref{tbh}) in $H^{\pm }$ searches clearly demands for the
appropriate inclusion of these dominant SUSY radiative corrections.\footnote{%
For the impact of this type of effects on the determination of the exclusion
plot in $(M_{H^{+}},\tan \beta )$ space at the Tevatron, see~\cite{tbmh}.}
Incidentally, let us recall that at present large \tb\ scenarios, such as
those derived from supersymmetric $SO(10)$ models with unification of the
top and bottom Yukawa couplings at high energies~\cite{Dmb,gut}, have become
more and more appealing since LEP searches for a light neutral Higgs boson, $%
h$, started to exclude the low-\tb\ region of the MSSM parameter space. 
The
latest analyses rule out the MSSM for \tb\ in the range 
$0.5<\tb<2.4$,
even with maximal stop mixing~\cite{LEPtb}.

We have already argued above that the charged Higgs boson production
process (\ref {tbh}) is expected to have large, and positive, QCD
corrections. This circumstance, together with the additional
cooperation from the SUSY corrections themselves, which can also
be large and positive, could be crucial
  to generate a substantial
``effective $K$-factor'' in the MSSM context :
\begin{equation}\label{KMSSM}
K^{\rm MSSM}=K^{\rm QCD}\,K^{\rm SUSY}\,.
\end{equation}
For $K^{\rm MSSM}\gtrsim 2$ (e.g. through $K^{\rm QCD}=1.4$ and
$K^{\rm SUSY}= 1.5$ ) the opportunity to chase a MSSM Higgs boson
already at the Tevatron would be open, while a more detailed
analysis could be performed at the LHC (see Section~\ref{sec:SUSYcorr}).

Concerning the method employed to compute the squared matrix elements, we
have made intensive use of the CompHEP package~\cite{CompHEP}, for both the
signal and background processes. Although CompHEP is only able (in
principle) to deal with tree-level calculations, we have managed to add the
supersymmetric corrections to the \tbH\ vertex and fermion propagators and
we have assessed the relevance of the off-shell contributions. The
importance and significance of the various corrections is explained in
Section~\ref{sec:SUSYcorr} where its numerical impact is evaluated for the signal (\ref{tbh}%
).  In the meanwhile, since the
background processes are completely insensitive to the leading
SUSY effects of the type we have mentioned, in the next section
we present a detailed signal versus background analysis in which
the cross-sections for all processes are computed 
 without supersymmetric corrections.

\section{Signal and background study}

\label{sect:svb}

\subsection{Preview}
\label{sec:sbspreview}

{We are interested in} the search for a heavy charged Higgs boson
with a mass larger than the top-quark mass. The case where the
$t\rightarrow bH^{+}$ decay is allowed has already been well
investigated for the upgraded
Tevatron~{\cite{HiggsRunII,Guchait,tev1-t-hb,tev2-t-hb}}. We focus on
the $t\bar{t}b\bar{b}$ signal
signature corresponding to the $H^{+}\rightarrow t\bar{b}$ Higgs boson
decay channel with the {largest} branching ratio. First we check
the potential of the Tevatron collider for the heavy Higgs boson
search, then we consider the LHC, whose detection region for
$\mH>\mt$ will obviously be enlarged~\cite{tev3-t-hb}.

Channels (\ref{gg-tbh}) and (\ref{gb-th}) have been studied in~\cite
{Gunion,Barger} for the  triple-$b$-tagging. Recently, channel (\ref
{gg-tbh}) has been considered for the four-$b$-tagging case~\cite{DP-4b}.
The four-$b$-tagging search improves the signal/background ratio but at the
cost of signal rate and significance. The study of channel (\ref{gb-th}) has
also been extended recently (for which both cross-sections (\ref{gb-th}) and
(\ref{gg-tbh}) have been combined) for the triple-$b$-tagging case~\cite
{DP-3b}. One should note the additional channels that have been suggested
recently for the charged Higgs boson search. In particular, the $H^{\pm
}W^{\mp}$~\cite{ppHW} and $H^{\pm}H^{\mp}$~\cite{ppHH} production modes
followed by
the decay $\Hpm\rightarrow hW^{\pm}$ (where $h=h^{0},H^{0},A^{0}$ is any of
the neutral MSSM Higgs boson allowed by phase space). However, none of these
modes is favoured at high $\tan\beta$ and therefore will not be considered
in our study.

{In this paper we consider the dominant} $t\bar{t}b\bar{b}$ signature for
the combined signal process~(\ref{tbh}) in detail and at a more realistic level than
it was done previously. {We focus here on} the triple-$b$-tagging case,
which gives the best possibility to measure the signal cross-section. This
is the crucial point, especially for the Tevatron, where the production rate
is too small to give any viable signal in the case of the four-$b$-tagging.
In the case of the LHC, a triple-$b$-tagging study allows the signal
cross-section to be measured more precisely ({as we shall show below}), even
though the signal/background ratio {can be} better for the four-$b$-tagging
case. Let us stress the following points before presenting the results of
our analysis:

\begin{itemize}
\item  To study the $t\bar{t}b\bar{b}$ signature, it is important to take
into account the effects of string fragmentation, initial- and final-state
radiation.

\item  As warned before, the processes (\ref{gg-tbh}) and (\ref{gb-th})
should be properly combined. This is important because in the high $%
p_{T}^{b} $ region those processes are qualitatively different (contrary to
the low-$p_{T}^{b} $ region). Of those events produced via $gg\rightarrow%
\tbH
$, only a $25(40)\% $ survive a $p_{T}^{b}>30\GeV$ cut 
 at the Tevatron
(LHC). However one should note that the efficiency of $b$-tagging is $p_{T}
$-dependent. 
After the $b$-tagging, the percentage of $gg $-initiated events
       under the  $p_{T}^{b}>30\GeV$ cut roughly doubles. That is why
       $p_{T} $-dependent $b$-tagging efficiency was used in our study.

\item  We present signal and background cross-sections and give the final
results in terms of the signal and background efficiencies. Based on this
information one can derive the reach of the Tevatron and LHC for a given
integrated luminosity.

\item  Another aspect of this work is the determination of the accuracy of
the signal cross-section measurement, which is the crucial point for the
measurement of the $H^{+}\bar{t}b$ coupling. This is very important in order
to understand how the signal cross-section is affected by SUSY corrections
in various regions of the parameter space. {Certainly this is the most
ambitious goal of the present study.}
\end{itemize}

\subsection{Signal and background rates}

For the signal and background  calculation, we use $\mt=175\GeV$, $m_{b}=4.6%
\GeV$, $\tb=50$, and the CTEQ4L set of PDFs~\cite{CTEQ4}. {Here }$%
m_{t},m_b$ {refer to the quark pole masses.} {Unless stated
otherwise, we assume }  $K^{\rm QCD}=1$ {to
compare signal and background rates. }Tables~\ref{tab:signal_tev}
and~\ref{tab:signal_lhc} present the tree-level total signal
rates as well as the rates for the subprocesses and the
subtraction term for the $2\TeV$ Tevatron collider and $14\TeV$
LHC
collider respectively. The total signal rates are illustrated also in Fig.~%
\ref{fig:tot-rate}. {Note
  that the computation using only the $\gb\rightarrow\bar{t}H^{+}$
  channel as done in~\cite{MoriondHtaunu} overestimates the cross-section by $\sim25 \%$ at the LHC.} The
subtraction term $g\rightarrow b\bar{b}\otimes \gb\rightarrow \bar{t}H^{+}$
is indeed not negligible as it involves a leading log, see eqs.(\ref
{modifiedPDF},\ref{subtract}).

\TABULAR[t]{|l|l|l|l|l|l|} {
\hline
$M_{A}\, (M_{H^{+}}) $ & $\gb\rightarrow\bar{t}H^{+} $ (fb) & $gg\rightarrow %
\tbH$ (fb) & Subt.term (fb) & $q\bar{q}\rightarrow\tbH$ (fb) & Total(fb) \\
\hline
200 (215) & 3.20 & 1.02 & 1.86 & 3.86 & 6.22 \\
250 (263) & 1.37 & 0.409 & 0.758 & 1.05 & 2.07 \\
300 (310) & 0.587 & 0.166 & 0.311 & 0.353 & 0.795 \\
350 (359) & 0.253 & 0.0688 & 0.128 & 0.125 & 0.319 \\
400 (408) & 0.110 & 0.0280 & 0.0531 & 0.0466 & 0.131 \\
500 (506) & 0.0209 & 0.00470 & 0.00914 & 0.00689 & 0.0234 \\
600 (605) & 0.00398 & 0.000778 & 0.00154 & 0.00105 & 0.00428 \\ \hline
} {Tree-level signal rates at the Tevatron ($2
\TeV$) for various subprocesses and at fixed
$\tan\protect\beta=50$,  $K^{\rm QCD}=1$.\label{tab:signal_tev} }

\TABULAR[t]{|l|l|l|l|l|l|} {
\hline
$M_{A}\, (M_{H^{+}}) $ & $\gb\rightarrow\bar{t}H^{+} $ (pb) & $gg\rightarrow %
\tbH$ (pb) & Subt.term (pb) & $q\bar{q}\rightarrow\tbH$ (pb) & Total(pb) \\
\hline
200 (215) & 5.55 & 3.03 & 4.22 & 0.101 & 4.46 \\
300 (310) & 2.53 & 1.30 & 1.83 & 0.0223 & 2.02 \\
400 (408) & 1.22 & 0.594 & 0.847 & 0.00742 & 0.973 \\
500 (506) & 0.625 & 0.294 & 0.422 & 0.00296 & 0.500 \\
600 (605) & 0.344 & 0.158 & 0.222 & 0.00133 & 0.281 \\
700 (704) & 0.194 & 0.0873 & 0.123 & 0.000641 & 0.159 \\
800 (804) & 0.114 & 0.0498 & 0.0705 & 0.000328 & 0.0938 \\ \hline
}{As in Table~\ref{tab:signal_tev} but for the LHC ($14 \TeV$).
\label{tab:signal_lhc}}

\FIGURE[t]{
\mbox{\resizebox*{0.5\textwidth}{0.4\textheight}{%
\includegraphics{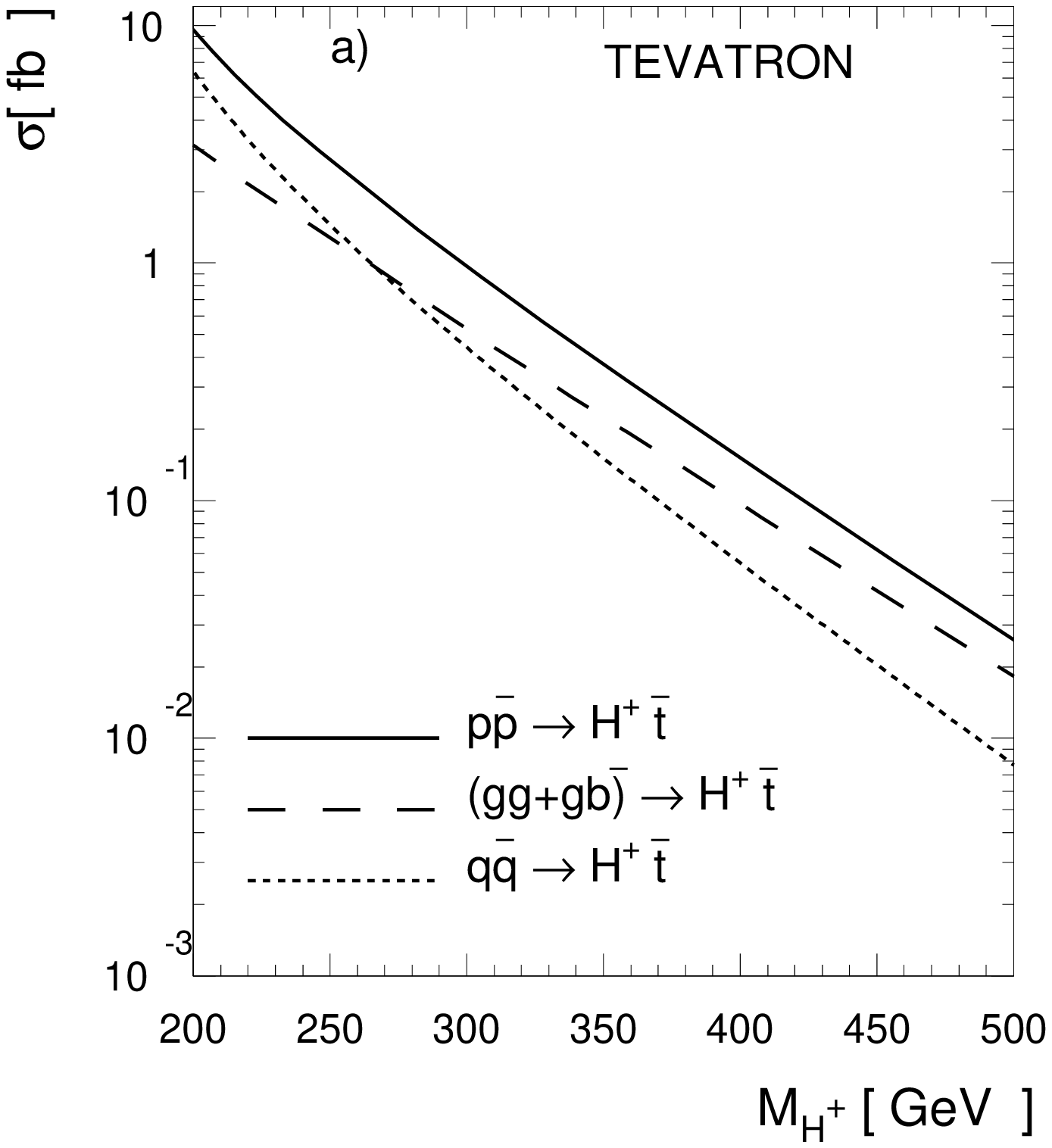}}
\resizebox*{0.5\textwidth}{0.4\textheight}{\includegraphics
{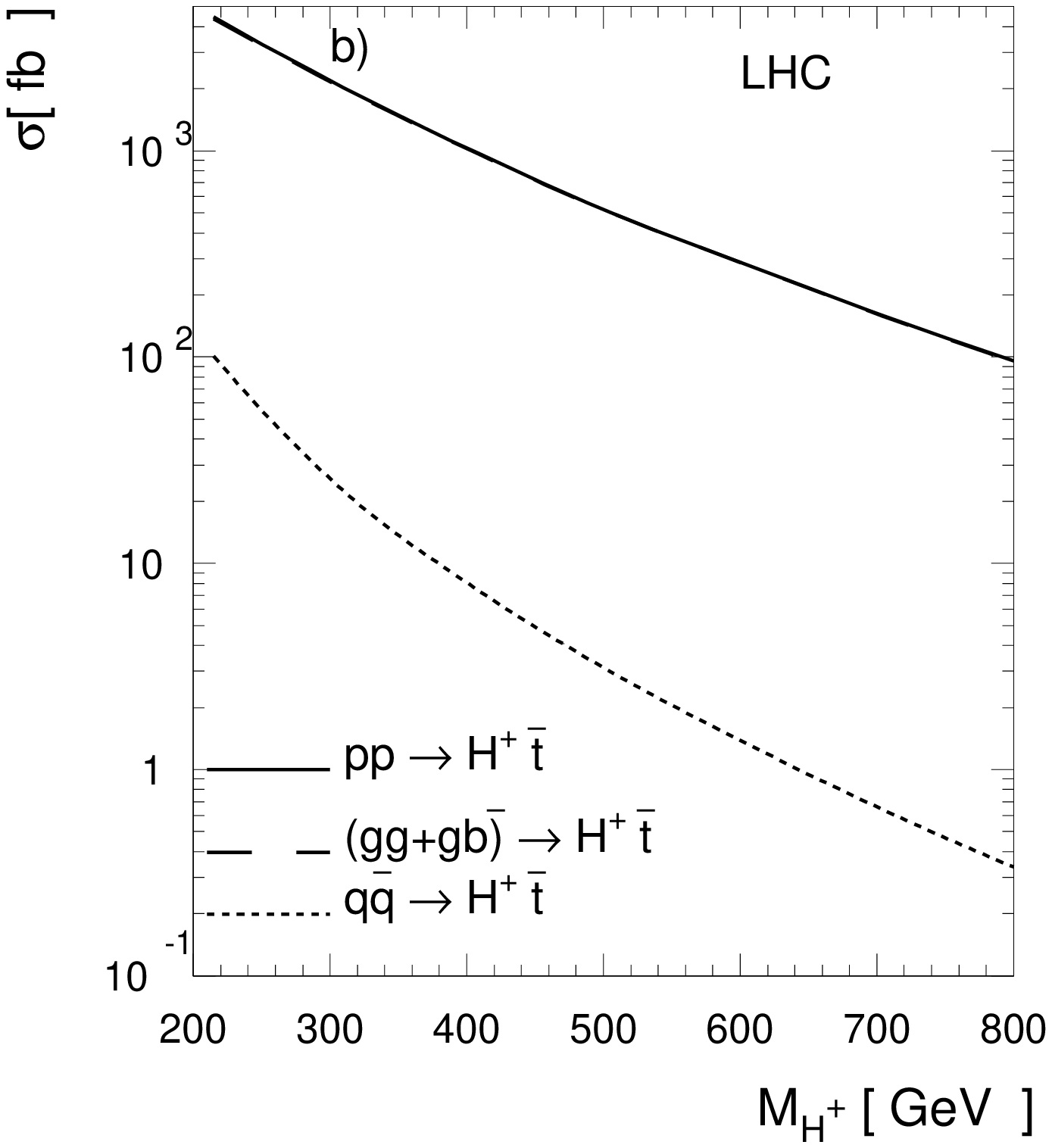}} } \caption{
{Signal rates as a function of the charged Higgs boson mass at \textbf{a)} the
Tevatron and \textbf{b)} the LHC (since the gluon initiated
processes~(\ref{gg-tbh}), (\ref{gb-th}) dominate  at the LHC,  the solid and
dashed lines are indistinguishable in this plot), for $\tb=50$,
$K^{\rm QCD}=1$.}} \label{fig:tot-rate}
}

The total cross-section is given only for the $H^{+}$ production; the
inclusion of the $H^{-}$ channel is just given by twice the result displayed
in tables and figures. In Fig.~\ref{fig:tot-rate} one can see that the signal rate is roughly $3$
orders of magnitude higher at the LHC than at the Tevatron. Also, the
production rate at the LHC drops with the mass slower than it does at the
Tevatron. For $\tan\beta=50$, $\mH=215(359)\GeV$ and $25\fb^{-1}$
integrated luminosity --- 174(8) $%
H^{+}$ bosons are produced at the Tevatron. For the same value of \tb, $%
M_{H^{+}}=200(600)\GeV$ and an integrated luminosity of $100\fb^{-1}$ ---
530k(32k) $H^{+}$ bosons
 are expected at the LHC. {The dependence of the signal cross-section with $%
\tan\beta $ can be appreciated in Fig.~\ref{fig:tree-tanb} for both the
Tevatron (at fixed $M_{H^{+}}=250\GeV$) and at the LHC ($M_{H^{+}}=500\GeV$%
). It is obvious that the signal remains negligible in the (allowed) low $%
\tan\beta$ segment, say $2<\tan\beta\lesssim20$, 
 as compared to its value for $\tan\beta\gtrsim 30$.}

\FIGURE[t]{
\begin{tabular}{cc}
\resizebox{!}{6cm}{\includegraphics{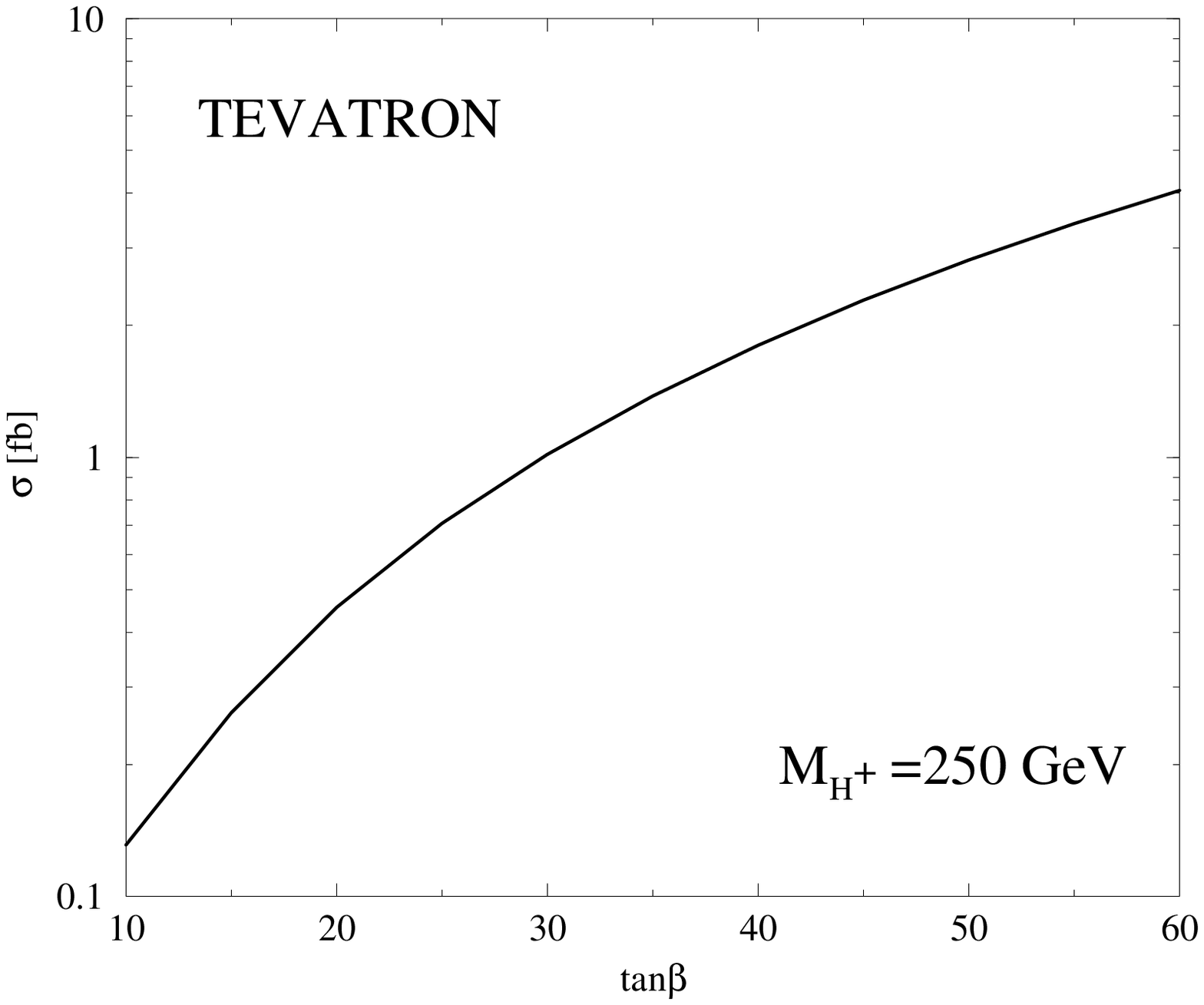}} & %
\resizebox{!}{6cm}{\includegraphics{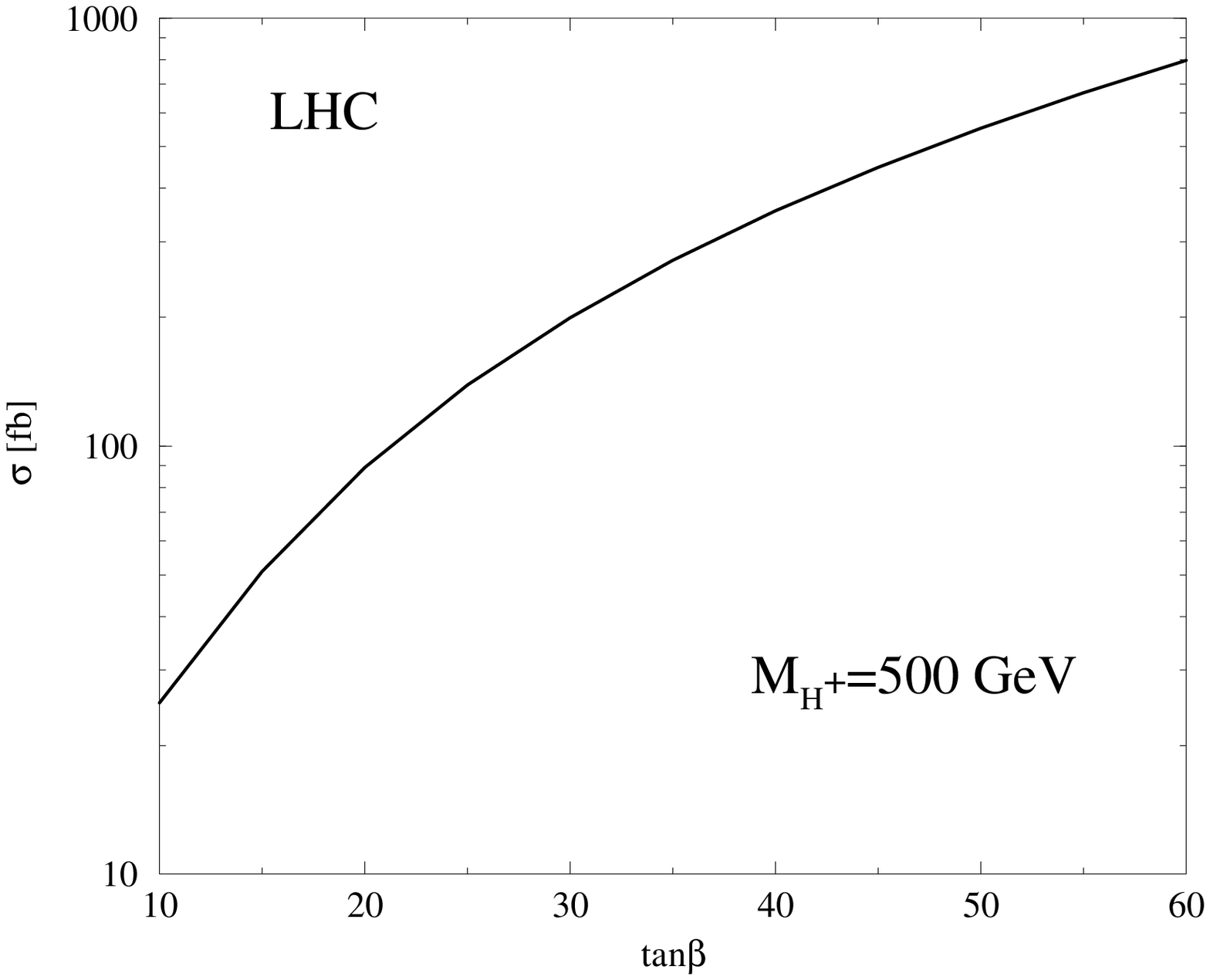}} \\
(a) & (b)
\end{tabular}
\caption{
{Signal rates as a function of \tb\ for
\textbf{a)} the Tevatron with $\mH=250\GeV$, and \textbf{b)} the
LHC with $\mH=500\GeV$,  $K^{\rm QCD}=1$. }}
\label{fig:tree-tanb}
}

{We have checked the uncertainty of the signal due to the
  choice of PDF sets. We have compared the results in
  Tables~\ref{tab:signal_tev} 
  and \ref{tab:signal_lhc} with the ones obtained with the MRST
  (central gluon) PDFs~\cite{MRST}. The results show a large
  deviation for some of the individual sub-channels (up to $\sim 50\%$
  deviation), but they are compensated in the sum, leaving a $5-10\%$
  uncertainty on the total cross-section. For example, for
  $\mH=215\GeV$ ($M_{A}=200\GeV$) we obtain the following set of cross-sections:
  $(\gb,gg,\textrm{Subt.term}, q\bar{q}, \textrm{Total})=(3.80,1.63, 
  2.95 , 4.52, 7.01) \fb$ at the Tevatron and $(
  6.54, 4.2, 5.68 , 0.136, 5.19)\pb$ at the LHC. These values correspond to an
  uncertainty in the cross-section of $(17, 46, 45,
  15, 11) \%$ at the Tevatron and $(8, 16, 15,
  15, 7.5)\%$ at the LHC.
  The substraction
  procedure plays a key role in reducing the uncertainty, since there
  is a compensation between the variation of the $gg$ and $\gb$ channels
  and the substraction term. This reduction is fully effective at the
  LHC. At the Tevatron the  compensation between the \textit{gluonic}
  channels also takes place, but the uncertainty is increased by that of
  the $q\bar{q}$-channel, which plays a key role at that machine.
  The cross-sections
  obtained using the MRST PDFs are usually larger than the ones of the
  CTEQ4L. For $\mH>300\GeV$ at the Tevatron they become smaller, and for
  $\mH=500\GeV$ the MRST prediction is a $25\%$ smaller than that of
  the CTEQ4L.}

{We have also checked the uncertainty in the signal due to
  the choice of renormalization and factorization scales (assumed both
  equal to $\mu_{R}$). 
  We have analyzed the dependence on $\mu_{R}$ inside the interval
  $\mH/2<\mu_{R}<2 \mH$. It is obvious that a larger value of $\mu_{R}$
  decreases the signal cross-section. Again, individual
  sub-channels show a stronger dependence than the total cross-section.
  The relative uncertainty is very weakly dependent on the charged
  Higgs boson mass. For the total cross-section it is around $\sim 28\%$
  at the Tevatron and $\sim 18\%$ at the LHC. For $\mH=215\GeV$
  ($M_{A}=200\GeV$)  we find
  the following sets of cross-sections   for $\mu_{R}=\mH/2$: 
  $( 4.07,   1.73,  2.52, 5.63,  8.91)\fb$ at the Tevatron; and 
  $(  5.91,  4.02, 4.66,  0.136,   5.41)\pb$ at the LHC. For $\mu_{R}=2
  \mH$ the results are: $( 2.68, 0.678, 1.44, 2.89, 4.81)\fb$ at the
  Tevatron and $( 5.38, 2.35, 3.87, 0.0895,  3.96)\pb$ at the LHC. The
  background channels in Table~\ref{tab:back} also present uncertainties
  due to the choice of PDFs and $\mu_{R}$.
{The central value of the $\mu_R$ scale for the
signal processes has been chosen equal to $\mH$, whereas that
of the background processes to $(2\mt)$.}
  } 

As we have already mentioned we will focus on the
$t\bar{t}b\bar{b}$ signature. We consider the case where one top
decays hadronically and the other leptonically (including
 \textit{only} electron and muon decay {channels})
in order to reduce the combinatorics when both top quarks are
reconstructed. The branching
ratio of $t\bar{t}b\bar{b}\rightarrow b\bar{b}b\bar{b}\ell^{\pm}\nu q\bar{%
q^{\prime}}$ is $2/9\times2/3\times2=8/27$.

In order to decide whether a charged Higgs boson cross-section leads to a
detectable signal, we have to compute the background rate. Since the
miss-tagging probability of light quark and gluon jets is expected to be $%
\lsim1\%$~\cite{d0tag,cmstag}, the main backgrounds leading to the same $t%
\bar{t}b\bar{b} $ signature and their respective cross-sections are {those}
shown in Table~\ref{tab:back}.

\TABLE[t]{\parbox{\textwidth}{ 
\rule{\textwidth}{.5pt}
$\mbox{(a) \ \ } p\bar{p}(pp)\rightarrow t\bar{t}b(\bar{b})$
\\\\
\begin{tabular}{crcl}
\multicolumn{4}{l}{Tevatron} \\
& $\sigma(qq\rightarrow t\bar{t}b\bar{b}) $ & = & 6.62 fb \\
& $\sigma(gg\rightarrow t\bar{t}b\bar{b}) $ & = & 0.676 fb \\
& $\sigma(gb\rightarrow t\bar{t}b) $ & = & 1.22 fb \\
& Subtr. term & : & 0.72 fb \\
& Total & : & 7.80 fb\\
\end{tabular}
\newline
\begin{tabular}{crcl}
\multicolumn{4}{l}{LHC} \\
& $\sigma(qq\rightarrow t\bar{t}b\bar{b}) $ & = & 0.266 pb \\
& $\sigma(gg\rightarrow t\bar{t}b\bar{b}) $ & = & 6.00 pb \\
& $\sigma(gb\rightarrow t\bar{t}b) $ & = & 4.33 pb \\
& Subtr. term & : & 2.1 pb \\
& Total & : & 8.50 pb
\end{tabular}
\newline
\rule{\textwidth}{.5pt}
$
\mbox{(b) \ \ } p\bar{p}(pp)\rightarrow t\bar{t}q{g} \mbox { \ \ when the
light quark or gluon are misidentified as a \( b \)-jet}$
\\\\
\begin{tabular}{crcl}
\multicolumn{4}{l}{Tevatron} \\
& $\sigma(q\bar{q}\rightarrow g\bar{t}\bar{t}) $ & = & 1890 fb \\
& $\sigma(gq\rightarrow qt\bar{t}) $ & = & 193 fb \\
& $\sigma(gg\rightarrow g\bar{t}\bar{t}) $ & = & 262 fb \\
& Total & : & 2345 fb
\end{tabular}
\begin{tabular}{crcl}
\multicolumn{4}{l}{LHC} \\
& $\sigma(q\bar{q}\rightarrow g\bar{t}\bar{t}) $ & = & 21 pb \\
& $\sigma(gq\rightarrow qt\bar{t}) $ & = & 122 pb \\
& $\sigma(gg\rightarrow g\bar{t}\bar{t}) $ & = & 371 pb \\
& Total & : & 514 pb
\end{tabular}
\newline
\rule{\textwidth}{.5pt}}%
\caption{
{The main background processes for the charged Higgs boson
production at the Tevatron and LHC. For the $t\bar{t}b\bar{b}$ and $t\bar t
qg$ processes we have applied the jet separation cut $\Delta_{R}^{jj}>0.5(%
\Delta_{R}=\protect\sqrt{\Delta\protect\theta^{2}+\Delta\protect\phi^{2}})$
and the cut $p_{T}^{j}>10\GeV$ ($p_{T}^{j}>20\GeV$) at the Tevatron (LHC).
For the $t\bar{t}j$ process the cut $p_{T}^{j}>10\GeV$ ($p_{T}^{j}>20\GeV$)
was applied at the Tevatron (LHC).}}
\label{tab:back}
}%
The above notations $q\bar{q} $, $gq $, $gb $ for background processes
assume that we have summed over initial-state quarks and anti-quarks. One
should notice that the double counting, as in the signal case, appears also
when one sums the $gg\rightarrow t\bar{t}b\bar{b} $ and $gb\rightarrow t\bar{%
t}b $ processes. Therefore one should subtract the overlap
between them, which we denoted by Subtr. term --
Cf. eq.(\ref{subtract}). 

In considering our final state signature $t\bar{t}b\bar{b}$ we should point
out that the partial width of the decay mode $H^{+}\rightarrow t\bar{b}$ can
be itself sensitive to important SUSY radiative corrections~\cite{SUSYHtoTB}%
. However, in the present instance it is only the branching ratio of this
process that enters the calculation. At high \tb, the only relevant mode
other than $H^{+}\rightarrow t\bar{b}$ is $H^{+}\rightarrow\tau^{+}\nu_{%
\tau} $ and the latter is of order 10\% at most. Therefore $%
BR(H^{+}\rightarrow t\bar{b})$ is not  too
sensitive to SUSY effects. By the same token
$BR(H^{+}\rightarrow\tau^{+}\nu_{\tau})$, though smaller, can be
quite sensitive~\cite{SUSYHtoTB} {and so with sufficient
statistics it could be used as an additional test of the
underlying SUSY physics.} {We factorize these corrections from
the production process itself, and we will take them into account
only in the combination of the signal/background analysis and the
radiative corrections in Section~\ref{sec:consequences}.}

\subsection{Simulation details}

To perform a realistic signal and background event simulation we complied to
the following procedure. The matrix element for the complete set of signal
and background processes has been calculated using the CompHEP package~\cite
{CompHEP}. The next step was the parton-level event simulation, also with
the help of CompHEP. Then we automatically linked the parton-level events
from CompHEP to the PYTHIA6.1 Monte Carlo generator~\cite{pythia}, using the
CompHEP--PYTHIA interface~\cite{comp-pyth}.

Therefore we took into account the effects of the final-state radiation,
hadronization and string-jet fragmentation using PYTHIA tools. The following
resolutions were used for the jet and electron energy smearing: $\Delta
E^{had}/E=0.8/\sqrt{E} $ and $\Delta E^{ele}/E=0.2/\sqrt{E} $. In our
analysis we used the cone algorithm for the jet reconstruction with a cone
size $\Delta R=\sqrt{\Delta\varphi^{2}+\Delta\eta^{2}}=0.7 $. The choice of
this jet-cone value is related to the crucial role of the final-state
radiation (FSR), which strongly smears the shape of the reconstructed
charged Higgs boson mass. We have checked that the value of 0.7 minimizes
the FSR effects.

The minimum $E_{T} $ threshold for a cell to be considered as a jet
initiator was chosen to be $5\GeV$ ($10\GeV$) for the Tevatron (LHC), while
the minimum threshold for a collection of cells to be accepted as a jet was
chosen as $10\GeV$ and $20\GeV$, respectively for the Tevatron and the LHC.

As already noted, we require three $b $-jets to be tagged. A realistic
description of the $b $-tagging efficiency is therefore very important. In
the case of the Tevatron, we use the projected $b $-tagging efficiency of
the upgraded D\O\ detector~\cite{d0tag}:
\begin{equation}
\epsilon_{b}=0.57\cdot\tanh\left( \frac{p_{T}}{35\GeV}\right) ,
\label{btag-tev}
\end{equation}
For the LHC, we parameterize numerical results from the CMS collaboration
\cite{cmstag}:
\begin{equation}
\epsilon_{b}=\left\{
\begin{array}{ll}
0.6, & \, \, \, \, \, \, {\text{for}}\, \, p_{T}>100\GeV \\
0.1+p_{T}/(200\GeV), & \, \, \, \, \, \, {\text{for}}\, \, 40\GeV \leq
p_{T}\leq100\GeV \\
1.5p_{T}/(100\GeV)-0.3, & \, \, \, \, \, \, {\text{for}}\, \, 25\GeV\leq
p_{T}\leq40\GeV
\end{array}
\right.
\label{btag-lhc}
\end{equation}
We assume that $b $-jets can be tagged only for pseudorapidity $%
|\eta_{b}|\leq2 $ by both Tevatron and LHC experiments.

\subsection{Combining the $\gb\rightarrow H^{+}\bar{t}$ and $gg\rightarrow
H^{+}\bar{t}b$ processes}

As we mentioned above, we apply the recipe of {Ref.}~\cite{TW} to combine $%
\gb\rightarrow H^{+}\bar{t} $ and $gg\rightarrow H^{+}\bar{t}b $ in
order to get the correct overall distributions. 

The tree level  $2\to 2$ and $2\to 3$ processes  reproduce correctly 
the $p_T$ distribution of the associated $b$-quark ($p_T^b$) only  in
certain  parameter regions.  The ($p_T^b$) distribution of  $2\to 2$
(for which  the $b$-quark comes from initial state radiation simulated by
PYTHIA) is correct only for small $p_T^b$ values,  because  the gluon
splitting function is not able to reproduce high $p_T^b$ region
correctly. Contrary, the $2\to 3$ processes  reproduces correctly
the distribution  at high $p_T^b$ since it includes the complete set of
respective diagrams,  but it fails to reproduce the correct $p_T^b$
in the low $P_T$ region where one should take care of the resummation  of
large values of  $\log[M_{H^+}/M_b]$.

We have compared various
kinematical distributions of the $2\rightarrow3 $ process $pp(gg)\rightarrow
H^{+}\bar{t}b $ and $2\rightarrow2 $ process $pp(\gb)\rightarrow H^{+}\bar
{t}+b_{split}$. We have found a proper matching between the resummed
contribution in the collinear region for the $b $-quark and the complete
tree-level contribution in the hard region. Figure~\ref{fig:HT} shows
transverse momenta and rapidity distributions for the final state $H^{+} $, $%
t $-quark and $b $-quark for the processes at the Tevatron.
For the LHC the distributions are qualitatively the same.
As expected the difference is clear in the $b $%
-quark distributions. For the $pp(\gb)\rightarrow H^{+}\bar{t}+b_{split} $
process, the $b $-quark is softer and less central than that for the $%
pp(gg)\rightarrow H^{+}\bar{t}b $ process. At the same time it is shown that
the $H^{+} $-boson and $t $-quark distributions are nearly the same.

We use the method of matching collinear and hard kinematical regions based
on the kinematical $p_{T}^{b}$ separation of the $pp(\gb)\rightarrow
H^{+}\bar {t}+b_{split}$ and $pp(gg)\rightarrow H^{+}\bar{t}b$ processes in
the regions $p_{T}^{b}<p_{T}^{cut}$ and $p_{T}^{b}>p_{T}^{cut}$,
respectively. We search for the value of $p_{T}^{cut}$ in order to satisfy
two requirements, namely: \newline
1) the common rate of $pp(\gb)\rightarrow H^{+}\bar{t}+b_{split}$ with $%
p_{T}^{b}<p_{T}^{cut}$ and $pp(gg)\rightarrow H^{+}\bar{t}b$ with $%
p_{T}^{b}>p_{T}^{cut}$ gives the combined total rate computed in the
previous section; in other words, one can normalize a rate in a collinear
region on the $\sigma_{total}-\sigma\lbrack pp(gg+q\bar{q})\rightarrow
H^{+}\bar {t}b,\,\,p_{T}^{b}>p_{T}^{cut}]$; \newline
2) the overall $p_{T}^{b}$ distribution should be smooth. \newline
The result is illustrated in Fig.~\ref{fig:sew}, where we show several
variants of the combination of these two processes for various values of $%
p_{T}^{cut}$. 
We found that the optimal $p_{T}^{cut}$ providing a smooth
sewing for these two processes at the Tevatron and LHC is equal to about $25%
\GeV$. This value gives physically reasonable answers on the main questions
of this section:\newline
a) in which kinematical regions should the $pp(\gb)\rightarrow H^{+}\bar{t}%
+b_{split}$ and $pp(gg)\rightarrow H^{+}\bar{t}b$ processes be considered
and how one should properly simulate them;\newline
b) how the subtraction term is distributed between $pp(\gb)\rightarrow H^{+}%
\bar{t}+b_{split}$ and $pp(gg)\rightarrow H^{+}\bar{t}b$ processes, and what
part of double counting should be subtracted from each subprocess. 
One should notice that  in some  particular cases, like the one chosen 
for  illustration in  Fig.~\ref{fig:sew}, there is practically no
difference in choosing $p_{T}^{cut}$   in the range of  $25-40\GeV$.
This can be  seen from Fig.~\ref{fig:sew} as well as confirmed by
our numerical results for the final  efficiencies.

\noindent
\FIGURE[tbph]{
\parbox{\textwidth}{\mbox{\resizebox*{0.45\textwidth}{0.29\textheight}{\includegraphics{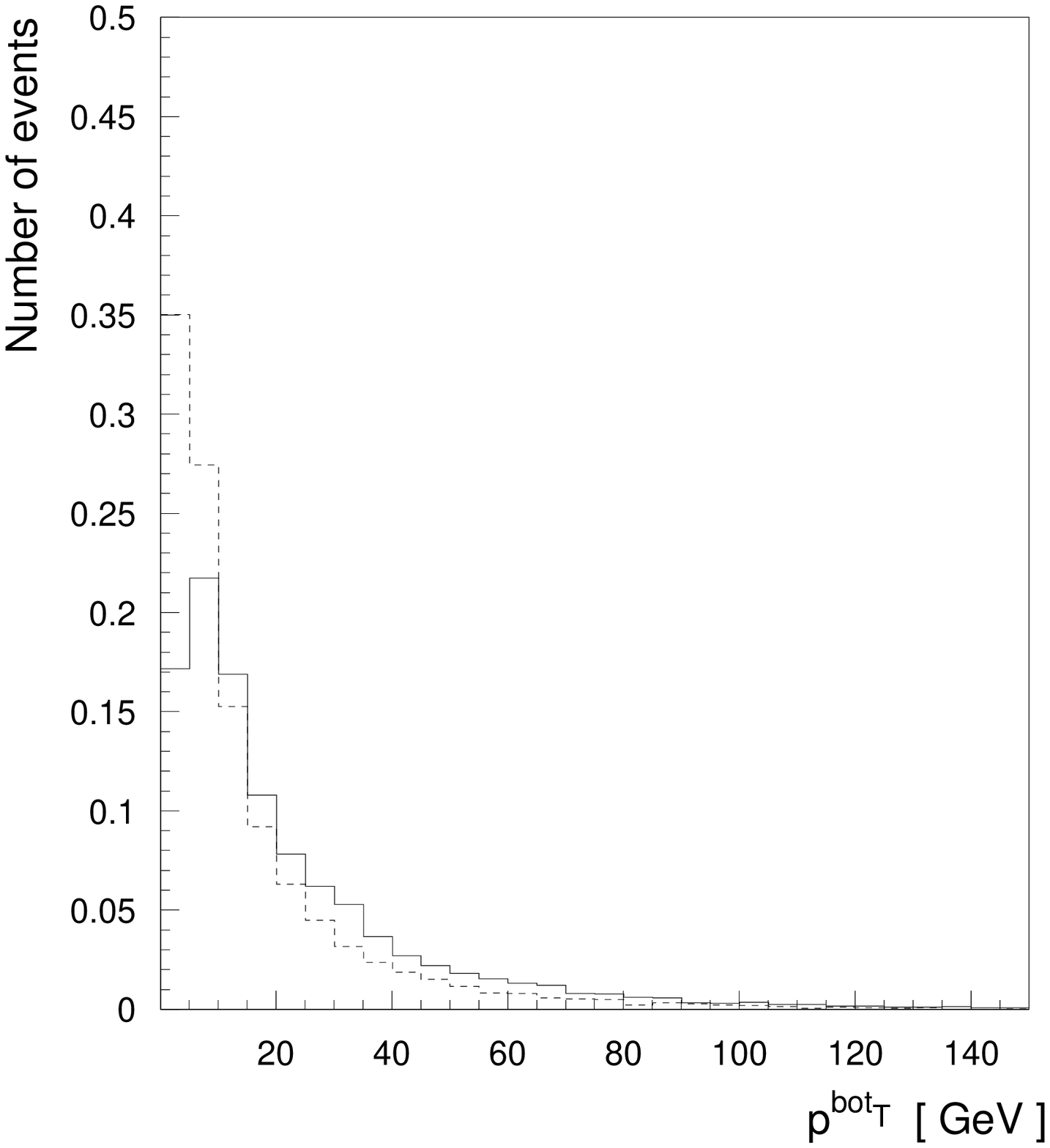}}
\resizebox*{0.45\textwidth}{0.29\textheight}{\includegraphics{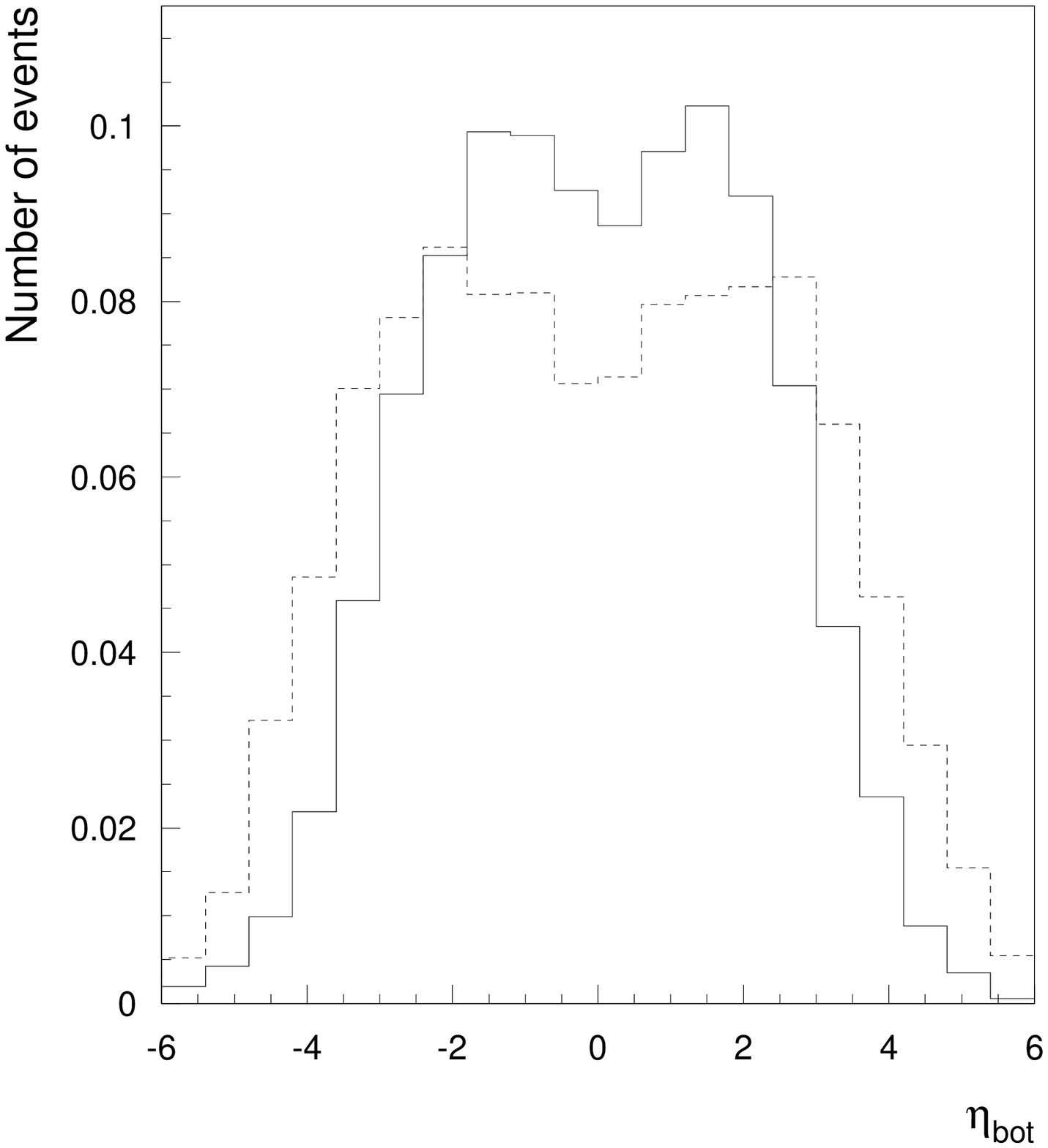}} }
\par
\vspace*{-0.6cm}
\par
\mbox{\resizebox*{0.45\textwidth}{0.29\textheight}{\includegraphics{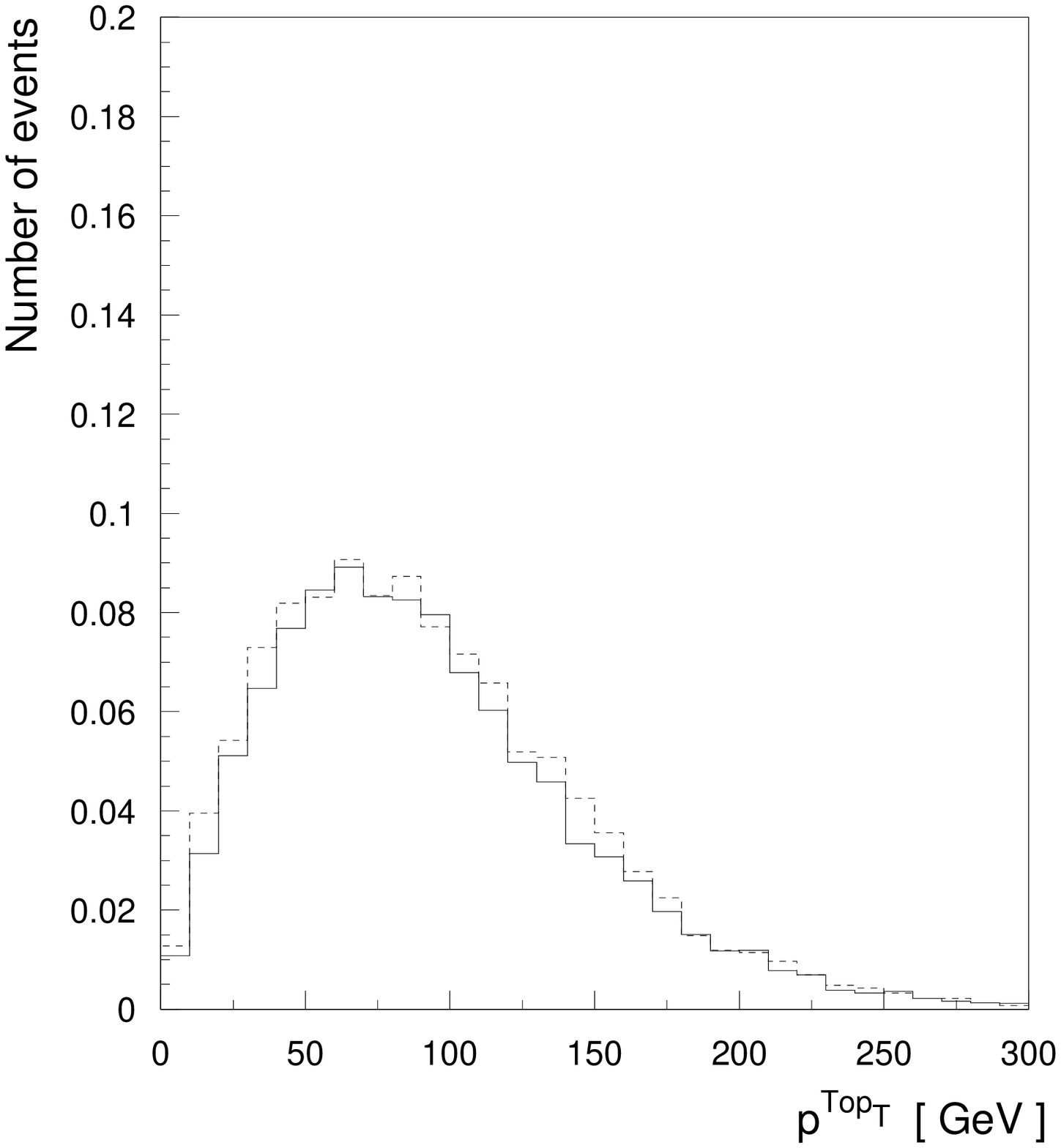}}
\resizebox*{0.45\textwidth}{0.29\textheight}{\includegraphics{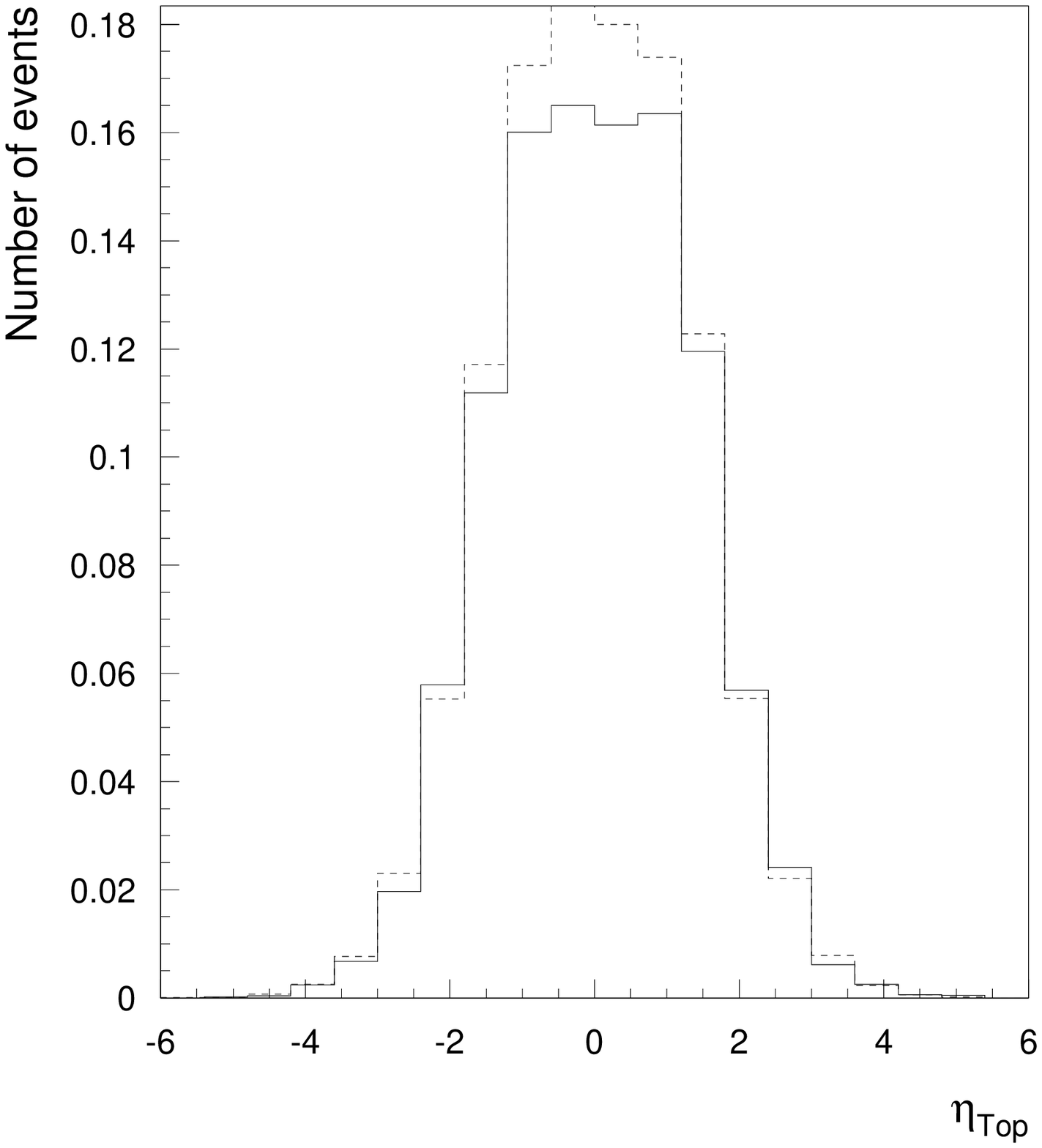}} }
\par
\vspace*{-0.6cm}
\par
\mbox{\resizebox*{0.45\textwidth}{0.29\textheight}{\includegraphics{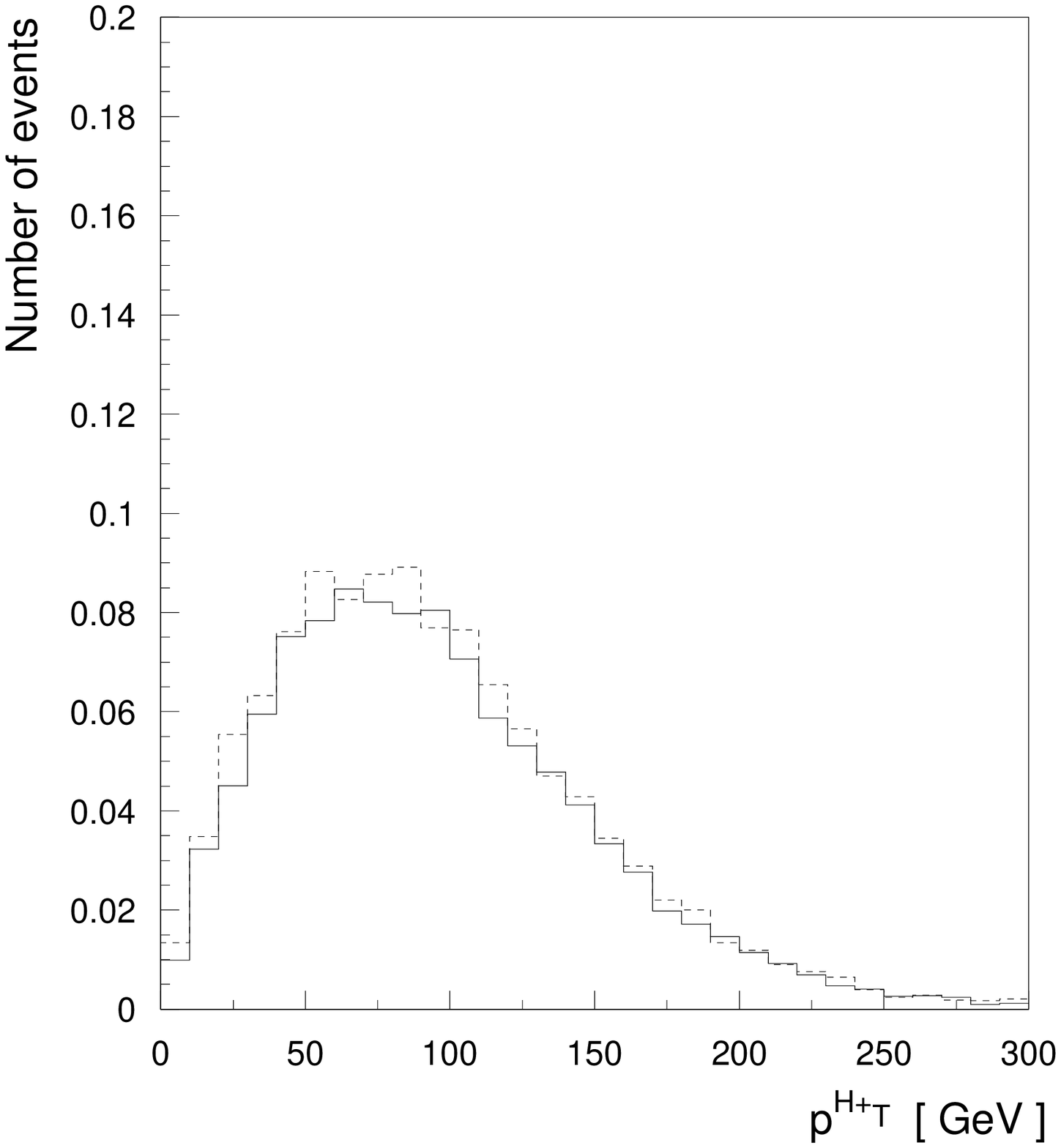}}
\resizebox*{0.45\textwidth}{0.29\textheight}{\includegraphics{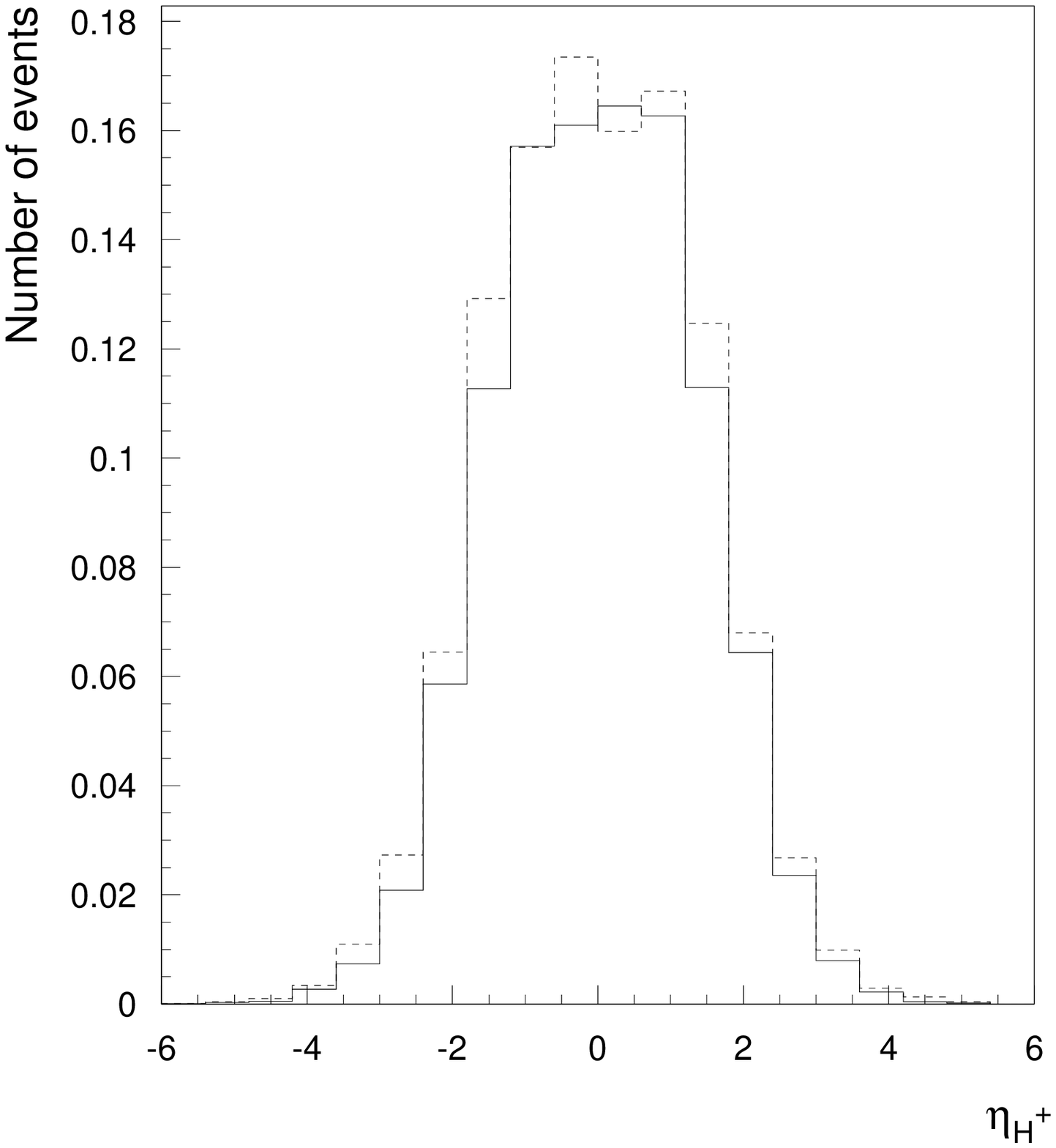}} }
\par
\vspace*{-0.6cm}}
\caption{
{Transverse momenta and rapidity distributions for the
final-state particles of the processes $pp(\gb)\rightarrow H^{+}\bar
{t}+b_{split}$ (dashed line), a $2\rightarrow2$ process with an additional $%
b $-quark from initial-state radiation, and the $2\rightarrow3$ $%
pp(gg)\rightarrow H^{+}\bar{t}b$ process (solid line) at the Tevatron
for $M_{H^+}=300\GeV$. }}
\label{fig:HT}
}
We conclude that the method of combining the $p_{T}^{b}$ distribution of $%
H^{+}\bar{t}+b_{split}$ and the complete tree-level $H^{+}\bar{t}b$ process
allows us to find the physically motivated $p_{T}$ cut on the $b$-quark,
which allows us to treat together those processes and simulate them in
different kinematical regions of $p_{T}^{b}$. Namely, we generate $pp(\gb%
)\rightarrow\bar{t}H^{+}$+ $b_{split}$ events using the PYTHIA generator,
and use them in the kinematical region $p_{T}^{b}<p_{T}^{cut}$ with the
weight, corresponding to the $\sigma_{total}-\sigma\lbrack pp(gg)\rightarrow
H^{+}\bar{t}b,\,\,p_{T}^{b}>p_{T}^{cut}]$ cross-section; in the region $%
p_{T}^{b}>p_{T}^{cut}$, instead, we use $pp(gg)\rightarrow H^{+}\bar{t}b$
events.
\FIGURE[t]{
\centerline{\mbox{\resizebox*{0.8\textwidth}{0.6\textheight}{\includegraphics
{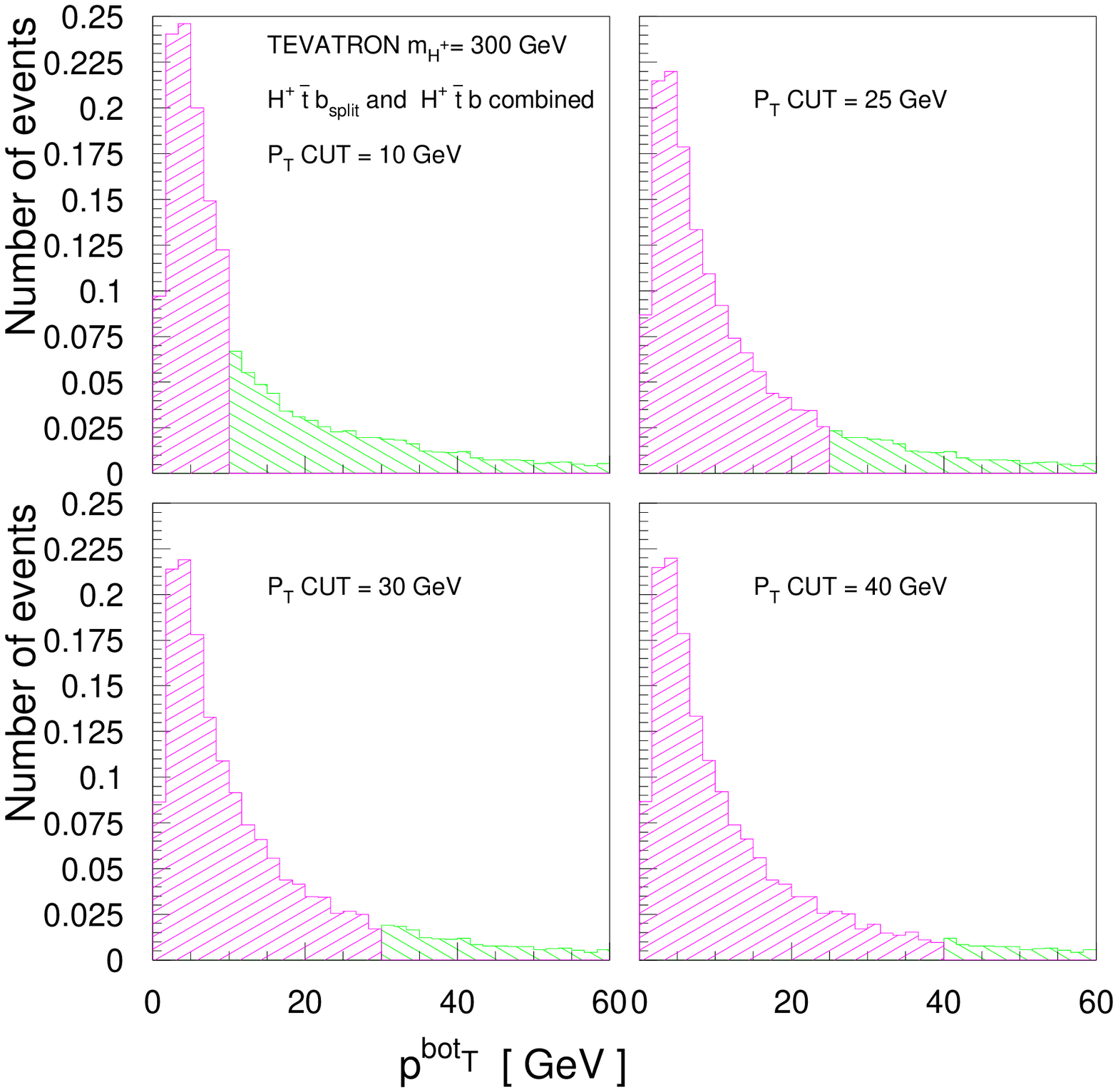}}}}
\caption{
{Transverse momenta distribution of the $b$-quark sewn for
various values of $p_{T}^{cut}$ for the $pp(\gb)\rightarrow H^{+}\bar
{t}+b_{split}$ and $pp(gg)\rightarrow H^{+}\bar{t}b$ processes at the Tevatron
for $M_{H^+}=300\GeV$. }}
\label{fig:sew}
}

\subsection{Kinematical analysis}

As we mentioned, the $b\bar{b}b(\bar{b})+\ell^{\pm}+2jets+p_{T}^{miss} $
signature of signal and background processes is the subject of this study.
One should reconstruct the $t\bar{t}b(\bar{b}) $ state from this signature
and then, the charged Higgs boson mass from all possible combinations of $tb $%
-invariant masses. At the same time one should work out an efficient set of
kinematical cuts for the background suppression.

For reconstructing the $t\bar{t}b(\bar{b}) $ final state, we follow the
procedure described below. 
\begin{itemize}
\item 
We reconstruct the W-boson mass from lepton and neutrino momenta: $%
M_{W1}^{rec}=(p_{\ell}+p_{\nu})^{2} $. The basic cuts for the
lepton (electron or muon) has been chosen as follows:
\begin{equation}
p_{T}^{\ell}>15\GeV,\, \, \, \, \, \, \, \, |\eta_{\ell}|<2.5,\, \, \, \,
p_{T}^{miss}>15\GeV\,\,.
\label{leptoncut}
\end{equation}
To find the $p_{z} $ of neutrino and therefore the neutrino 4-momentum, one
should solve the quadratic equation $m_{\ell\nu}=M_{W} $, which can have two
solutions. We reject events if this equation has no solutions, while in case
it has two solutions we keep both of them.
\item 
We reconstruct the mass of the second W-boson ($M_{W2}^{rec} $):
we keep all dijet combinations for the moment: the effects of the initial-
and final-state radiation and $b $-miss-tagging, the number of light jets is
almost always larger than two. The following basic cuts for the jets were
chosen:
\begin{equation}
p_{T}^{j,b}>20\, \, (30)\GeV\, \, [\mbox{Tevatron(LHC)}]\, \, \, \, ,\, \,
\, \, |\eta_{j}|<3,\, \, |\eta_{b}|<2\, .
\label{jetcut}
\end{equation}
\item 
Then we form and keep all $m_{t_{1}}=M_{W1b} $ and $%
m_{t_{2}}=M_{W2b} $ combinations for the first and second top-quarks.
\item 
In the final step we form the $\chi$ function
\begin{equation}
\chi=\sqrt{(M_{W1}^{rec}-M_{W})^{2}+(M_{W2}^{rec}-M_{W})^{2}+(m_{t_{1}}-\mt%
)^{2}+(m_{t_{2}}-\mt)^{2}}
\label{chidef}
\end{equation}
for all combinations of $b $-jets, jets, lepton and neutrino and choose the
combination giving the smallest (best) value of the $\chi$ function. The
optimized value of the cut on the $\chi$ function was found to be
\begin{equation}
\chi<\chi_{max}=100\GeV\, \, \mbox{at the Tevatron},\, \, \, \,
\chi<\chi_{max}=60\GeV\, \, \mbox{at the LHC}\, .
\label{chicut}
\end{equation}
\end{itemize}
It should be noted that the reconstruction thus made is independent of the
order in which $M_{W1}^{rec},M_{W2}^{rec},m_{t_{1}} $, $m_{t_{2}} $ were
formed. This leads to a better signal efficiency, a better probability of a
correct reconstruction and a better control of the efficiency through the
only relevant parameter, $\chi_{max} $. At the Tevatron, the cut on $\chi$
is quite relaxed by the lack of statistics, while at the LHC this cut could
be tightened further, gaining both signal/background ($S/B $) ratio and
significance ($S/\sqrt{B} $).

After the reconstruction of the $t\bar{t}b\bar{b} $ state one should
reconstruct the charged Higgs boson mass for the signal and the continuous $tb $
mass for the background. We assume that the  $b $-jet with the highest $%
p_{T} $ in $\bar{t}bt(\bar{b}) $ signature comes from the $H^{+} $ decay.
The probability of that being correct is directly related to the value of \mH%
: for $\mH\simeq200\GeV$ it is only about 50\% while for $\mH\simeq300\GeV$
the $b $-jet coming from the $H^{+} $ decay has the highest $p_{T} $ in
already 75\% of the cases. Since we chose just one $b $-jet, there will be
two $tb $-invariant mass combinations with two top-quarks that will enter
the $m_{tb} $ invariant mass plot.

Like in {Ref.}~\cite{DP-3b}, we found that $p_{T}^{b} $ from $%
H^{+}\rightarrow t\bar{b} $ decay is a good variable for the separation of
signal from background. However, instead of the fixed $p_{T}^{b} $ cut we
apply here the \mH-dependent $p_{T}^{b} $ cut:
\begin{equation}
p_{T}^{b}>[M_{H^{+}}/5-15]\GeV\, ,
\label{bottomcut}
\end{equation}
since the peak of this $b $-jet distribution is just proportional to
$M_{H^{+}} $.
 This $M_{H^{+}} $ dependence of the $p_{T}^{b} $ cut
allows us to increase the efficiency of the kinematical cut and
selection for the signal. 
The cut depends on the Higgs mass and
should be understood as the cut chosen with the respect to the  mean
value of the mass window where we are looking for the Higgs signal.
After the fitting of  the Higgs mass   using  the 'rough'
$p_T^{b}$, cut value, one can use the 
fitted Higgs mass as the input for $p_T^b(M_{H^+})$.

The window around the selected values of $M_{H^{+}} $ was also chosen $%
M_{H^{+}} $-dependent:
\begin{equation}
|m_{tb}-M_{H^{+}}|<5\sqrt{M_{H^{+}}}\,\,.
\label{masscut}
\end{equation}

One could think of an additional set of kinematical cuts, such as the
correlation angle between the Higgs boson direction and its decay products
(in the Higgs boson rest frame), the angle between top and bottom from
Higgs boson decay
(since they tend to be more back-to-back). It turns out that already at the
PYTHIA simulation level the difference in those distributions between signal
and background is quite blurred. The application of the respective cuts
would lead to some increase of the $S/B $ ratio, but at the same time to a
definite decrease of the significance and of the accuracy of the signal
cross-section measurement.

\FIGURE[t]{
\noindent%
\mbox{
\resizebox*{0.45\textwidth}{!}{\includegraphics{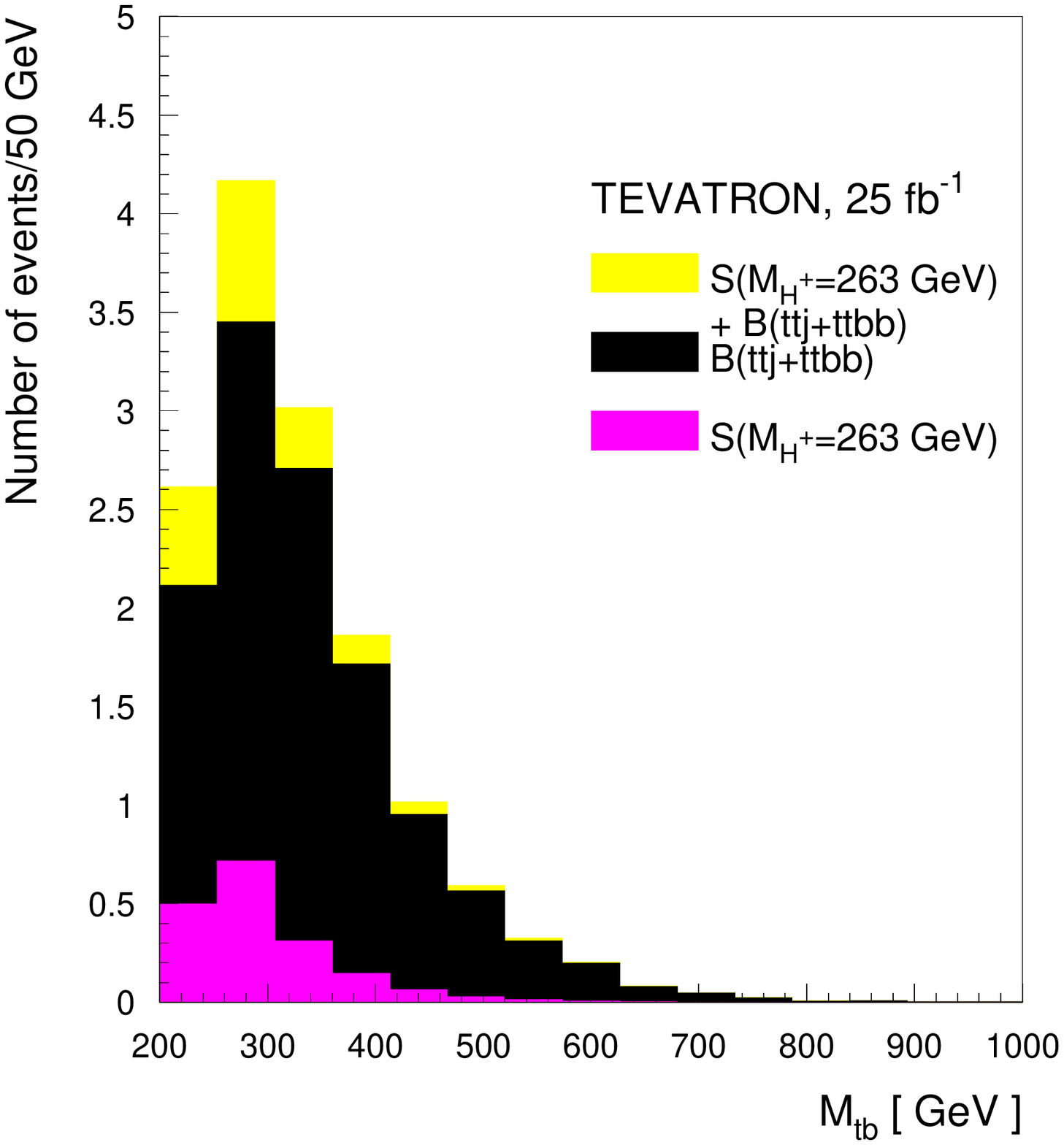}} 
\resizebox*{0.45\textwidth}{!}{\includegraphics{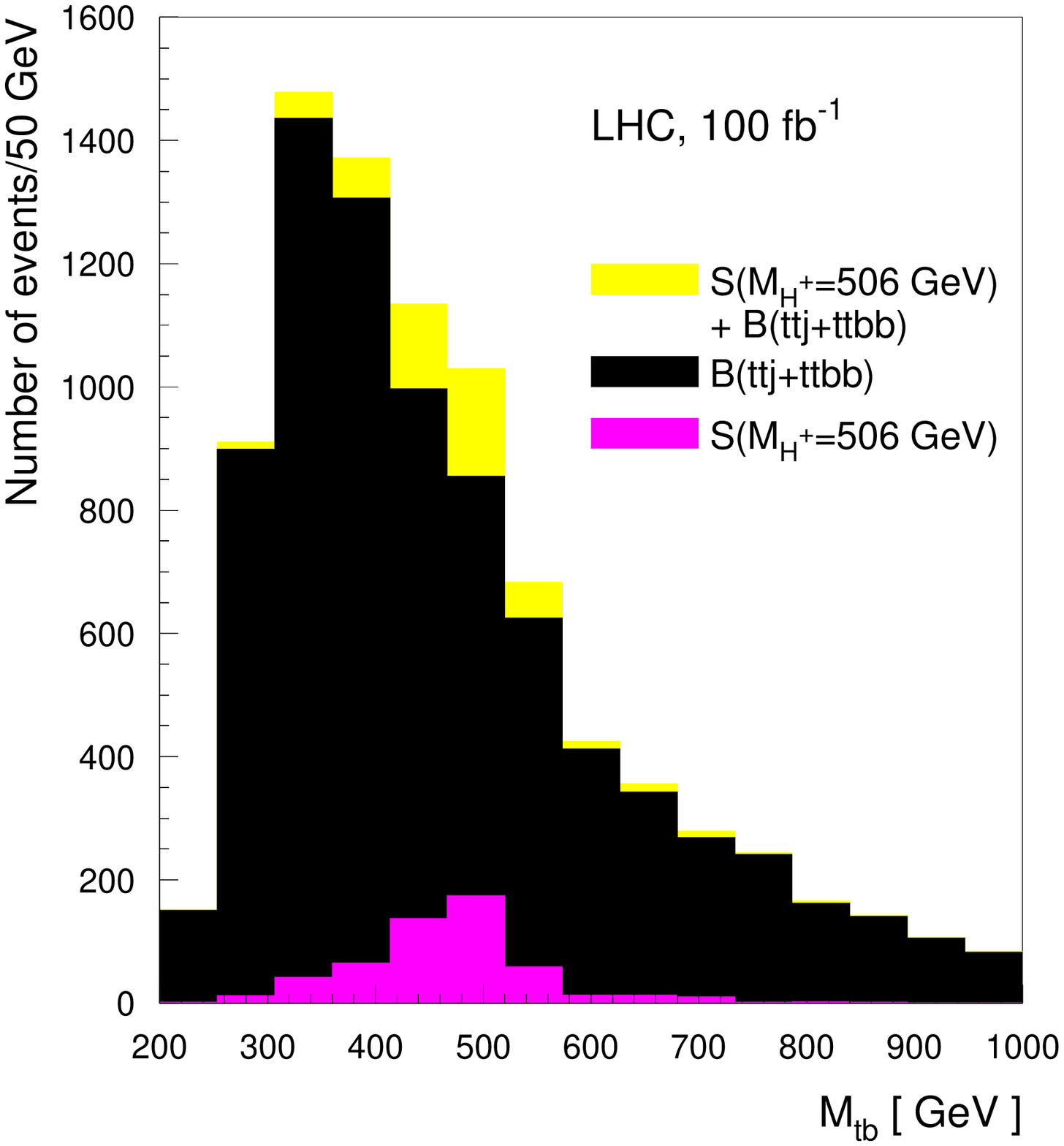}}
}
\caption{
{Reconstructed $tb$ invariant-mass distribution for signal,
background and signal plus background events, for \textbf{a)} $M_{H^{+}}=263
\GeV$ ($M_{A^0}=250\GeV$), at the Tevatron and \textbf{b)}
$M_{H^{+}}=506 \GeV$ ($M_{A^0}=500\GeV$), at the LHC,
$K^{\rm QCD}=1$.}}
\label{fig:tb}
}

After all cuts are set up we are ready to present signal and
background efficiencies, $S/B$ ratio and significance. The
results are summarized in Tables~\ref{tab:res-tev}
and~\ref{tab:res-lhc} for the Tevatron and the LHC respectively.
We present there the number of
signal~{(\ref{qq-tbh})-(\ref{gb-th})} $[S]$ events and of
background (Table~\ref{tab:back})
$[B]$ events, as well as their ratio and significance after the
reconstruction procedure,
$b$-tagging~{(\ref{btag-tev})-(\ref{btag-lhc})}
and set of cuts~{(\ref{leptoncut})-(\ref{masscut})}.
Numbers are given for $\tb=50$ for an integrated luminosity
 of  $L\equiv\int
\mathcal{L}dt=25\fb^{-1}$ at
 the Tevatron and an integrated luminosity  of $100\fb%
^{-1}$ at the LHC. 
The number of signal events corresponds the
tree-level   cross-section ($K^{\rm QCD}=1$).
Tables \ref{tab:res-tev}-\ref{tab:res-lhc}
also present the various efficiencies  ($\epsilon$)
of the cuts~{(\ref{leptoncut})-(\ref{masscut})}
and reconstruction (including $b$-tagging) for the signal and
backgrounds.  The last columns of these tables give
the 95\% C.L. (resp. $5\sigma$) discovery  limits
of the total signal cross-section in femtobarns (resp. picobarns) for  a $25\fb%
^{-1} $ (resp. $100\fb^{-1}$)  of total integrated
luminosity for the given efficiencies at the Tevatron (resp. LHC).

As an example, Figure~\ref{fig:tb} shows the reconstructed $tb$
invariant-mass distribution for signal and background events at the Tevatron
and the LHC. Owing to the much higher signal rate, the significance of the
signal at the LHC is considerably higher than at the Tevatron. In addition,
the difference in shape of the signal and background distributions could be
clear only for $M_{H^{+}}>400\GeV$, which will be accessible only at the
LHC.

\TABULAR[t]{|l|l|l|l|l|l|l|l|l|}{
\hline
$M_{H^{+}} $(GeV) & $S $ & $B $ & $S/B $ & $S/\sqrt{B} $ & $\eps%
_{signal}(\%) $ & $\eps_{ttbb}(\%) $ & $\eps_{ttj}(\%) $ & 95\%  C.L. (fb) \\
\hline
215 & 9.8 & 14.0 & 0.70 & 2.62 & 7.10 & 6.20 & 0.060 & 7.50 \\
263 & 3.5 & 7.7 & 0.46 & 1.27 & 7.70 & 3.10 & 0.034 & 5.39 \\
310 & 1.6 & 7.7 & 0.21 & 0.58 & 8.90 & 3.10 & 0.034 & 4.66 \\
359 & 0.7 & 6.4 & 0.10 & 0.26 & 9.20 & 2.70 & 0.028 & 4.29 \\
\hline
} {Number of signal~(\ref{qq-tbh})-(\ref{gb-th})
$[S]$ and background 
{--Table~\ref{tab:back}--}
events $[B]$, their ratio and significance after reconstruction procedure, 
$b$-tagging~{(\ref{btag-tev})-(\ref{btag-lhc})}
and set of cuts~{(\ref{leptoncut})-(\ref{masscut})}
at the Tevatron. The last column gives the 95$\%$
C.L. discovery limit ($\sim 2\,\sigma$) on the
total signal cross-section in femtobarns for $25\fb^{-1}$
 of total integrated luminosity and given
efficiencies. 
See text for details. $K^{\rm QCD}=1$.\label{tab:res-tev}}

\TABULAR[t]{|l|l|l|l|l|l|l|l|l|}{
\hline
$M_{H^{+}} $(GeV) & $S $ & $B $ & $S/B $ & $S/\sqrt{B} $ & $\eps%
_{signal}(\%) $ & $\eps_{ttbb}(\%) $ & $\eps_{ttj}(\%) $ & $5\sigma$(pb) \\
\hline
215 & 847 & 1490 & 0.57 & 21.9 & 0.32 & 0.29 & 0.005 & 1.02 \\
310 & 2386 & 5890 & 0.41 & 31.1 & 2.00 & 1.25 & 0.018 & 0.32 \\
408 & 1560 & 6210 & 0.25 & 19.8 & 2.70 & 1.50 & 0.016 & 0.25 \\
506 & 773 & 3770 & 0.21 & 12.6 & 2.60 & 0.97 & 0.009 & 0.20 \\
605 & 433 & 2070 & 0.21 & 9.5 & 2.60 & 0.56 & 0.004 & 0.15 \\
704 & 217 & 1170 & 0.19 & 6.3 & 2.30 & 0.33 & 0.002 & 0.13 \\
804 & 117 & 666 & 0.18 & 4.5 & 2.10 & 0.19 & 0.001 & 0.10 \\
\hline}
{%
{Number of signal~{(\ref{qq-tbh})-(\ref{gb-th})}
$[S]$ and background
{--Table~\ref{tab:back}--}
events $[B]$, their ratio and significance after reconstruction procedure, $%
b $-tagging~{(\ref{btag-tev})-(\ref{btag-lhc})}
and set of cuts~{(\ref{leptoncut})-(\ref{masscut})}
at the LHC. The last column gives the 5$\protect\sigma$ discovery
limit on the total signal cross-section in picobarns for
$100\fb^{-1}$  of total integrated luminosity and
given efficiencies. 
See text for details. $K^{\rm QCD}=1$.}\label{tab:res-lhc}}

\FIGURE[t]{
\mbox{\resizebox*{0.5\textwidth}{!}{\includegraphics{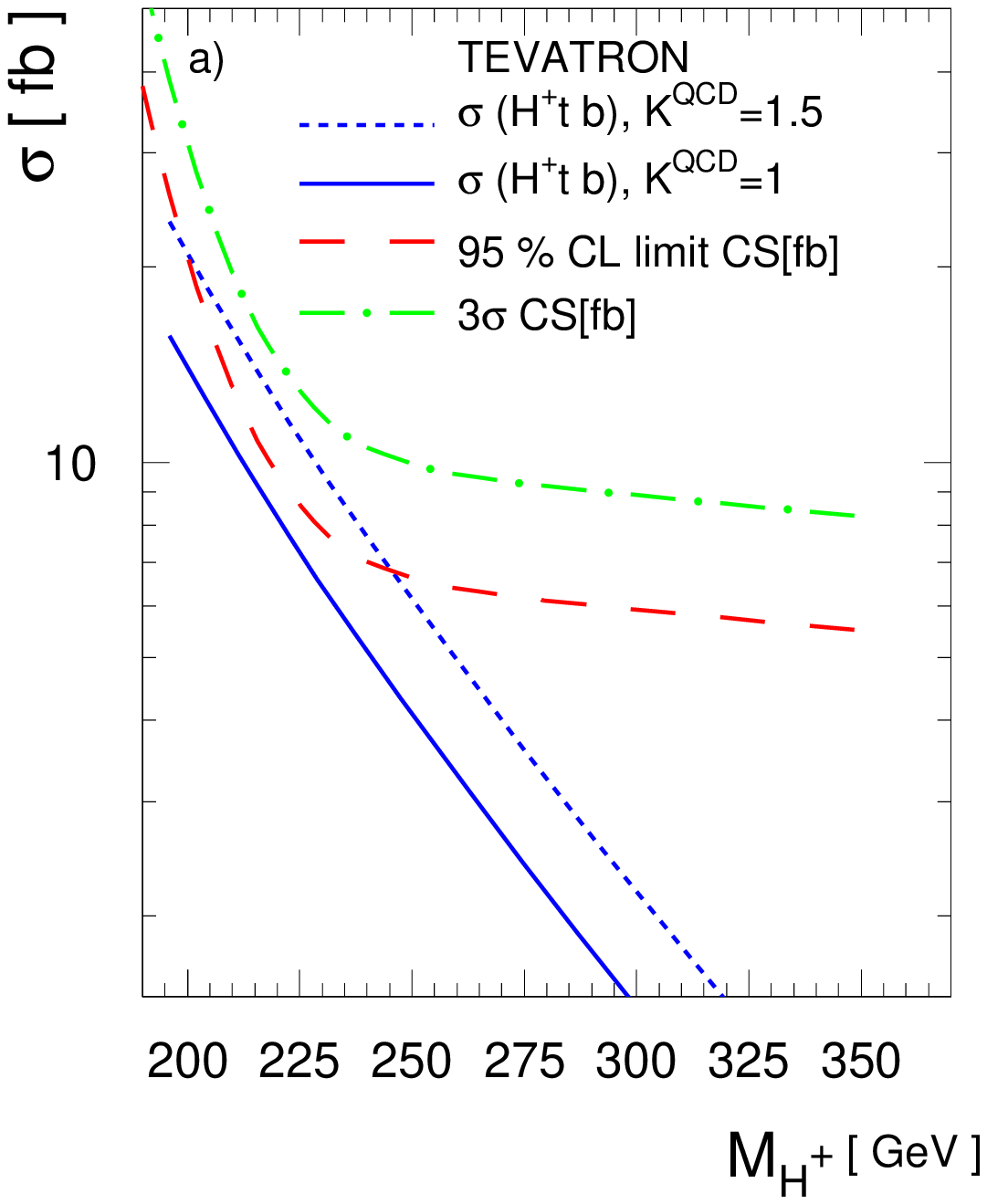}} 
      \resizebox*{0.5\textwidth}{!}{\includegraphics{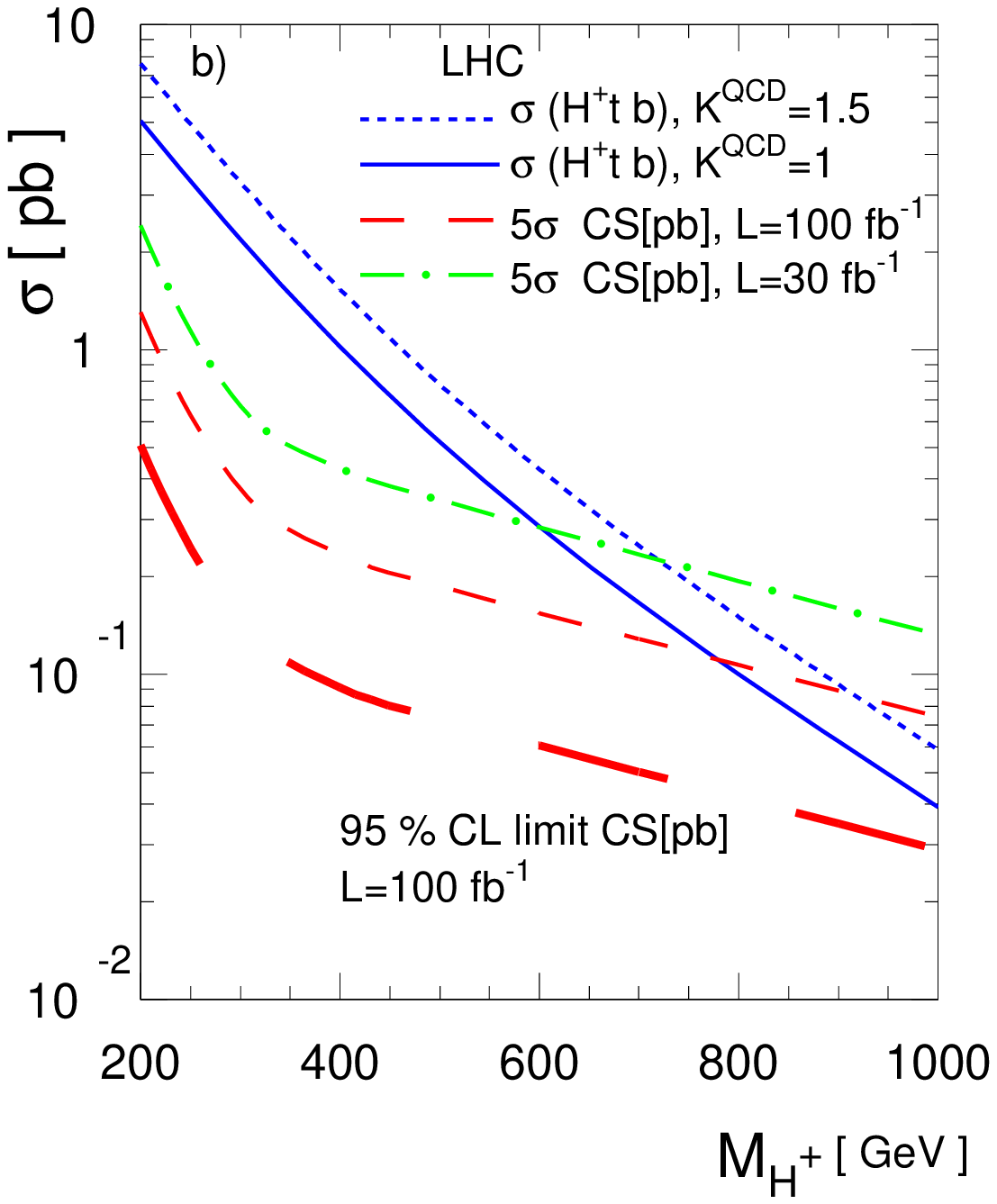}}} \\
\caption{%
The total cross-section $\sigma(\pptbH)$
for fixed 
$\tb=50$ including a QCD factor  $K^{\rm QCD}=1$ (solid) and 
$K^{\rm  QCD}=1.5$ (dotted): \textbf{a)} at the Tevatron, dashed and
dot-dashed lines correspond to the  
   cross-section values necessary for the $2\sigma$ ($95\%$ C.L.)
   exclusion and the $3\sigma$ evidence respectively, at the integrated
   luminosity $L=25\fb^{-1}$; \textbf{b)} at
the LHC, dot-dashed and dashed lines corresponding to the cross-section
necessary for the $5\sigma$ discovery at the integrated luminosity
$L=30\fb^{-1}$ and $100\fb^{-1}$ respectively; the long-dashed line at the
bottom corresponds to the $95\%$ C.L. limit at $L=100\fb^{-1}$.
}
\label{fig:lhc-eff}
}

In spite of the fact that the S/B ratio is quite high at the Tevatron for $%
M_{H^{+}}=215\GeV$, 
the signal rate itself is small, even for $\tb%
=50$. That is why we try to keep the signal efficiency as high as possible
by relaxing the acceptance cut on $p_{T}^{j,b}$ and the cut on $\chi$. Table~%
\ref{tab:res-tev} shows that {at the Tevatron there is no $5\,\sigma$
discovery limit available for Higgs boson masses above $200\GeV$, although there
is a narrow window $(175-195)\GeV$ for Higgs boson masses just above the
kinematical limit for the top quark decay into a charged Higgs boson. In general,
however, at the Tevatron one can only hope to exclude charged Higgs
boson masses up
to the certain value } defined by the efficiencies, the value of $\tan\beta$
(and the other MSSM parameters beyond the tree level) and the total
integrated luminosity. Even the exclusion region is apparently small. The
signal significance is only $3.4$ for the charged Higgs boson with mass $215%
\GeV$, $\tan\beta=50$ and $K$-factor  $K^{\rm
QCD}=1.5$. For this value of \tb\ the 95\%  C.L.
limit on the charged Higgs boson mass is $240\GeV$. We use the Poisson
statistics when putting this limit at the Tevatron, since the
number of events is small for both the signal and the background.
{On the other hand, for $K^{\rm QCD}\leq 1$ it is not possible to
obtain any limits using this channel, as it is obvious from the
comparison of Tables~\ref{tab:signal_tev} and \ref{tab:res-tev}.}

At the LHC, the situation is much better, with higher luminosity ($30-300\fb%
^{-1}$) and a signal rate higher by 3 orders of magnitude with
respect to the Tevatron. Here we can allow ourselves to use the
tighter kinematical cuts
(\ref{leptoncut})-(\ref{chicut}). We try to reach the best
 significance ratio $S/\sqrt{B} $ while keeping
the $S/B $ ratio above the 10\% level. Based on the efficiencies of Table~%
\ref{tab:res-lhc}, we {find} that a charged Higgs boson can be discovered at
the $5\sigma$ level, with mass up to about $900\GeV$ for $\tb=50 $ and  $%
K^{\rm QCD}=1.5 $.

It is important also to understand the role of the $K$-factor.  In
   Fig.~\ref{fig:lhc-eff} we present the cross-section for the $\pptbH$
   process for $K^{\rm QCD}=1.5$ and $K^{\rm QCD}=1$ (dotted and solid
   lined, respectively), and the cross-section for necessary for the
   discovery/exclusion of a charged Higgs boson.  In
   Fig.~\ref{fig:lhc-eff}a we show the $95\%$ C.L. limit (dashed) and
   the $3\sigma$ evidence cross-section at the Tevatron for an
   integrated luminosity of $25\fb^{-1}$.  Fig.~\ref{fig:lhc-eff}b
   shows the cross-section for the $5\,\sigma$ discovery {(i.e.
   $5\,\sqrt{B}$)} at the LHC for an integrated luminosity of
   $L=30\fb^{-1}$ (corresponding to $3$ years of low-luminosity regime)
   and $100\fb^{-1}$ ($1$ year high-luminosity regime).  Notice that the
   discovery curves also evolve with $M_{H^{+}}$ because the
   efficiencies are $M_{H^{+}}$-dependent (see Tables~\ref{tab:res-tev}
   and \ref{tab:res-lhc}).
Even at low
luminosity,
the $5\,\sigma$ discovery limit for the LHC is~$600\GeV$ for  $K^{\rm
   QCD}=1$,  $700\GeV$ and for $K^{\rm QCD}=1.5$. {Of course, if
   the $K^{\rm QCD}$ factor turns out to be below 1, the mass reach is
   lowered accordingly, e.g. $450\GeV$ for $K^{\rm QCD}=0.5$}. One can
   see that the value of the 
$K$-factor is really important because a variation of $K^{\rm
QCD}$ from $1$ to $1.5$ causes a $100\GeV$
shift of the limit on the charged Higgs boson mass. For  $L=100\fb%
^{-1}$ the discovery limit is $790\GeV$ if
$K^{\rm QCD}=1$, and $900\GeV$ for $K^{\rm QCD}=1.5$ {($600\GeV$
  for $K^{\rm QCD}=0.5$)}. {In this
figure we also show the $2\,\sqrt{B}$ curve, from which one
obtains immediately that} the charged Higgs boson can be excluded at the
$95\%$ C.L. up to a mass of about $1\TeV$ even for $K^{\rm
QCD}=1$ at $100\fb^{-1}$ of integrated luminosity.
{In Section~\ref{sec:SUSYcorr} we shall come back to this figure
  and we will show that there are
regions of the MSSM parameter space where the total effective
 $K^{\rm MSSM}$ factor in eq.(\ref{KMSSM}) could
be $\sim (40-50)\%$ larger than the
 typical $K$-factor expected from pure QCD alone.}

One can assume that systematic uncertainties due to the parton
distribution uncertainties as well as to the uncertainty from the
higher-order QCD corrections for the background (or from a choice
of the QCD scale for the background) could be significantly
reduced by normalizing the MC $tb$ distribution to the data.
Under the assumption that normalization to the data will reduce
systematic uncertainties up to the level of $7\%$, one can also
estimate the accuracy of the measurement of the signal
cross-section using efficiencies from Table~\ref{tab:res-lhc}.
For example, for $M_{H^{+}}=408\GeV$ the accuracy of the
cross-section measurement will be:
\begin{equation}
\frac{\delta S}{S}=\frac{\sqrt{(\sqrt{B})^2+
\delta B_{syst}^{2}}}{S}
=\frac {\sqrt{%
6210+(0.07\times 6210)^{2}}}{1560}\simeq28\%
\end{equation}
This means that $\tbH$ coupling could be measured with an accuracy of $%
(1/2)\delta S/S\simeq14\%$. In the next section we show that the
SUSY corrections could be substantially higher than $14\%$,
therefore being potentially measurable!

\section{The role of the SUSY corrections}

\label{sec:SUSYcorr}

\subsection{The leading effects. Theoretical discussion}
\label{sec:SUSYcorrleading}

In Section~\ref{sect:svb} we have concentrated on the results for the cross-section and
background for the charged Higgs boson production process (\ref{tbh}) without
including the various sources of potentially important SUSY corrections.
However, in Section~\ref{sec:Xseccomp} we have already warned about the possibility of
non-negligible SUSY effects in the signal and we have correctly identified
the region of the parameter space where they can be optimized, namely at
high $\tan\beta$. It is now time to study this issue in detail because, as
we shall see below, the supersymmetric loops can be very sizeable and may
play a momentous role in the production of a charged Higgs boson. In fact,
they can not only enhance the signal versus the background, but they can
also be used as a means to characterize the supersymmetric nature of the
charged Higgs boson potentially found in hadron colliders through the mechanism (%
\ref{tbh}).

Among the plethora of possible SUSY corrections we disregard
virtual supersymmetric effects on the $gqq$ and $ggg$ vertices
and on the gluon propagators. We expect those to be of order
$(\aS/4\pi)\cdot(s/M_{SUSY}^{2})$ and thus suppressed by a
non-enhanced (i.e. \tb-independent) MSSM form factor coming from
the loop integrals. Therefore, we can neglect these contributions
as we are only considering effects of the form
$(\aS/4\pi)^n\cdot\tan^n\beta$ at large \tb\,. {As previously
emphasized, the cross-section for the signal increases steeply
with \tb\ (see Fig.~\ref{fig:tree-tanb}) and becomes highly significant for 
\tb $>30$, while it is much smaller for \tb\ in the low interval
$2-20$ where the remaining SUSY corrections are of the same order
or even dominant}, so our approximation is well justified.
 Similarly, we neglect  all those electroweak
corrections in vertices and self-energies which are proportional
to pure $SU(2)_L\times U(1)_Y$ gauge couplings; in particular,
vertices
involving electroweak gauge bosons and those involving electroweak gauginos%
\footnote{{We have performed the full calculation in the squark
and chargino-neutralino mass-eigenstate basis, but we have
cross-checked the identification of the leading parts using
 also the weak-eigenstate basis, i.e. in terms of
diagrams involving squarks, gauginos and higgsinos}.}.
Furthermore, we have checked that vertices involving Higgs bosons
exchange yield a very tiny overall contribution, due to automatic
cancellations arranged by the underlying supersymmetry. Finally,
there are the strong gluino-squark diagrams and the
$\tan\beta$-enhanced higgsino-squark vertices implicit in
chargino-neutralino loops. We have extracted (see below) the
parts of these interactions which are (by far) the more relevant
ones at high $\tan\beta$ and confirmed that the remaining
contributions are negligible.  In
practice this means that we will concentrate our analysis on the
interval $\tan\beta>20$ where we can be sure that our
approximation does include the bulk of the MSSM corrections while
at the same time the  cross-section of process (\ref{tbh}) starts
to be sufficiently large to consider it as an efficient mechanism
for charged Higgs boson production (Cf. Fig.\,\ref{fig:tree-tanb}). To
be more precise, we shall hereafter confine our study of process
(\ref{tbh}) to within the relevant interval $20<\tb<60$, where
the upper limit on $\tb$ is approximately fixed by the condition
of perturbativity of the Yukawa couplings.

As a consequence we can just concentrate on the leading quantum effects. The
latter can be conveniently described through an effective Lagrangian
approach that contains effective couplings absorbing both the leading SUSY
contributions and the known part of the QCD corrections~\cite{eff}. {At high
$\tan\beta$ the most relevant piece is} the effective $tbH^{+}$-coupling as
it carries the leading part of the quantum effects. Indeed, on the one hand
the quantity $\Dmb$ in (\ref{hb}) contains the bulk of the supersymmetric
contributions. On the other by trading the bottom quark mass by the
corresponding running quantity one can also include a sizeable part of the
QCD effects, {namely the (universal) renormalization group effects}. As a
result the relevant piece of the effective Lagrangian at high $\tan\beta$
involving the $tbH^{+}$ vertex can be cast as follows:
\begin{equation}
\mathcal{L\,}\mathbb{=\,}\frac{gV_{tb}}{\sqrt{2\,}M_{W}}\,\frac{\overline
{m}_{b}(\mu_{R})\,\tan\beta}{1+\Delta m_{b}}\,H^{+}\overline{t}%
_{L}\,b_{R}+h.c.  \label{effecLag}
\end{equation}
where $\mu_{R}$ is the renormalization scale for the bottom quark
running mass $\overline{m}_{b}(\mu_{R})$ at the NLO in the
$\overline{MS}$ scheme. One usually assigns a value to $\mu_{R}$
given by the characteristic energy scale of the process under
consideration. In our case (Cf. Fig. 1)
 it would be natural to set
$\mu_{R}=m_{t}+M_{H^{+}}$.
{{However, following the discussion in Section~\ref{sec:Xseccomp},} we will use
the
 (on-shell) pole quark masses in the tree-level cross-section, and will parameterize
the QCD corrections by means of a $K^{\rm QCD}$-factor}.

On the SUSY side the effects encoded in $\Dmb$ are related to the bottom
mass counterterm $\delta m_{b}$ in the on-shell scheme. If $m_{b}$ is the
pole mass, then its relation with the corrections resummed into the bottom
quark Yukawa coupling leads to the consistency formula
\begin{equation}
m_{b}+\delta m_{b}{=}\frac{m_{b}}{1+\Delta m_{b}}\,;  \label{mcounter}
\end{equation}
in particular $\Delta m_{b}=-\,\delta m_{b}/m_{b}$ at one loop. The SUSY
effects on $\Delta m_{b}$ can be of two types, SUSY-QCD and SUSY-EW. The
strong part of $\Delta m_{b}$ originates from finite corrections induced by
mixed LH and RH weak-eigenstate sbottoms and gluino loops~\cite{SUSYtbH,Dmb}
(Fig.~\ref{fig:leadingdiags}a):
\begin{align}
\left( \Delta m_{b}\right) _{\mathrm{SUSY-QCD}} & =-\,C_{F}{\frac
{\alpha_{S}(M_{\mathrm{SUSY}})}{2\pi}}\,m_{\tilde{g}}\,M_{LR}^{b}\,I(m_{%
\tilde {b}_{1}},m_{\tilde{b}_{2}},m_{\tilde{g}})  \notag \\
& \rightarrow+{\frac{2\alpha_{S}(M_{\mathrm{SUSY}})}{3\pi}}\,m_{\tilde{g}%
}\,\mu\,\tan\beta\,I(m_{\tilde{b}_{1}},m_{\tilde{b}_{2}},m_{\tilde{g}})\,,
\label{eq:dmbQCD}
\end{align}
{where $C_{F}=(N_{c}^{2}-1)/2N_{c}=4/3$ is a color factor (with $N_{c}=3$)}.
Moreover we have $\alpha_{S}(M_{\mathrm{SUSY}})\simeq0.09$, $%
M_{LR}^{b}=A_{b}-\mu\,\tan\beta$ and {the last expression in (\ref{eq:dmbQCD}%
) holds only for sufficiently large $\mu\,\tan\beta\gg A_{b}$. Here $\mu$ is
the higgsino mass parameter in the superpotential, and should not be
confused with the renormalization scale $\mu_{R}$. Finally, $A_{b}$ is the
bottom quark trilinear coupling in the   soft-SUSY-breaking part of the
superpotential}. As for the electroweak effects on $\Delta m_{b}$ they stem
from similar loops involving mixed LH and RH-stops and mixed charged
higgsinos $\tilde{H}_{1}-\tilde{H}_{2}$ and read (Fig.~\ref{fig:leadingdiags}%
b) 
\begin{align}
\left( \Delta m_{b}\right) _{\mathrm{SUSY-Yukawa}} & =+{\frac{h_{t}\,h_{b}}{%
16\pi^{2}}}\,\,{\frac{\mu}{m_{b}}}\,m_{t}\,M_{LR}^{t}I(m_{\tilde {t}_{1}},m_{%
\tilde{t}_{2}},\mu)  \notag \\
& \rightarrow+{\frac{Y_{t}}{4\pi}}\,A_{t}\,\,\mu\,\tan\beta\,\,I(m_{\tilde
{t}_{1}},m_{\tilde{t}_{2}},\mu)\,\,.  \label{eq:dmbEW}
\end{align}
{Here $A_{t}$ is the top quark counterpart of $A_{b}$; we have defined $%
Y_{t}=h_{t}^{2}/4\pi\simeq1/4\pi$ and $M_{LR}^{t}=A_{t}-\mu\cot\beta$, and
again the last expression holds for large enough $\tan\beta$ only.} {%
Gaugino-higgsino 
($\tilde{W}-\tilde{H}_2$) 
mixing also contributes with \tb\ enhanced terms to \Dmb\
(Fig.~\ref{fig:leadingdiags}c):}
\FIGURE[t]{
\centerline{\includegraphics{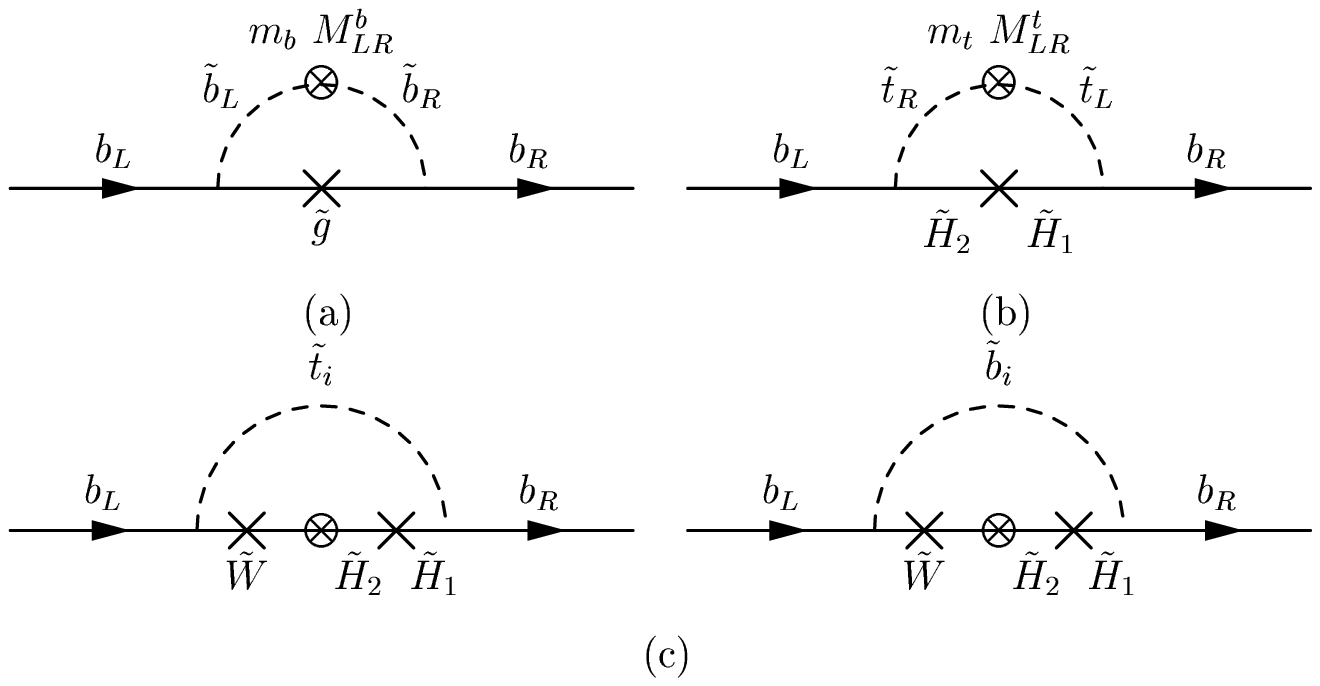}}
\caption{%
Leading contributions to $\Dmb$ -- eq.~(\ref{mcounter}).
   \textbf{a)} SUSY-QCD contributions (\ref{eq:dmbQCD}); \textbf{b)}
   SUSY-Yukawa contributions (\ref{eq:dmbEW}); \textbf{c)} SUSY-EW gauge
   contributions (\ref{eq:dmbEWgauge}). Each cross represents a
   mass-insertion; each cross with a circle represents a field mixing
   term in the electroweak eigenstate basis for squarks or gaugino-higgsinos.
}
\label{fig:leadingdiags}
}
\begin{align}
\left( \Delta m_{b}\right) _{\mathrm{SUSY-gauge}} = - & \frac{g^2}{16 \pi^2}
\mu\tb M_2 \bigg[ \cos^2\theta_{\tilde t}\, I(\mst1^2,M_2^2,\mu^2)+
\sin^2\theta_{\tilde t}\, I(\mst2^2,M^2_2,\mu^2)\,  \notag \\
& + \frac{1}{2} \left(\cos^2\theta_{\tilde b}\, I(\msb1^2,M_2^2,\mu^2)+
\sin^2\theta_{\tilde b}\, I(\msb2^2,M^2_2,\mu^2)\right)\bigg]\,.
\label{eq:dmbEWgauge}
\end{align}
In the above formulae we have introduced the (positive-definite) three-point
function
\begin{align}
I(m_{1},m_{2},m_{3}) & \equiv16\,\pi^{2}i\,C_{0}(0,0,m_{1},m_{2},m_{3})
\notag \\
& ={\frac{m_{1}^{2}\,m_{2}^{2}\ln{\frac{m_{1}^{2}}{m_{2}^{2}}}%
+m_{2}^{2}\,m_{3}^{2}\ln{\frac{m_{2}^{2}}{m_{3}^{2}}}+m_{1}^{2}\,m_{3}^{2}%
\ln {\frac{m_{3}^{2}}{m_{1}^{2}}}}{(m_{1}^{2}-m_{2}^{2})%
\,(m_{2}^{2}-m_{3}^{2})\,(m_{1}^{2}-m_{3}^{2})}}\,.  \label{eq:I123}
\end{align}
This function is well defined even when all masses are equal, which is
precisely the situation where it attains its minimum. Thus, if we roughly
assume that all masses are equal to $M_{SUSY}$, it boils down to
\begin{equation}
I(m_{i}=M_{SUSY})=\frac{1}{2\,M_{SUSY}^{2}}\,.  \label{eq:Immm}
\end{equation}
Furthermore, by setting $\left| \mu\right| =\,m_{\tilde{g}}=M_{SUSY}$ and $%
\tan\beta=50$ one immediately sees that eq.(\ref{eq:dmbQCD}) gives a large
correction of $\pm\mathcal{O}(50\%)$ depending on the sign of $\mu$. {Notice
that in the on-shell renormalization scheme defined in~\cite{SUSYtbH}, the
SUSY-EW gauge contributions from eq.~(\ref{eq:dmbEWgauge}) are
compensated for
by those equivalent to Fig.~\ref{fig:leadingdiags}c but with a external $%
\tau $-lepton and sneutrinos/$\tau$-sleptons circulating in the loop.%
\footnote{{See Fig.~21b and eq.~(86) in the first reference of~\cite{SUSYtbH}%
.}} However the resummation in~(\ref{effecLag}) can only be done in a
mass-independent renormalization scheme~\cite{eff}, and the diagrams in Fig.~%
\ref{fig:leadingdiags}c must be taken into account. The difference between
the two renormalization schemes amounts to a finite shift in the definition
of \tb\ which, however, is not significant in the interesting region $\tb%
\gg1 $.}

{Of course, the top quark Yukawa coupling receives  $\Delta\mt$
corrections   which are the
counterpart to those for the bottom quark,
  eqs.~(\ref{eq:dmbQCD})-(\ref{eq:dmbEWgauge}%
), but they are suppressed (rather than enhanced) by \tb} and so
they will be ignored.

It should be emphasized that both types of SUSY contributions to $\Delta
m_{b}$ can be sizeable; naively one could think that the SUSY-EW
contributions induced by higgsino-gaugino exchange are suppressed with
respect to the (gluino induced) SUSY-QCD contribution by a factor of $%
\alpha_{W}/\aS \sim10^{-2}$, but this is not necessarily so. {In fact, in
regions of the parameter space where the $3$-point functions on the RHS of (%
\ref{eq:dmbQCD}) and (\ref{eq:dmbEW}) are approximately
equal\,\footnote{Notice that the
function $I(m_{1},m_{2},m_{3})$ in (\ref{eq:I123}) is slowly
varying unless there are large hierarchies among the
masses.} the two types of corrections are roughly of order $%
\alpha_{S}\,\,\mu\,\tan\beta/M_{SUSY}$ and $Y_{t}\,\mu\,\tan\beta/M_{SUSY}$
respectively, and since $\alpha_{S}\simeq Y_{t}$ they should be comparable.
However, from a more accurate estimation including the color factor $C_{F}$
the ratio between the SUSY-QCD and SUSY-EW effects yields }
\begin{equation}
\frac{8}{3}\left( \frac{\alpha_{S}}{Y_{t}}\right) \left( \frac{m_{\tilde {g}}%
}{A_{t}}\right) \simeq3\,\left( \frac{m_{\tilde{g}}}{A_{t}}\right) \,.
\label{qcd-ew}
\end{equation}
This ratio can be smaller than $1$ if $A_{t}\gtrsim
3\,m_{\tilde{g}}$.  However in the numerical analysis below we
typically choose $m_{\tilde{g}}>A_{t}$ (Cf. Table
\ref{tab:SUSYpar}), and this makes the SUSY-QCD effects generally
dominant. There are indeed reasons to envision this kind of
scenario. First, the gluino mass is expected to be rather large,
of the order of $1\TeV$.  Second,  we cannot choose $A_{t}$ too
large, say above $1\TeV$,  because this could disrupt the
$SU(3)_{c}$ and $U(1)_{em}$ symmetries of the electroweak vacuum
state. An approximate necessary condition for this not to happen
is~\cite{colorbreak}
\begin{equation}
A_{t}^{2}<3\,(M_{Q}^{2}+M_{U}^{2}+M_{H_{2}}^{2}+\mu^{2})\,.
\label{vacbreak}
\end{equation}
Here $M_{Q,U}$ are the LH and RH stop SUSY-breaking mass
parameters, $M_{H_{2}}^{2}$ (with $M_{H_{2}}^{2}<0$, due to the
radiative breaking of the gauge symmetry) is the
soft-SUSY-breaking mass associated to the $H_{2}$ doublet (the
one that couples the top quark) and $\mu$ is again the
supersymmetric higgsino mass parameter. In a typical scenario
where $m_{\tilde{g}}$ is larger than any squark mass and any
electroweak mass parameter, one may assume $A_t\lesssim
M_{Q,U}\lesssim m_{\tilde{g}}$ (Cf. Table \ref{tab:SUSYpar}). In
this case the RHS of eq.(\ref{qcd-ew}) should be greater than $1$
and the SUSY-QCD effects are dominant. Third, these effects have
the remarkable property that they decouple very slowly with the
gluino mass\thinspace\cite{SUSYtbH}. Indeed, for sbottom masses of a few
hundred  $\mathrm{GeV}$ the gluino mass can be as high as $1\TeV$ and the
correction is still of order $(20-30)\%$. This is due to the
helicity flip in the gluino line which generates the gluino mass
term in the numerator of eq.(\ref{eq:dmbQCD}).  For
$m_{\tilde{g}}\sim1\TeV$ it would be contrived to assume
$A_{t}>3\,m_{\tilde{g}}$, and hence the SUSY-EW effects must
surely have gone away in the heavy gluino regime while the strong
supersymmetric effects can still remain. Of course the correction
(\ref{eq:dmbQCD}) eventually dies out with $m_{\tilde{g}}$, but
very slowly. Notice that in the opposite limit,
$m_{\tilde{g}}\rightarrow 0$, the SUSY-QCD effects
(\ref{eq:dmbQCD}) also tend to zero and one is left with the
subleading SUSY-QCD contributions from the renormalized
three-point functions. We shall not focus on the light gluino
case here, although in practice it is equivalent to assume that
the bulk of the SUSY corrections is just given by the electroweak
part (\ref{eq:dmbQCD}), which can still be rather substantial.
These remarks show that the leading type of SUSY effects on our
process can have different origin and remain significant in large
portions of the MSSM parameter space.

Ultimately the origin of the  exceptional SUSY quantum effects (\ref{eq:dmbQCD}) and (%
\ref{eq:dmbEW}) stems from the breaking of SUSY through
dimensional soft terms like gaugino masses and trilinear
couplings, and therefore  {they cannot be
described -- unlike the conventional QCD effects considered
before -- through the universal renormalization group running of
the parameters}. Technically, they are finite threshold effects
which must be included explicitly at energies above
 $ M_{SUSY}$, the characteristic SUSY breaking
scale at low energies~\cite{Dmb}.

Remarkably, the finite threshold effects do not vanish when the 
 soft-SUSY-breaking parameters, together with the higgsino mass parameter $\mu$,
are scaled up simultaneously to arbitrarily large values. This is
immediately seen from (\ref{eq:dmbQCD}) and (\ref{eq:dmbEW}) by letting
simultaneously $\mu\rightarrow\lambda\,\mu$ and $\widetilde{m}%
\rightarrow\lambda \,\widetilde{m}$ in these equations, where
$\lambda$ is an arbitrary dimensionless scale factor and
$\widetilde{m}$ is any  soft-SUSY-breaking parameter. However, it
is important to notice that the apparent non-decoupling behaviour
exhibited by this kind of SUSY effects is rendered innocuous
after resumming all the corrections of the form $\left(
\alpha_{S}\,\,\mu \,\tan\beta/M_{SUSY}\right) ^{n}$ and $\left(
Y_{t}\,\mu\,\tan\beta /M_{SUSY}\right) ^{n}$ to all orders
$n=1,2,\ldots$~\cite{eff}. The resummed result is indeed given by
the effective bottom quark Yukawa coupling (\ref{hb}), which
shows that there are no higher order disturbing effects of the
form  $\Dmb{}^{n}\sim
(\aS/4\pi)^n\cdot\tan^n\beta$ when $\Dmb\ge 1$ 
and the expansion 
\begin{equation}
\frac{1}{(1+\Dmb)}=1-\Dmb+\Dmb{}^{2}-\ldots  \label{higherDmb}
\end{equation}
is not possible.  On the other hand for $\left|
\Dmb\right| \ll1$ the expansion is of course allowed and the
higher order powers $\Dmb{}^{n}$ give small corrections to
the linear approximation $1-\Dmb$ used in~\cite{SUSYtbH}.
However, for $\left| %
\Dmb\right| \lesssim1$ and $\Dmb<0$ the resummed SUSY contribution
is seen to substantially reinforce the one loop result. For
example, if $\Dmb=-50\%$,
the one-loop correction is $+50\%$ whereas the resummed result gives a $%
+100\%$ effect! Therefore, the SUSY radiative corrections can be quite
large, and may remain so even for very high values of $M_{SUSY}$, but their
resummation is eventually kept under control when $\left| \Dmb\right| \gg1$.

The effective Lagrangian approach just outlined is also very useful for the
practical calculation of the cross-sections. In fact, by appropriately
modifying the MSSM Feynman rules using the effective vertex interactions
described above we have managed to automatically generate the leading
quantum effects from the CompHEP calculation of the cross-sections.

Although $\Dmb$ in the above equations is
 the
only correction of order $(\alpha/4\pi)\tb$ ($\alpha=\alpha_{S},%
\alpha_{W} $) that dominates for large \tb, we have also tested
the effect from the off-shell SUSY-QCD and SUSY-EW corrections to
the $H^{+}\bar{t}b$ vertex and to the fermion propagators. As
already mentioned above, some of these off-shell vertices (in
particular those involving chargino-neutralino exchange) carry
some $\tan\beta$-enhanced Yukawa couplings  that
could be important.  Nevertheless it turns out that
these contributions are not comparable to the leading $\Dmb$ ones.
{The point is that the effects entering $\Dmb $ at high
$\tan\beta$ are of the type $\alpha_{S}\tan\beta$ or
$Y_{t}\tan\beta$ whereas those from the aforementioned vertices
are of the type $\left( g\,m_{b}\,\tan\beta /M_{W}\right)
^{2}/4\pi\sim Y_{t}$, and so they are down by an extra power of
$\tan\beta$}. At the end of the day they are subleading effects
on equal footing to the previously dismissed SUSY corrections to
pure QCD vertices. Actually, to have full control on the
quantitative size of these effects we have explicitly checked
that their numerical contribution is  immaterial.
To this end we have evaluated the full set of one-loop SUSY
diagrams for the relevant $tbH^{+}$ vertex. The same set was
considered in detail Ref.\cite{SUSYtbH} in the case where all
external particles are on-shell. In the present case, however, at
least one of the quarks in that vertex is off shell (see Fig. 1).
Therefore, we can use the same bunch of diagrams as in the
on-shell case but we have to account for the off-shell external
lines, which is a non-trivial task. We have studied this issue in
detail by expanding the off-shell propagators
\begin{equation}
\frac{1}{(p+q)^{2}-M^{2}}=\frac{1}{q^{2}-M^{2}}\left[ 1-\frac{p^{2}+2pq}{%
q^{2}-M^{2}}+\left( \frac{p^{2}+2pq}{q^{2}-M^{2}}\right) ^{2}+\ldots\right]
\label{propag}
\end{equation}
in the expression of the one-loop functions. 
The result is that the off-shell corrections are 
 generally small
for they are of
order or below the non-leading one-loop effects that we have 
neglected.  
For the computation of the SUSY-corrected matrix elements
we have proceeded in the following way: first, we have modified CompHEP's
Feynman rules to allow for the most general 
$\bar{t}bH^{+}$
vertex; then we have let CompHEP reckon the squared matrix elements and dump
the result into REDUCE code. 
At this point, we have inserted expressions
for the coefficients of the 
off-shell $\bar{t}bH^{+}$ vertex that include the one-loop off-shell
supersymmetric corrections to the vertex itself and to the off-shell fermion
propagators and fermionic external lines\footnote{%
We shall not write down here the analytic expressions for the renormalized
vertex and propagators. They can easily be derived by just generalizing
previous on-shell calculations, such as those for $t\rightarrow bH^{+}$~\cite
{SUSYtbH}.}. Only half the renormalization of an internal fermion line has
to be included, the other half being associated to the $gqq$ vertex. 
Higher order terms not included in the effective
Lagrangian~(\ref{effecLag}) have been discarded.
{The subsequent numerical computation of the
one-loop integrals has been done using the package LoopTools~\cite{LoopTools}%
.} 
This
procedure has allowed us to estimate the relative size of the off-shell
effects in the signal cross-section, 
 whose effects are analyzed in the next section.

The upshot is that the approximation of neglecting 
$q\bar{q}g$
vertex and
propagator corrections  in the cross-section computation of
process (\ref{tbh}), which may be called ``improved Born''
approximation  (see next section), is really
justified in the relevant region of parameter space.

\subsection{Numerical analysis of the SUSY corrections}

We have seen above that the SUSY corrections are potentially large and can
be of both signs. The leading correction to the cross-section of our process
(\ref{tbh}) goes roughly as $\sigma\rightarrow\sigma(1-2\Dmb)$ at one loop,
so that the sign of the SUSY-QCD part carries just the sign of $-\mu$
whereas the SUSY-EW one depends on the sign of the combined parameter $-\mu
A_{t}$. {Recall that the radiative B-meson decays (based on the $%
b\rightarrow s\,\gamma$ transition) prefer the sign $\mu A_{t}<0$
(see e.g Refs.~\cite{Bsg2l,bsgamm}) if the charged Higgs boson is not
exceedingly heavy. The reason is that if stops and charginos are
of similar mass to that of the Higgs boson then they can compensate for
the Higgs boson contribution, which by itself would overshoot the
experimentally allowed range for $BR(b\rightarrow s\,\gamma)$. In
this case the $\mu A_{t}<0$ sector of the MSSM parameter space
offers the Tevatron II a chance to observe a supersymmetric charged
Higgs boson of around $200\GeV$. For a much heavier Higgs boson,
however, namely that accessible only to LHC, $b\rightarrow
s\,\gamma$ does not place any restriction on the sign of $\mu\,
A_{t}$.}\footnote{The precise
measurement of the anomalous magnetic moment of the
muon~\cite{gminus2ex} could favor $\mu>0$~\cite{gminus2SUSY},
but the impact on our case is marginal because the $(g-2)_{\mu}$
value in the MSSM critically depends on the values of some slepton
masses whose influence on our calculation is completely
negligible.}

{Obviously there is a preferred sign for our process: the optimal situation
for the charged Higgs boson production mechanism (\ref{tbh}) to be maximally
sensitive to the SUSY effects is when $\mu<0$ \textit{and} $A_{t}>0$
corresponding to both corrections, SUSY-QCD and SUSY-EW, giving a positive
contribution to the cross-section. The next-to-interesting case is when $%
\mu<0$ and $A_{t}<0$. Here the electroweak loops try to balance
the SUSY-QCD ones, but the dominance of the latter still leaves a
sizeable net outcome -- some $60\%$ of the previous case. Of
course there is also the odd possibility that $\mu>0$ and
$A_{t}>0$, but  in this circumstance
the SUSY quantum effects could  still manifest themselves
(if $\tb$ is large) through an effective $K$%
-factor  (\ref{KMSSM})  sensibly
smaller than the one predicted by QCD expectations:
 $K^{\rm MSSM}< K^{\rm QCD}$. The worse possible
situation would occur if $\tan\beta$ is small, say in the
intermediate range $2<\tan\beta<20$, since no positive nor
negative effect whatsoever could be detected; in fact, in this
case not even the tree-level signal would be available! In the
following we are going to concentrate our numerical analysis on
the  case $\mu<0$ and $A_t>0$, which is the one
phenomenologically most appealing and therefore defines the scope
of the present study.}

{In Fig.~\ref{fig:corrtb1} we present the total cross-section as
a function of \tb\ at the LHC (with $\mH=500 \GeV$) and the
Tevatron (with $\mH=250 \GeV $)  for $K^{\rm QCD}=1$. We show the
tree-level result and four additional cases that include SUSY
corrections: they correspond to the four sets of parameters
displayed in Table~\ref {tab:SUSYpar}. The SUSY-corrected curves
contain all the known corrections discussed in the previous
section, including the resummation of the SUSY effects.  As can
be seen the numerical effect of the SUSY contributions can be
dramatic; for the Set A there is a 100\% positive enhancement of
the cross-section. However the fact that we have resummed all the
leading terms of the form $(\alpha\tb)^n$ ensures that the
prediction is robust under the inclusion of additional higher
order effects.}

{Let us comment briefly on the treatment of the strong structure constant $%
\alpha_S$. We have used the SM value of $\alpha_S(Q)$ for the
gluon-gluon and quark-gluon vertices. We have to do so, since the
value of $\alpha_S$ is related to the determination of the PDFs
 and moreover the SUSY effects on these vertices
are, as we have discussed before, non-leading in our case.
However, in order to compute the SUSY corrections, we need to
compute $\alpha_S(M_{SUSY})$, and it would be a
very bad approximation to use the SM value. So we use the MSSM value $%
\alpha_S^{\mathrm{MSSM}}(Q)$ to compute the corrections to the $\tbH$ vertex.%
}

\TABULAR[t]{|c|c|c|c|c|c|c|c|c|} {
\hline\hline
& $\tb$ & $\mu$ & $M_2$ & 
$m_{\tilde g}$ 
& $\mst1$ & $\msb1$ & $A_t$ & $A_b$ \\ \hline
Set A & 50 & -1000 & 200 & 1000 & 1000 & 1000 & 500 & 500 \\ \hline
Set B & 50 & -200 & 200 & 1000 & 500 & 500 & 500 & 500 \\ \hline
Set C & 50 & 200 & 200 & 1000 & 500 & 500 & -500 & 500 \\ \hline
Set D & 50 & 1000 & 200 & 1000 & 1000 & 1000 & -500 & 500 \\ \hline\hline
}{Sets of SUSY parameters used in the computation of the
SUSY-corrections to the charged Higgs boson associated 
production (all  masses and trilinear couplings in~GeV).\label{tab:SUSYpar}}

\FIGURE[t]{
\begin{tabular}{cc}
\resizebox{!}{6cm}{\includegraphics{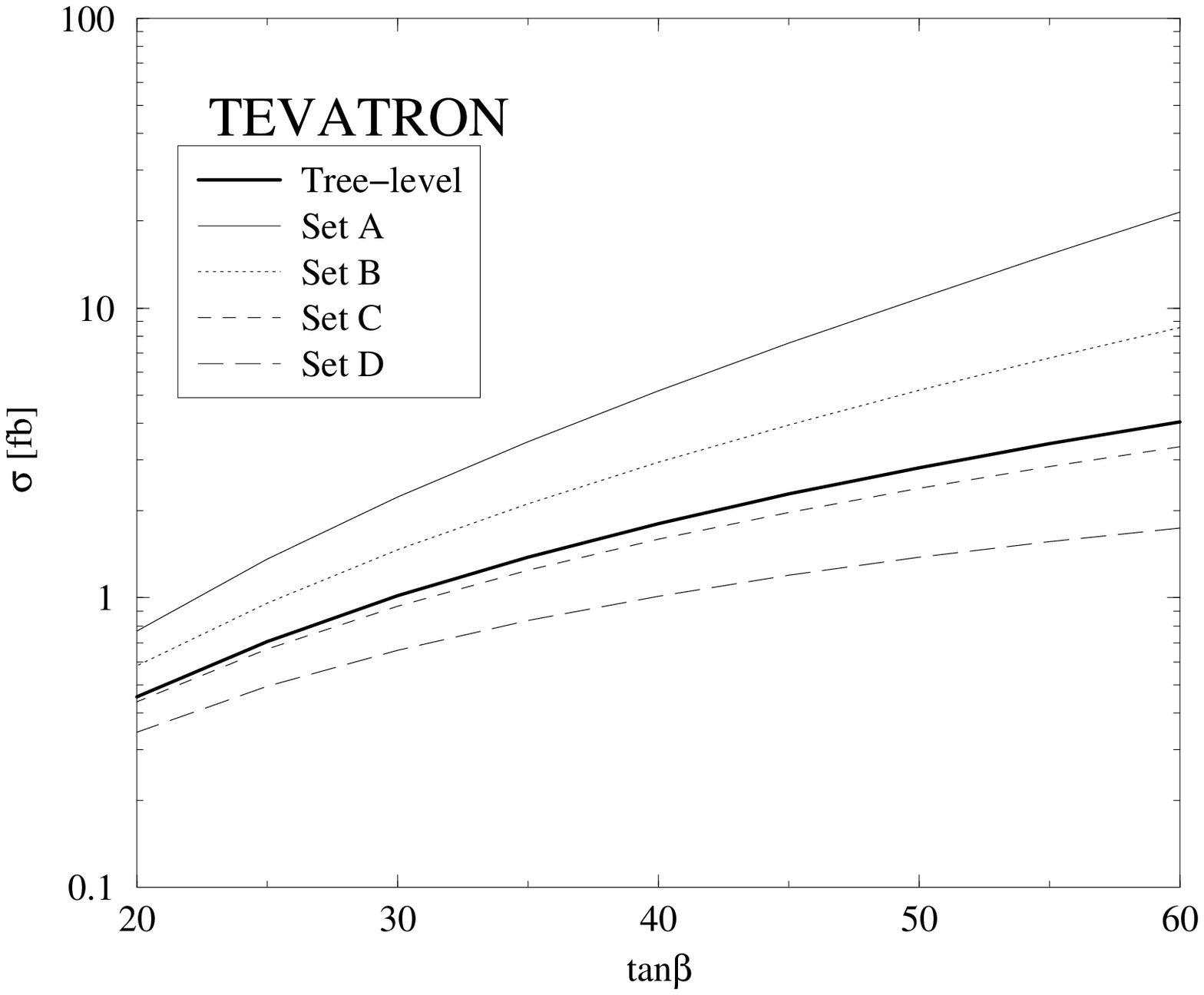}} & %
\resizebox{!}{6cm}{\includegraphics{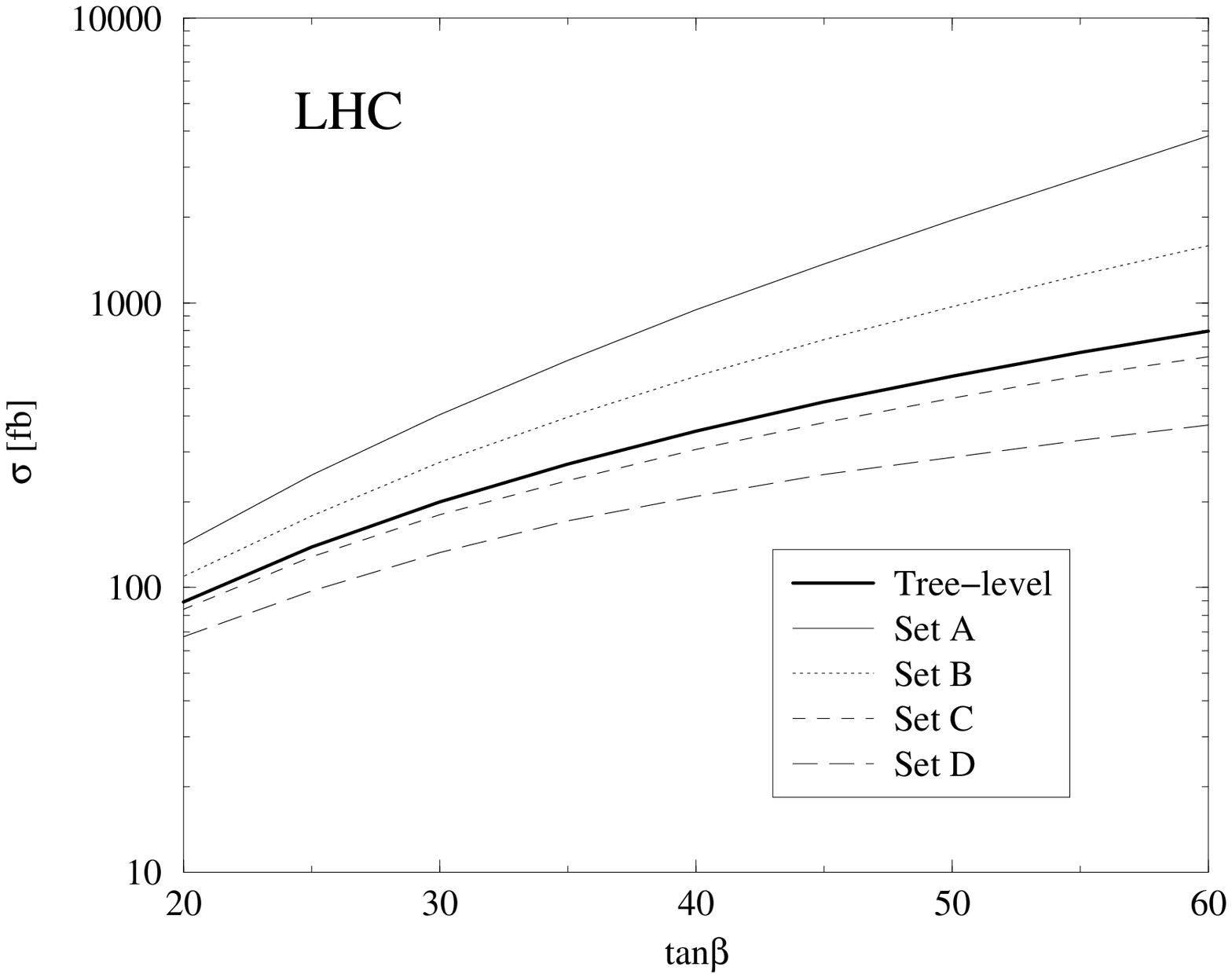}} \\
(a) & (b)
\end{tabular}
\caption{%
{Total cross-section for the process \pptbH\ as a
function of \tb\  in the relevant range of this
parameter.  The tree-level result and the
SUSY-corrected result are shown (with $K^{\rm QCD}=1$) for the
four sets of  parameters from Table~{\ref{tab:SUSYpar}}, for
\textbf{a)} the Tevatron with $\mH=250\GeV$, and \textbf{b)} the
LHC with $\mH=500\GeV$. }} \label{fig:corrtb1}
}

{Of course, one would like to know which is the relative
importance of each effect in the final corrected cross-section
for the process (\ref{tbh}).  Let us compare the
following set of corrections: }

\begin{itemize}
\item  the full set of  SUSY corrections, including  resummation and the off-shell two-
and three-point functions (labeled \textit{Full});

\item  the corrections including  resummation, but using on-shell values for
the irreducible three-point functions (\textit{On-Shell});

\item  the one-loop (without resummation)
  corrections, using also on-shell values for
the irreducible three-point functions
(\textit{OS-nr});

\item  the tree-level result with the pole quark masses replaced by the effective
Yukawa couplings eq.~(\ref{effecLag}) (\textit{Improved-Born}).
\end{itemize}

{In Fig.~\ref{fig:corrtb2} we show the total cross-section
 of process (\ref{tbh}) using the four
approximations.  We use the intermediate Set B of
Table~\ref{tab:SUSYpar} which represents a moderate case.
The following hierarchy of quantum
effects is observed:
\begin{equation}\label{CShiearchy}
\sigma({\rm Full)}\gsim\sigma({\rm Improved-Born})>\sigma({\rm
On-shell})>\sigma({\rm OS-nr})\,.
\end{equation}
The effect of the  resummation is clearly visible.
The difference between the resummed and non-resummed results
grows with the magnitude of the corrections; in sets A and D this
difference is much larger than that of sets B and C. On the other
hand the difference between the \textit{Full} and the
\textit{On-Shell} results is nearly
indistinguishable.  So the
\textit{Off-Shellness} effect is generally small in our case and
it may be neglected in a first approximation, although it should
be taken into account in a detailed study.}

\FIGURE[t]{
\begin{tabular}{cc}
\resizebox{!}{6cm}{\includegraphics{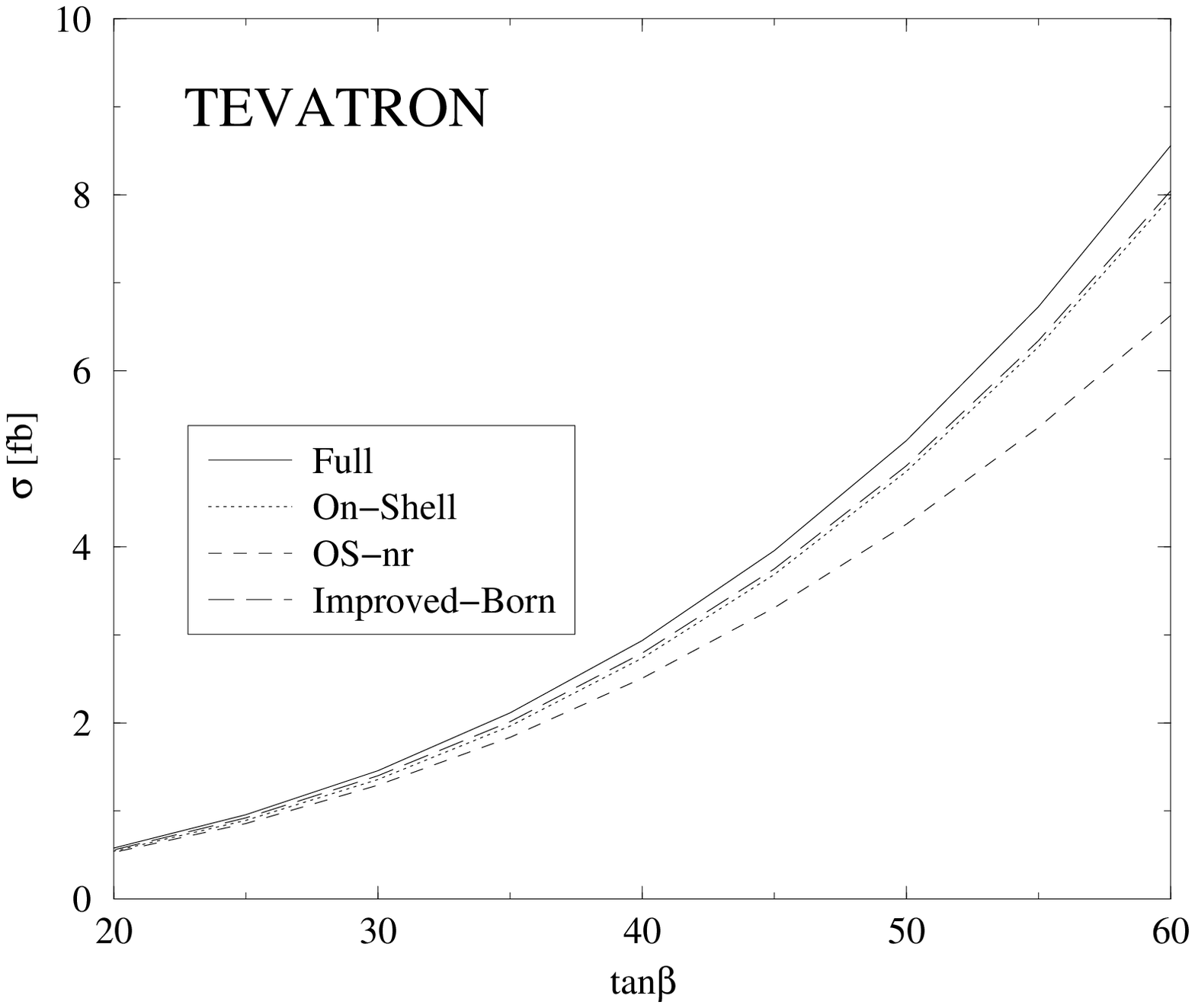}} & %
\resizebox{!}{6cm}{\includegraphics{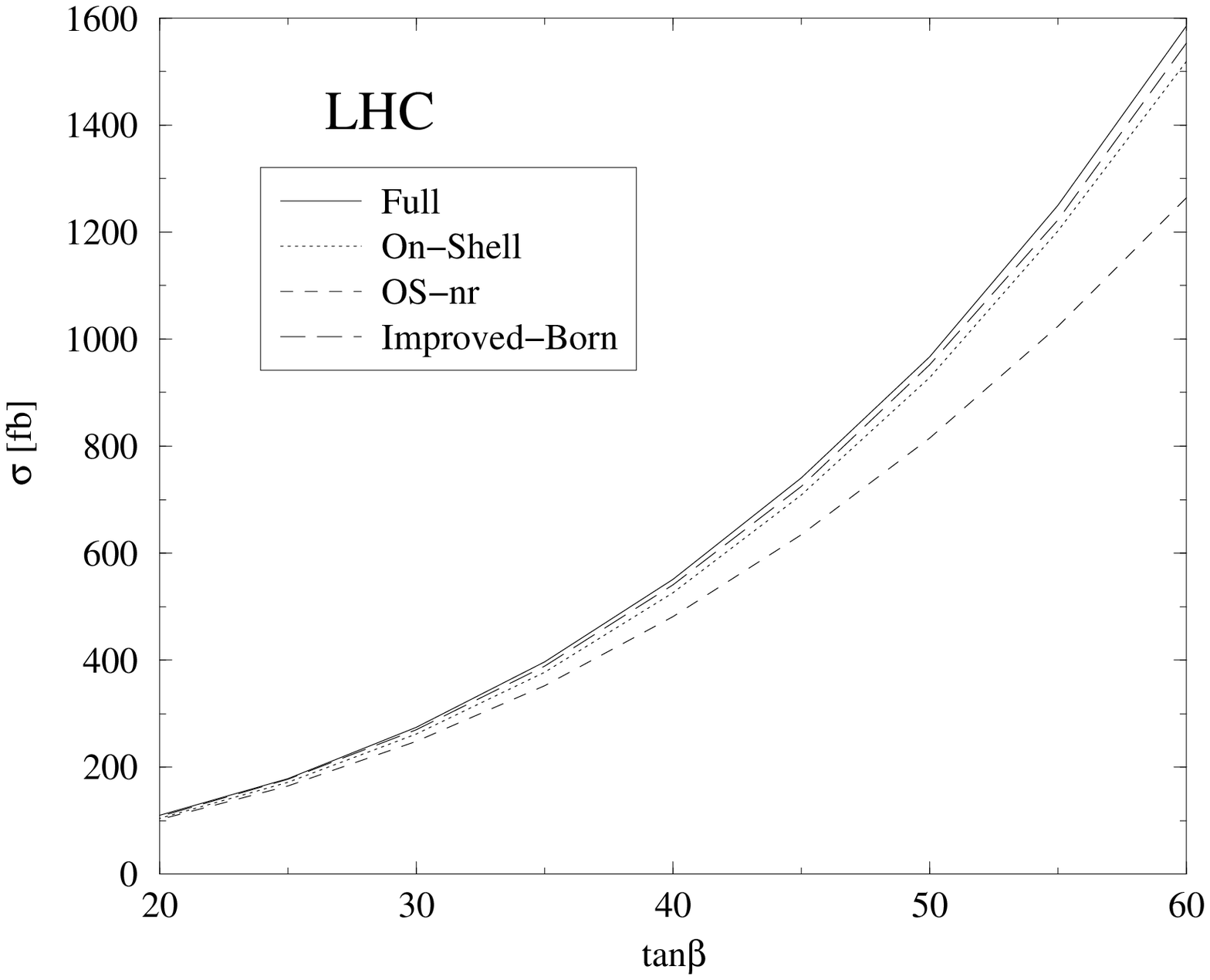}} \\
(a) & (b)
\end{tabular}

\caption{%
{As in Fig. 9, but
now showing the different approximations described in the text
for the intermediate Set B in Table~{\ref{tab:SUSYpar}},
\textbf{a)} for the Tevatron with $\mH=250\GeV$, and \textbf{b)}
for the LHC with $\mH=500\GeV$.
 }} \label{fig:corrtb2}
}

{In the following we restrict ourselves to plot the \textit{effective SUSY $%
K $-factor}  ($K^{\rm SUSY}$) involved in the definition
(\ref{KMSSM})}.  This factor embodies
the main results of our calculation as it gauges directly the
potential impact of the genuine SUSY corrections. It is defined
as follows:
\begin{equation}  \label{eq:effSUSYK}
K^{\mathrm{SUSY}}=\frac{\sigma^{\mathrm{SUSY-corrected}}}{\sigma^{\mathrm{%
Tree-level}}}\, ,
\end{equation}%
\FIGURE[t]{
\parbox{\textwidth}{\begin{tabular}{cc}
\resizebox{!}{6cm}{\includegraphics{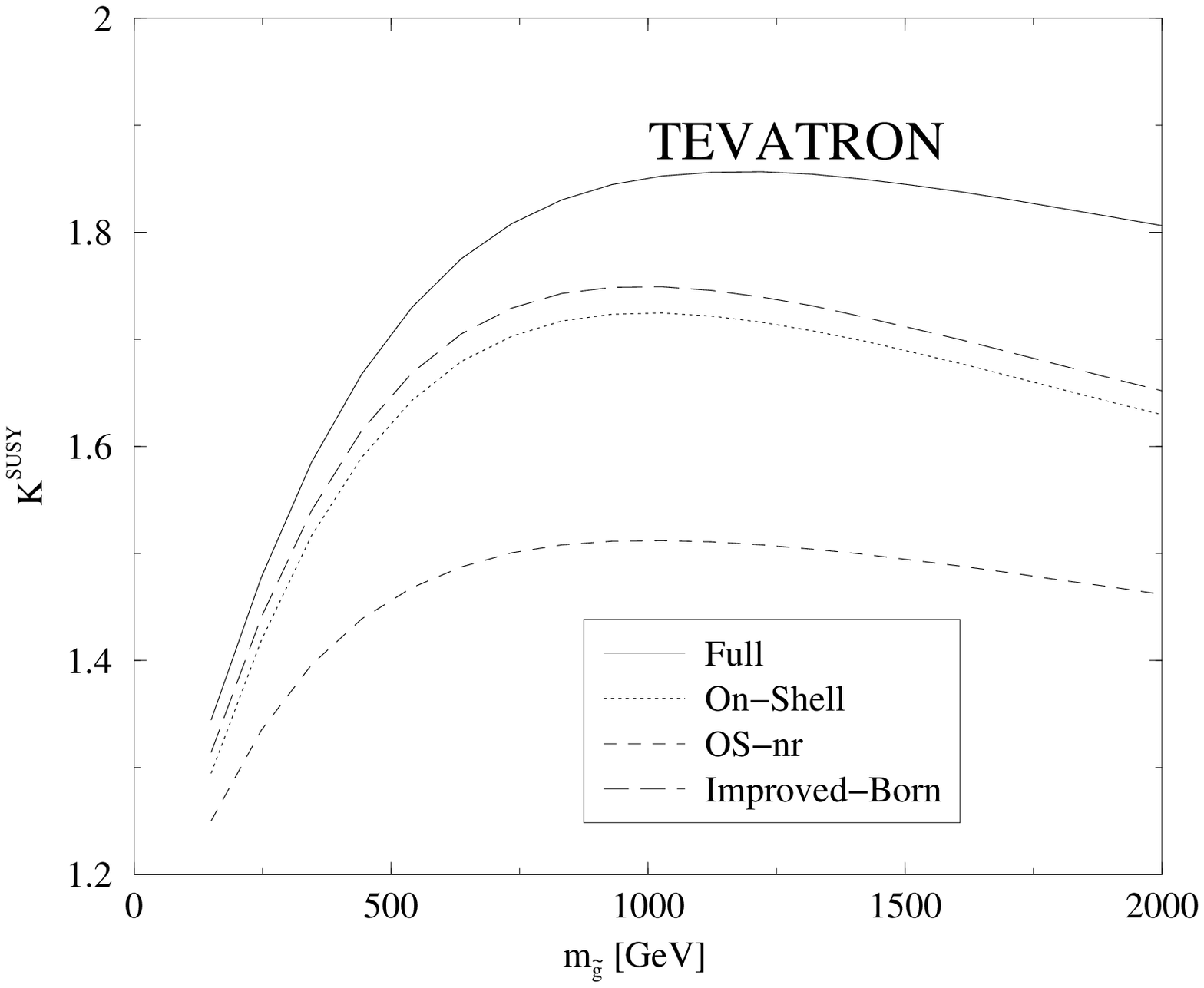}} & %
\resizebox{!}{6cm}{\includegraphics{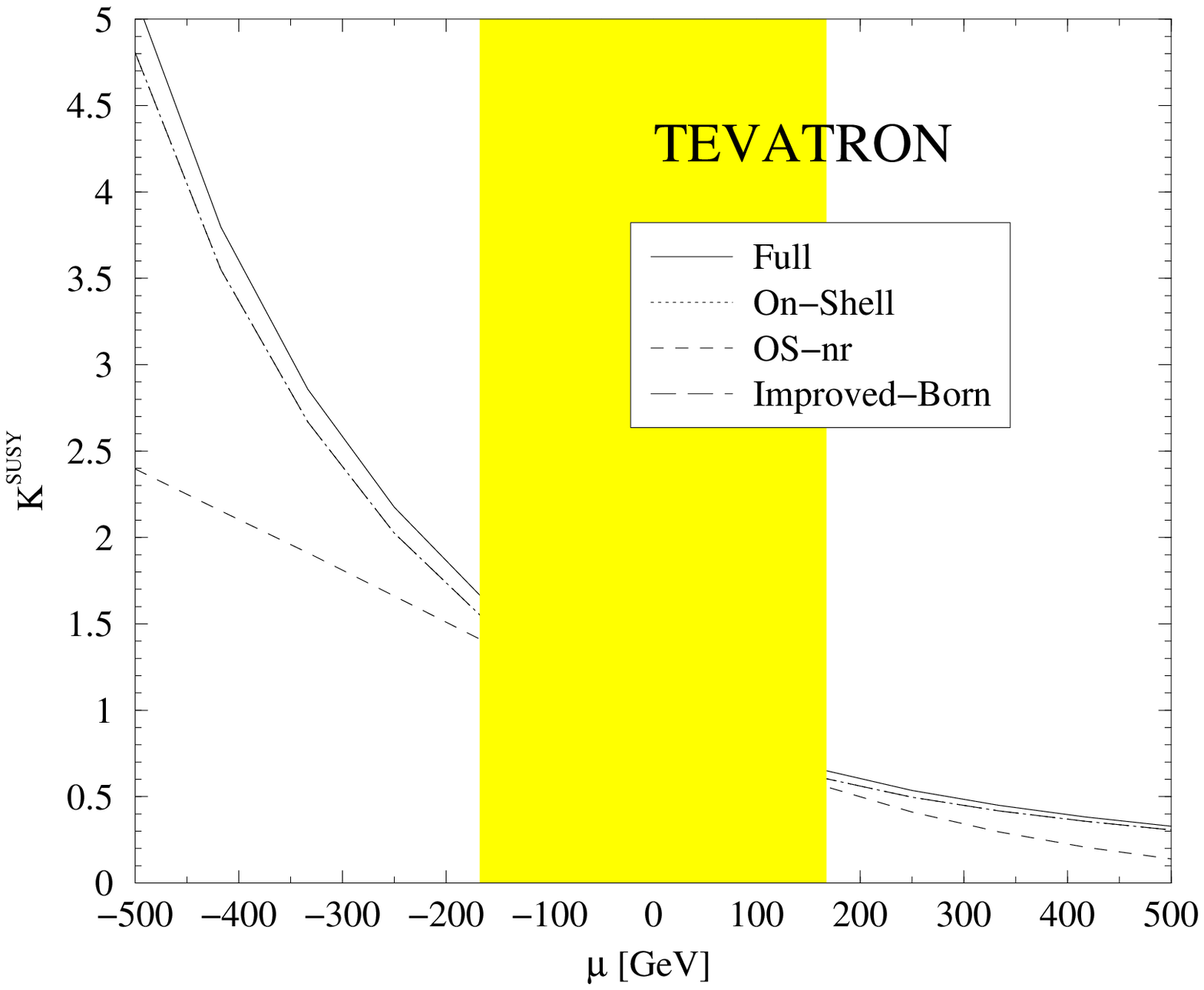}} \\
(a) & (b) \\
\resizebox{!}{6cm}{\includegraphics{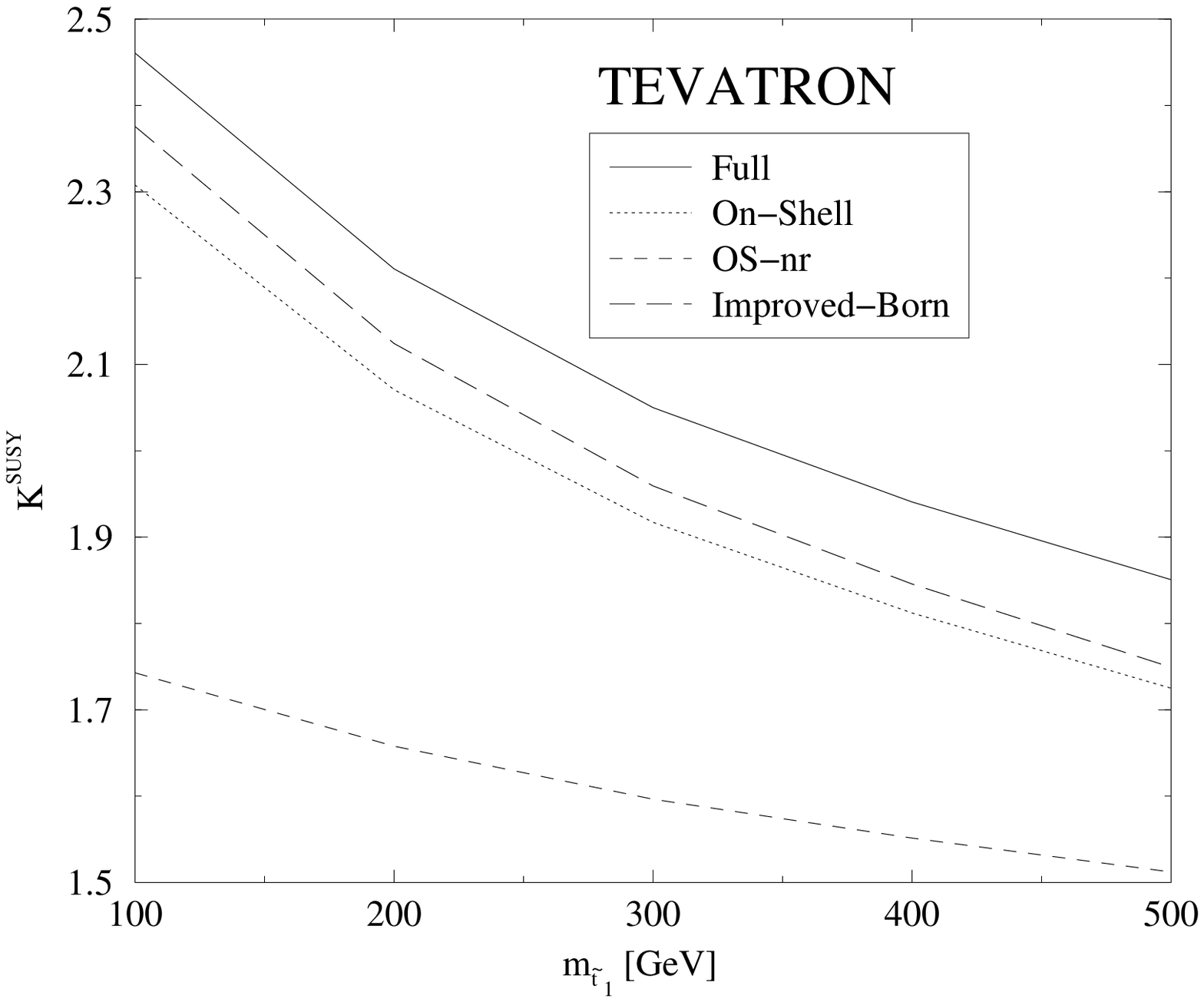}} & %
\resizebox{!}{6cm}{\includegraphics{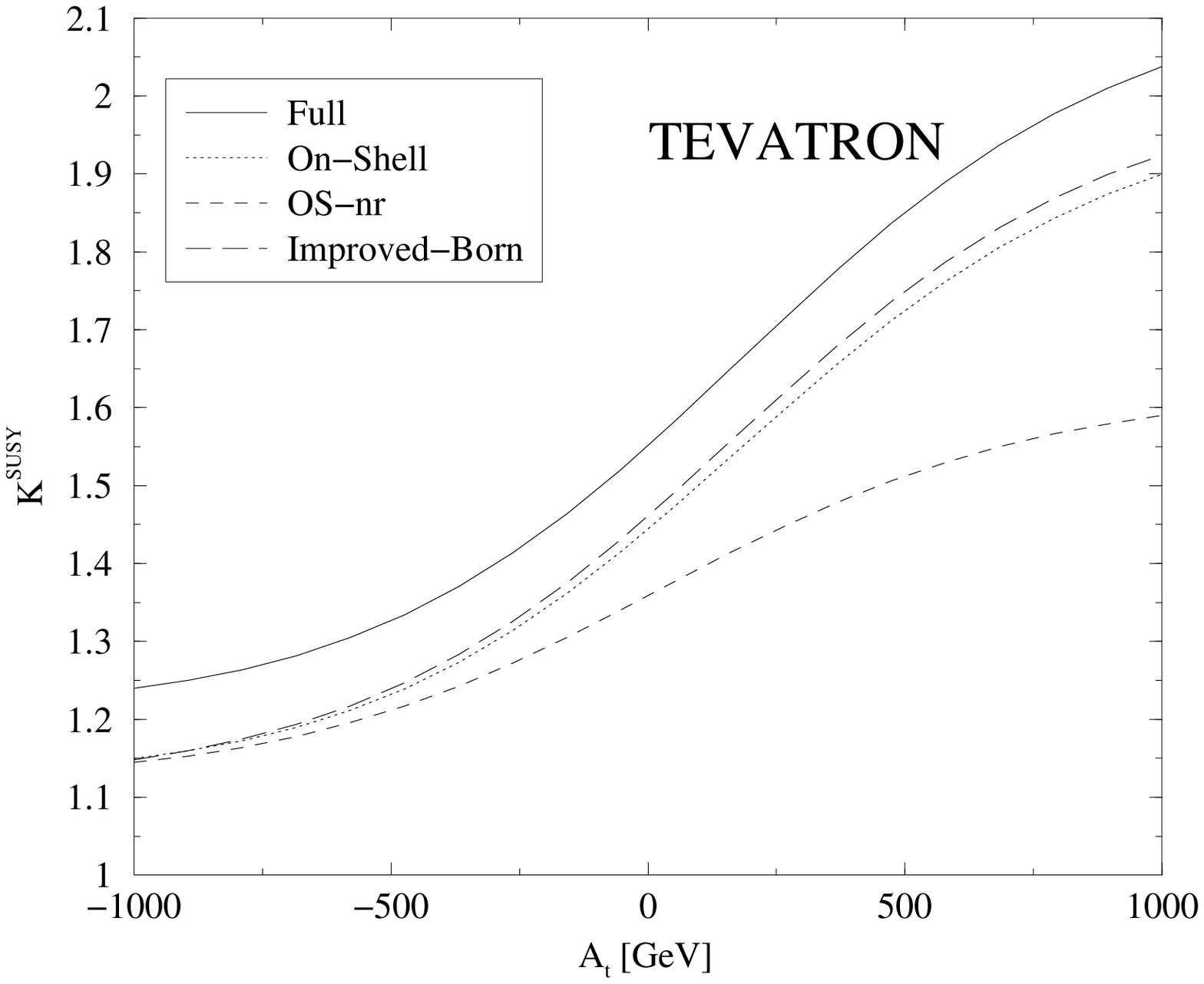}} \\
(c) & (d)
\end{tabular}}
\caption{%
{$K^{\rm SUSY}$-factor --
eq.~(\ref{eq:effSUSYK}) -- for the parameter Set B
--Table~\ref{tab:SUSYpar}-- and $\mH=250\GeV$ at the
Tevatron, as a function of \textbf{a)} the gluino mass; \textbf{b)} the $%
\protect\mu$ parameter; \textbf{c)} the lightest stop mass and \textbf{d)}
the top-squark  soft-SUSY-breaking trilinear parameter $A_t$. Shown are the
four approximations explained in the text.}}
\label{fig:TeVcorr}
}%
\FIGURE[t]{
\parbox{\textwidth}{\begin{tabular}{cc}
\resizebox{!}{6cm}{\includegraphics{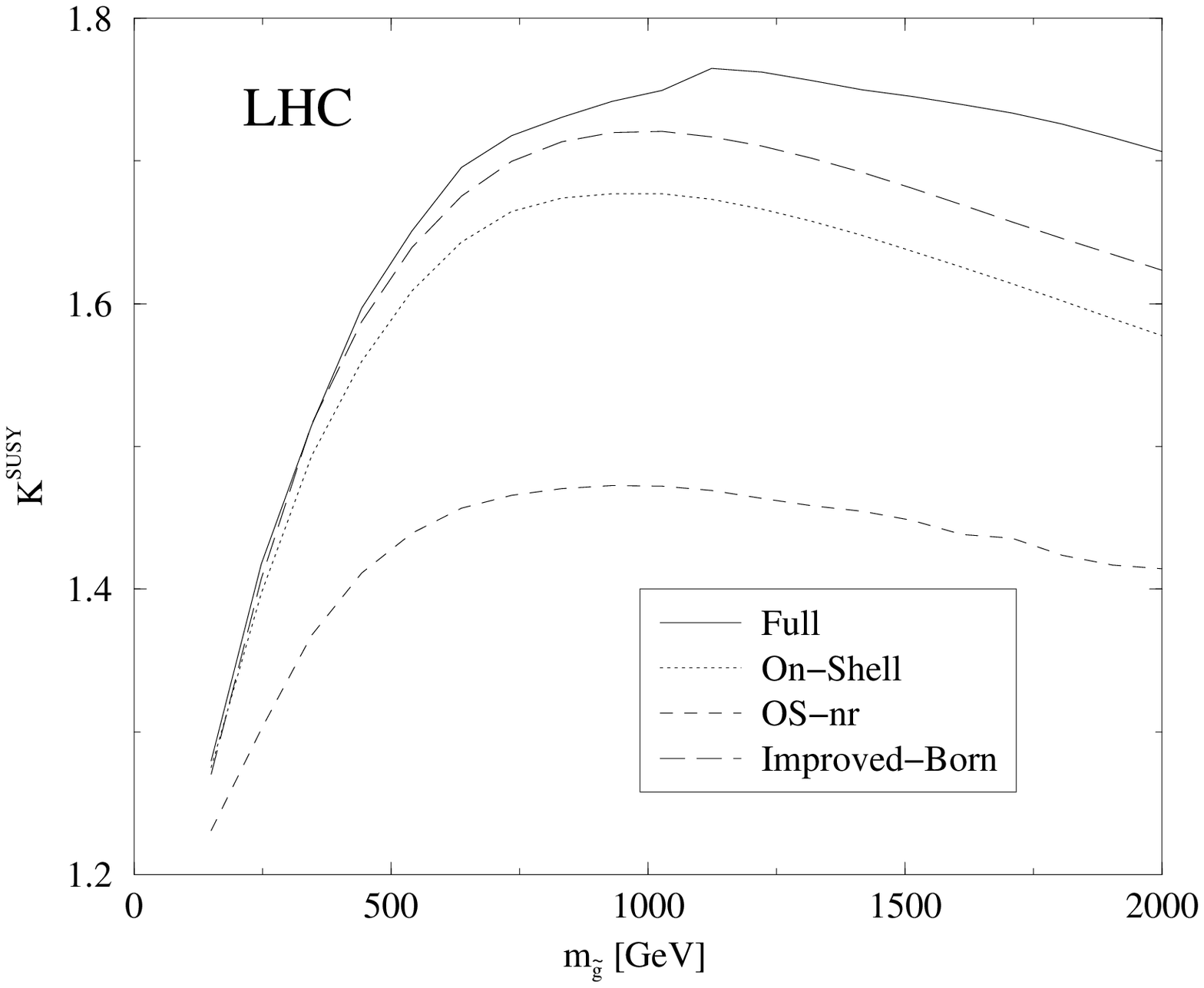}} & %
\resizebox{!}{6cm}{\includegraphics{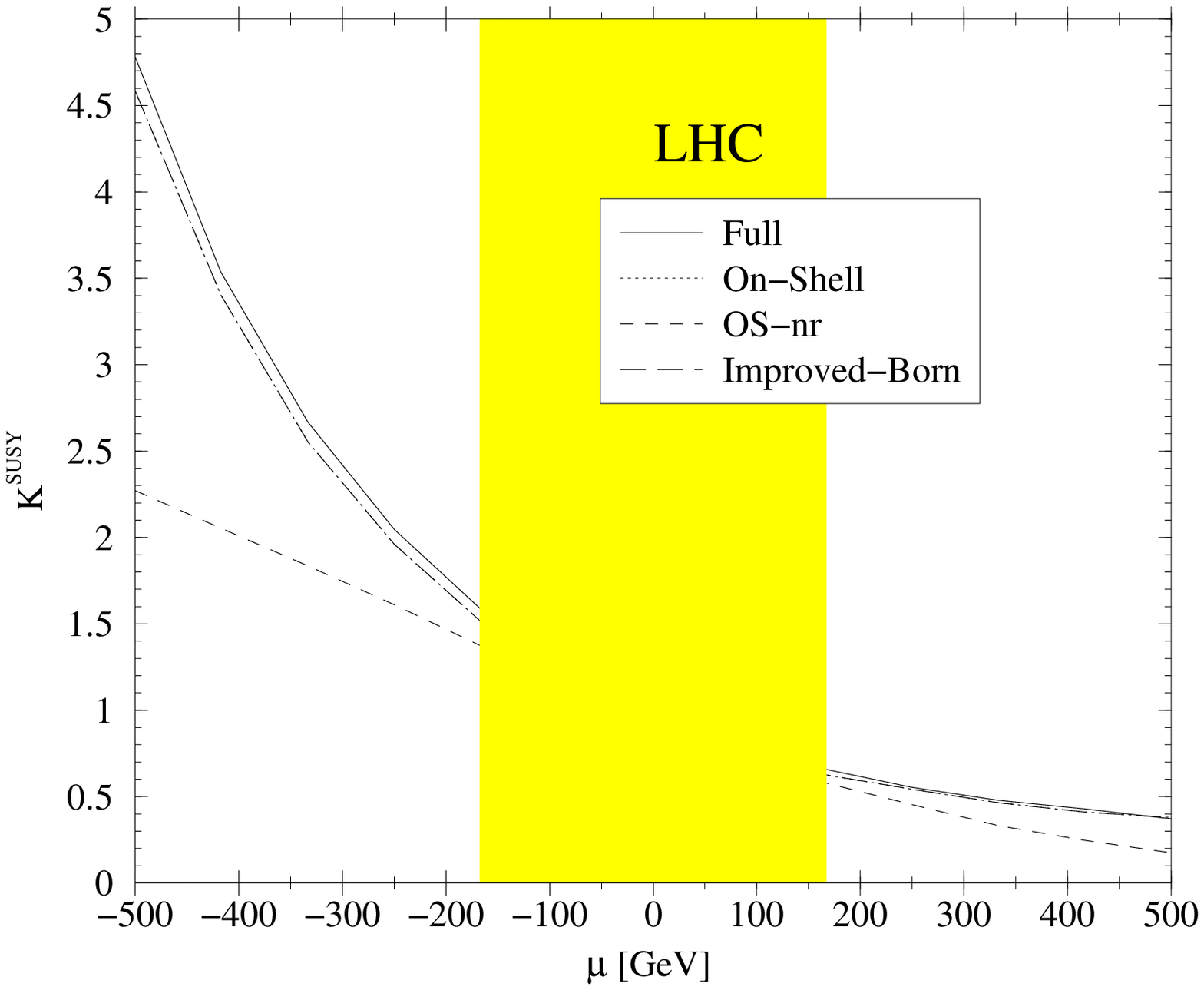}} \\
(a) & (b) \\
\resizebox{!}{6cm}{\includegraphics{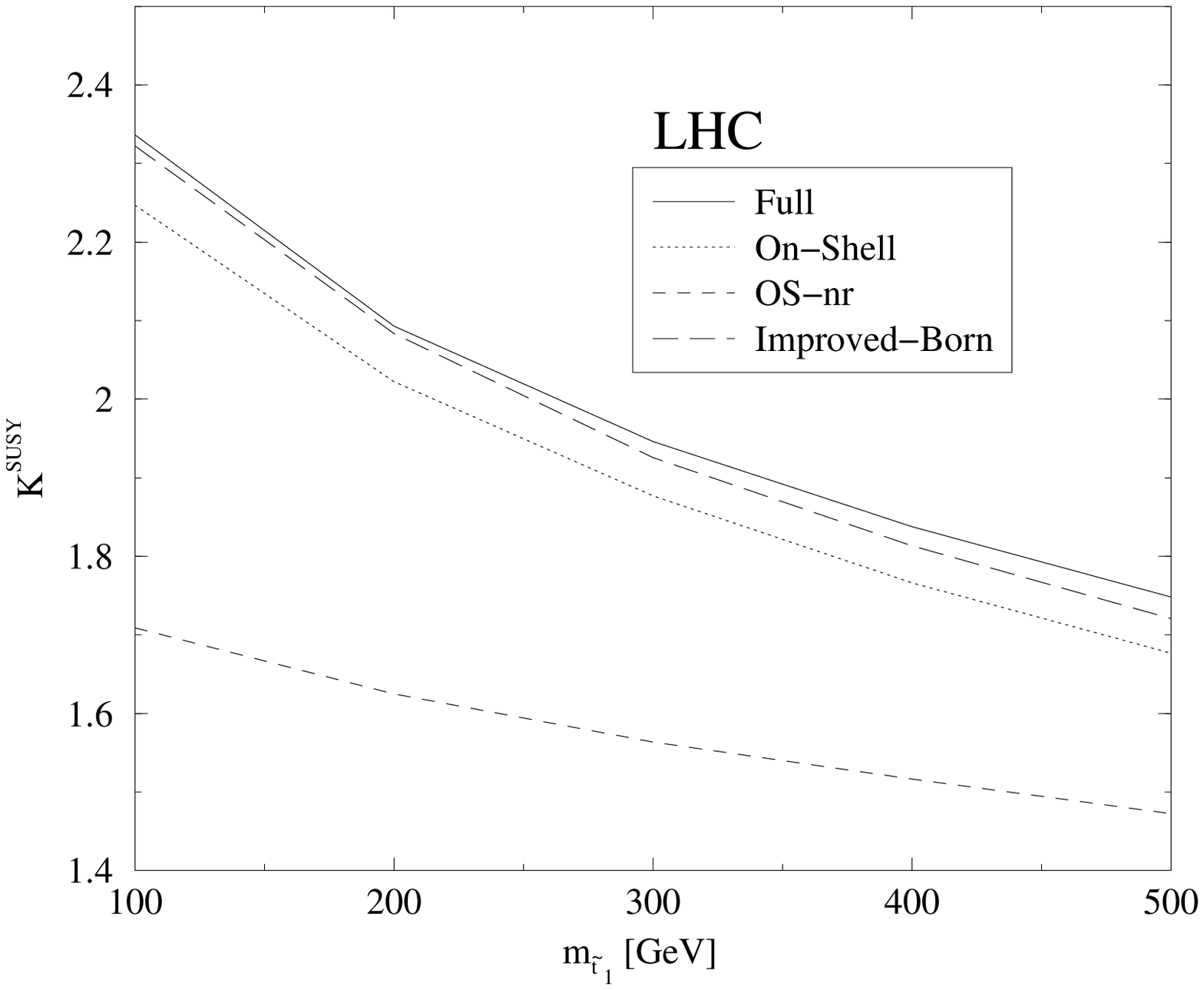}} & %
\resizebox{!}{6cm}{\includegraphics{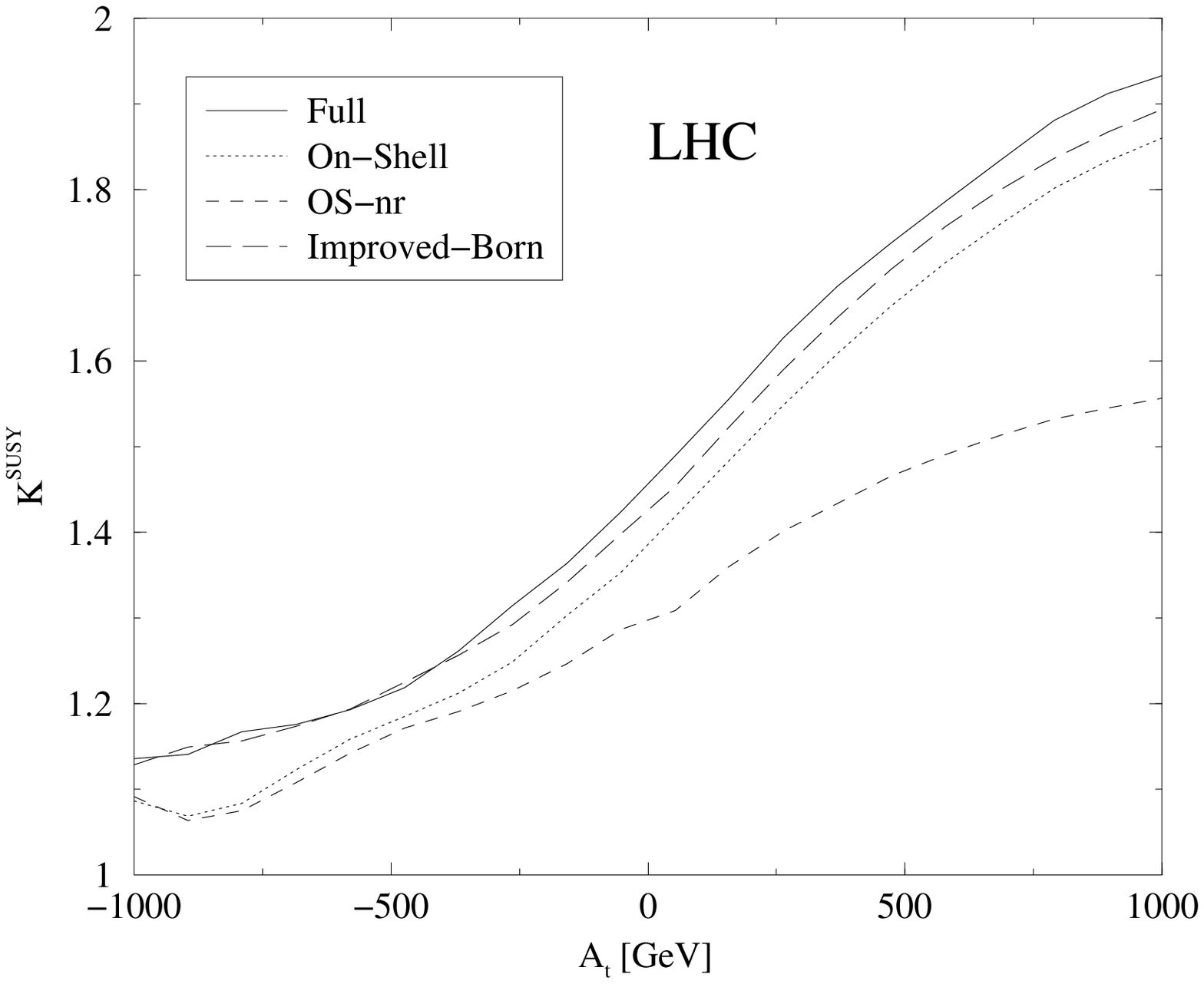}} \\
(c) & (d)
\end{tabular}}
\caption{%
{$K^{\rm SUSY}$-factor --
eq.~(\ref{eq:effSUSYK}) -- for the parameter Set B
--Table~\ref{tab:SUSYpar}-- and $\mH=500\GeV$ at the
LHC, as a function of \textbf{a)} the gluino mass; \textbf{b)} the $\protect%
\mu$ parameter; \textbf{c)} the lightest stop mass; and \textbf{d)} the
top-squark  soft-SUSY-breaking trilinear parameter $A_t$. Shown are the four
approximations explained in the text.}}
\label{fig:LHCcorr}
}%
{for the different contributions mentioned
above.\footnote{{The pole quark
masses are used in the ratio~(\ref {eq:effSUSYK}). 
Notice that although the total cross-section changes
significantly by the choice of running or on-shell masses, the
change in $K^{\mathrm{SUSY}}$ is rather mild, since it affects
both  the numerator and denominator of~(\ref {eq:effSUSYK}).}}
In Fig.~\ref{fig:TeVcorr} we present the evolution of the
 $K^{\rm SUSY}$-factor~(\ref{eq:effSUSYK})
as a function of various MSSM
parameters for the case of the Tevatron and for a charged Higgs
boson mass of $250\GeV$.  In Fig.~\ref{fig:LHCcorr} we show the
same effective
$K^{\rm SUSY}$%
-factor for the LHC and for a charged Higgs boson mass of $500\GeV$. Both cases
look pretty similar, the reason being that the leading corrections (eqs.~(%
\ref{eq:dmbQCD}-\ref{eq:dmbEWgauge}),
Fig.~\ref{fig:leadingdiags}) are insensitive to the charged Higgs
boson mass  and the energy of the process. For the
same reason the main features of the corrections follow closely
 the pattern already observed in
the partial decay widths $\Gamma(t\to H^+b)$~\cite{SUSYtbH} and $%
\Gamma(H^+\to t\bar{b})$~\cite{SUSYHtoTB}. In
these figures we have concentrated on the moderate Set B,
 although larger squark masses will not
 decrease significantly the corrections, unless
there exists a large hierarchy between the masses of the squarks,
the gluino and the $\mu$ parameter.}

{For the parameter Set B, the corrections are always positive.
They decouple very slowly with the gluino mass, as seen in
Figs.~\ref{fig:TeVcorr}a  and \ref {fig:LHCcorr}a, so a very
heavy gluino does not prevent to have large corrections. In
Figs.~\ref{fig:TeVcorr}  and \ref{fig:LHCcorr} the role of the
resummation is very clear; failing to include it would lead to
completely underestimated cross-sections. The shaded regions in
Figs.~\ref{fig:TeVcorr}b  and \ref{fig:LHCcorr}b correspond to a
chargino mass below the LEPII mass exclusion limit. The $\mu$
parameter is also a key parameter in the process under study,
since the SUSY-QCD corrections grow \textit{linearly} 
with it~--~Cf. Eq.~(\ref{eq:dmbQCD}).  In the $\mu<0$ scenario the
SUSY-QCD corrections  to the cross-section are
 positive: roughly,
$\delta\sigma/\sigma\sim -2\,(\Delta m_b)_{SUSY-QCD}>0$. Notice,
however, that in Figs.~\ref{fig:TeVcorr}b  and
\ref{fig:LHCcorr}b, $A_t>0$, and  so the SUSY-EW corrections add
constructively to the SUSY-QCD ones
$\delta\sigma/\sigma\sim -2\,(\Delta m_b)_{SUSY-EW}>0$. In the
 $\mu<0$ and  $A_t<0$ scenario, instead, the SUSY-EW corrections would
 partially
compensate the  positive SUSY-QCD loops, giving a
$K^{\mathrm{SUSY}}$ factor slightly smaller.} Finally, there is
the case  $\mu>0$ and
 $A_t>0$, where the two SUSY effects on the
cross-section would be negative: $K^{\rm SUSY}< 1$.
 As we have already mentioned, this
could lead to $K^{\rm MSSM}\simeq 1$  in
eq.(\ref{KMSSM}) and then the signature of the underlying SUSY
would be that the QCD corrections are ``missing'' or even
negative!  While this is possible for an $Ht\bar{t}$ final
state\,\cite{SpiraNLO}, for example due to Coulomb exchange of
gluons between quasistatic top quarks, this is not allowed in the
presence of $b$-quark in final states at high energies.

{We come now to the effects  from the squark sector. In
Figs.~\ref{fig:TeVcorr}c and \ref{fig:LHCcorr}c we show the
evolution with the corrections with the lightest stop-quark mass.
The plots end at $\mst1=500\GeV$ because for larger stop-quark
masses there is no physical solution for the mixing angle (given
$\msb1=500\GeV$)  in this particular set of inputs.
We see that, although the corrections have a dependence on the
squark masses, this parameter is not  the most critical one.
 In fact, we emphasize that larger squark masses
(as e.g. $\msb1\simeq\mst1\simeq 1\TeV$) do not necessarily lead
to smaller corrections -- see Fig.~\ref{fig:corrtb1}, parameter
 Set A.

 We finally focus on the influence of the  soft-SUSY-breaking trilinear stop-quark coupling $%
A_t$ (Cf. Figs.~\ref{fig:TeVcorr}d and~\ref{fig:LHCcorr}d),
 which is another key parameter for the
corrections under study. We see that, again, the preferred range
 ($\mu<0,A_t>0$) gives the  largest
positive corrections, thus enhancing the total cross-section
 dramatically. }
\FIGURE[t]{
\begin{tabular}{cc}
\resizebox{!}{6cm}{\includegraphics{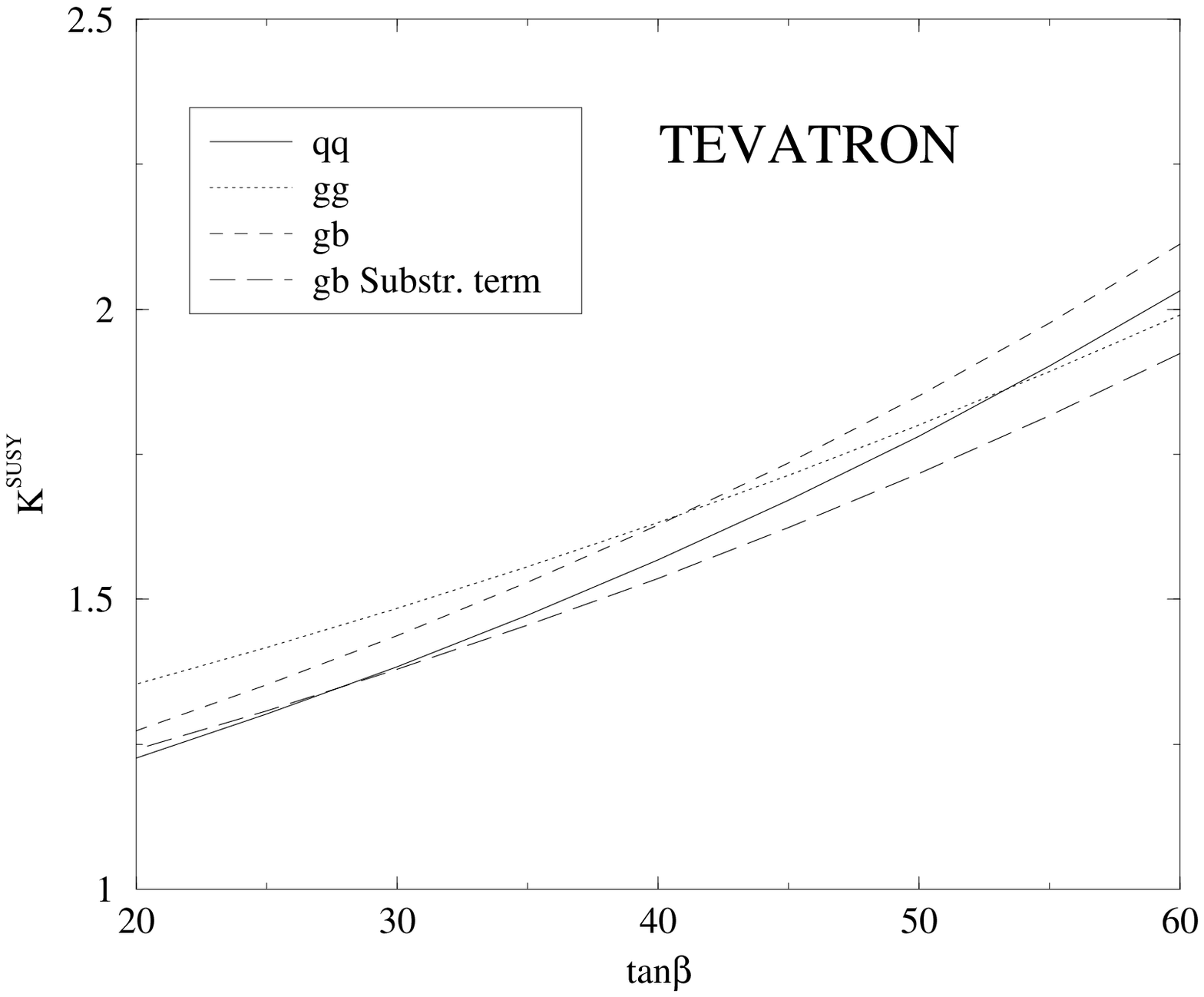}} & %
\resizebox{!}{6cm}{\includegraphics{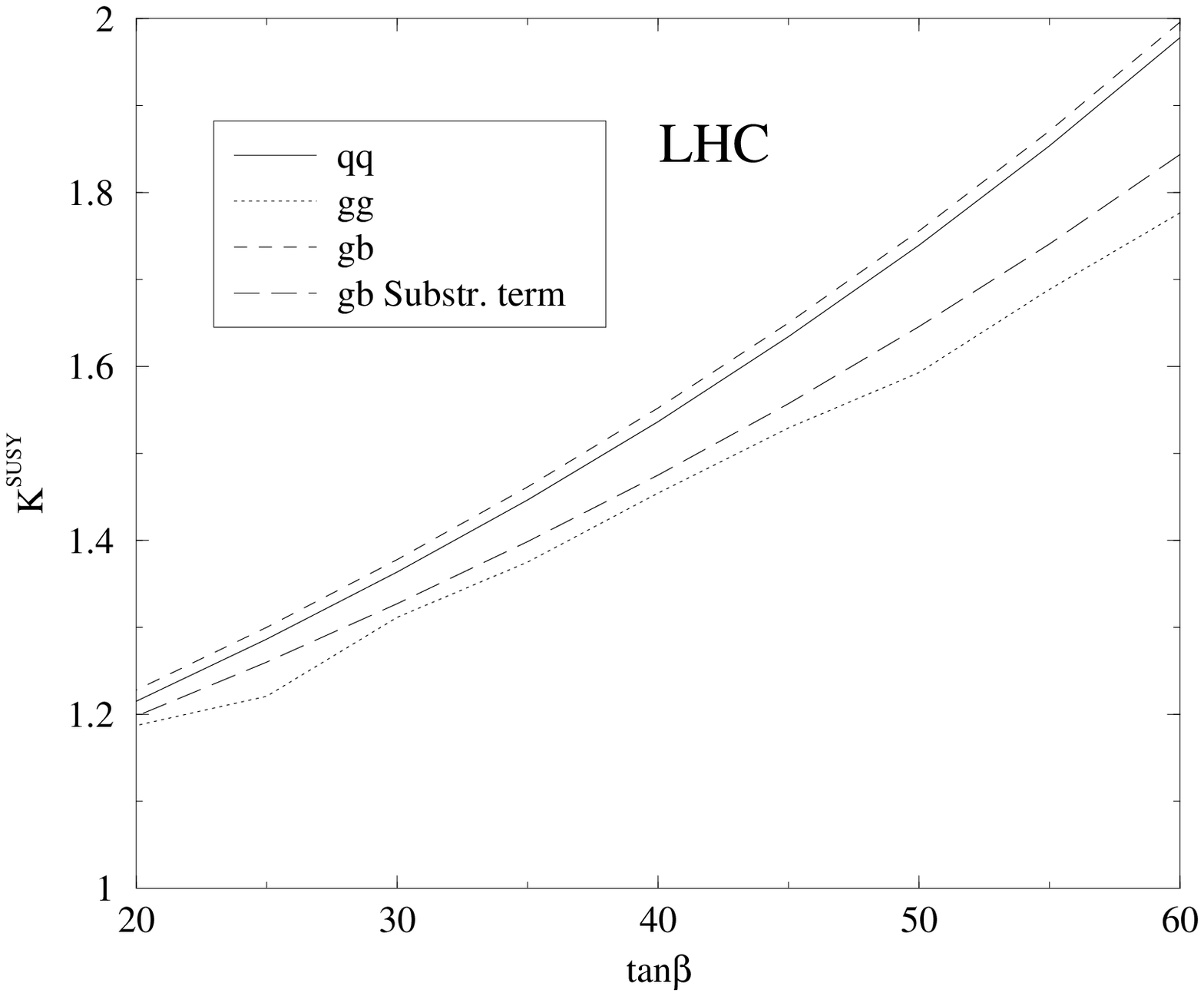}} \\
(a) & (b)
\end{tabular}
\caption{%
{$K^{\rm SUSY}$-factor -- 
eq.~(\ref{eq:effSUSYK}) -- for the parameter Set B
--Table~\ref{tab:SUSYpar}-- for the different production channels
as a function of \tb: \textbf{a)} For the Tevatron, and $\mH=250
\GeV $ and \textbf{b)} the LHC and $\mH=500\GeV$. }}
\label{fig:detcorr}
}

{The final point that we wish to address is the
study of the relative contributions from  the different
channels involved in the production cross-section --eqs.~(\ref{qq-tbh}%
)-(\ref{gb-th}). Since the main effect comes from the
 $\Delta m_b$ term, we expect that the
SUSY-corrections will be very similar in all the channels. In
Fig.~\ref{fig:detcorr} we  display the corrections
from the different channels for the Tevatron and the LHC.  We
see, indeed, that the corrections to the various channels are
very similar.  This means that to
obtain the discovery potential (or exclusion region) for the
different accelerators in the presence of the SUSY-corrections,
we may directly scale up the previous results from
Fig.~\ref{fig:lhc-eff} by a factor $K^{\rm SUSY}$ without having to
remake the full kinematical analysis. }

\subsection{Consequences for the charged Higgs boson search}

\label{sec:consequences}

 In the previous
figures~\ref{fig:corrtb1}-\ref{fig:LHCcorr} we have exemplified a
wide variety of MSSM effects on the  cross-section of process
(\ref{tbh}) based on a typical set of MSSM inputs (sparticle
masses and  soft-SUSY-breaking parameters). The new contributions
entail  cross-section enhancements at the level of $50-70\%$ beyond
standard QCD expectations.  In other words, the SUSY loops  have
the capability to modify the largest expected QCD $K$-factor from
$K_{\max}^{\rm QCD}=1.6$ up to an effective MSSM value
 (\ref{KMSSM}) reaching
 $K^{\rm MSSM}=2.4-2.7$. Surely
enough, if such a dramatic effect is really there, it could not
be experimentally missed.
In turn the discovery and exclusion limits on the charged Higgs
boson inferred from the  previous
analysis will get significantly modified as compared to the
tree-level case (Cf. Fig.\,\ref{fig:lhc-eff}), as we shall see
next.

{As already warned in Section~\ref{sec:sbspreview},
one should also keep in mind the SUSY corrections to the
branching ratio $BR(H^+\to t\bar{b})$. They are certainly much
tamed than those potentially affecting the partial decay width
  $\Gamma(H^+\to t\bar{b})$~\cite{HtbQCD}, but even so they can have some impact on
  the charged Higgs boson mass
  discovery/exclusion limits. Recall that the QCD corrections decrease
  significantly the partial decay width
  $\Gamma(H^+\to t\bar{b})$~\cite{HtbQCD}.  The genuine SUSY corrections,
  on the other hand, are governed by the same kind of contributions
  discussed in this paper, and can therefore have both
  signs~\cite{SUSYHtoTB,eff}.  As the sign of the SUSY quantum
  effects on the branching ratio is the same as for the production
  cross-section, they just mutually enhance or suppress the
  signal. Here we use the \textit{improved}
  expression for this partial decay width given in Ref.~\cite{eff}. The
  only other decay channel that we consider is
  $H^+\to \tau^+\nu_\tau$, but we neglect the SUSY corrections to its
  partial decay width.  Notice that, since $BR(H^+\to t\bar{b})$ is much
  larger than $BR(H^+ \to \tau^+\nu_\tau)$, the quantum corrections to
  the branching ratio $BR(H^+\to t\bar{b})$ will  be
  small in general.}
{We remark that in the light charged Higgs boson mass scenario
  relevant 
to our $t\bar{t}b\bar{b}$ signature (viz. $m_t+m_b<
M_{H^{\pm}}<300\GeV$), the partial decay channel $ H^+\to\tau^+\nu_\tau$
has a sizeable branching ratio which raises very fast with decreasing
$M_{H^{\pm}}$. For instance, for $M_{H^{\pm}}\sim (230-240)\GeV$ and a
moderately high $\tb=30$ one gets $BR(H^+\to\tau^+\nu_\tau)\gtrsim 20\%$
(without SUSY corrections). In some cases (specifically when $\mu>0,
A_t<0$, like in Sets C and D in Table \ref{tab:SUSYpar}) this branching
ratio can be further enhanced by SUSY effects due to the corresponding
decrease of $\Gamma(H^+\to t+\bar{b})$\footnote{See J.A. Coarasa
\textit{et al.} in~\cite{SUSYHtoTB} for the SUSY analysis of
$BR(H^+\to\tau^+\nu_\tau)$.}. Thus for charged Higgs masses in the heavy
range ($M_{H^{\pm}}> 300\GeV$), where the uncorrected branching ratio of
the $\tau$-mode would be below $10\%$, the latter can actually remain at
the level of $15\%$ (even for $M_{H^{\pm}}\gtrsim 500\GeV$) thanks to
the SUSY effects;  this scenario would be realized e.g. for
$(\mg,\mst1,\msb1,\mu, A_t, A_b)\simeq (500,150,300,200,-200,200)\GeV$
at $\tb=30$ as before.  Hence, it is clear that there are situations for
$M_{H^{\pm}}\gtrsim m_t-m_b$ where the $\tau$-mode could be successfully
used to explore the light charged Higgs mass region, as it was already
the case in Run I for $M_{H^{\pm}}<m_t-m_b$ through the study of the
combined decay $t\rightarrow H^{+}+b\rightarrow \tau^+\nu_\tau+b$\,\cite
{TeVHpm,CDF,DPR1,tbmh}. Nevertheless, the $\tau$-mode can alternatively
be highly suppressed by the SUSY effects themselves -- if they turn out
to greatly enhance $\Gamma(H^+\to t+\bar{b})$. Indeed, for
$(\mg,\mst1,\msb1)\simeq (500,150,500)\GeV$ and assuming $(\mu, A_t,
A_b)$ as in Sets A and B, the $\tau$-decay of the charged Higgs narrows
down to a $5\%$ or less. Therefore, the SUSY suppression of the
$\tau$-mode turns out to occur in the regions of the MSSM parameter
space where the production processes \pptbH\ are maximized. We conclude
that in the parameter space relevant for \pptbH\ the
$t\bar{t}b\bar{b}$ signature is in general the most suited one, and
indeed it is the only one used in the present study.}
{We warn the reader that at the opposite end of the spectrum, where $\mH \gsim
1\TeV$, the branching ratio $BR(H^+\to t\bar{b})$ would significantly decrease
if $\mH>\msb1+\mst1$, due to the opening of the squark decay channels $H^+\to
\tilde{b}_i
\tilde{t}_j$ -- which quickly become dominant at high $\tb$.  Given the parameters
in Table~\ref{tab:SUSYpar}, these squark channels are relevant only for
$\mH>1\TeV$ (resp. $2\TeV$) for the parameter Sets B and C (resp. A and D). 
Since these squark channels have not been considered in the present study, the
conclusions on the charged Higgs boson discovery/exclusion limits are strictly
valid only for $\mH<\msb1+\mst1$.
While we could easily include the new $H^+$ decay modes, the typical results
from our study would not change at all. In fact, we have emphasized 
that the leading SUSY corrections relevant for our process do not necessarily
decrease with the SUSY mass scale, so that our numerical results can be
extended to $\mH > 1\TeV$ since we can always find a modified particle spectrum
satisfying $\mH<\msb1+\mst1$ and producing the same radiative corrections as
the corresponding parameter set in Table~\ref{tab:SUSYpar}.}

{In Fig.~\ref{fig:xsecmhc} we show the equivalent of
Fig.~\ref{fig:lhc-eff}, but using the full SUSY-corrected
cross-section, {and the corrected branching ratios,} for
$\tb=50$.  In the Tevatron case, whose tree-level  cross-section
borders the discovery limits from below, we have applied a QCD
correction factor  $K^{\rm QCD}=1.5$ to account for the (yet
unknown) standard QCD corrections to the production
cross-section.  {Without large QCD corrections, the Tevatron
cannot discover a heavy charged Higgs boson $M_{H^+}>200\GeV$ (not
even at a $3\sigma$ level of significance) unless the SUSY
corrections are as high as e.g. those corresponding to Set A in
Table~\ref{tab:SUSYpar}. In this
case a charged Higgs boson in the range  $M_{H^+}\lesssim 255\GeV$ can
still be discovered at $3\sigma$ at the Tevatron. 
This is certainly a welcome possibility as it would throw light
on the SUSY nature of that charged Higgs boson, in case of being
detected, and in addition it would strongly hint at regions of
the MSSM parameter space very much akin to those defined by Set A.
 On the other hand, for intermediate SUSY sets
(like Set B), the concurrence of the QCD corrections is
indispensable  to get a similar charged Higgs boson
signal. At the LHC, in contrast,
  large  discovery/exclusion regions exist already using the tree-level
  approximation ($K^{\rm QCD}=1$),  so we are more conservative in this case and we can afford  not
 to apply a $K^{\rm QCD}$ correction factor}.
 The band in Fig.~\ref{fig:xsecmhc} interpolates
the parameter space between the parameter sets A and D
--Table~\ref{tab:SUSYpar}--,  which can be
regarded as plausible parameter sets defining the typical range
of maximum and minimum cross-sections expected in the MSSM.
 The interpretation of this figure is as follows:
in the context of the MSSM,  and within the
representative region of parameter space discussed in Section~\ref{sec:SUSYcorrleading}, the Tevatron should be able to discover a relatively heavy
charged Higgs boson ($M_{H^+}\lsim 250 \GeV$) at the canonical
$5\sigma$ significance only
 in an scenario of very large
positive SUSY corrections.  Nonetheless, the
Tevatron  can still find some reasonable evidence
(at $3\sigma$, i.e. $99.7\%$ C.L.) of a charged Higgs boson
with a mass  $200-250\GeV$ if only the condition $%
\mu<0$ applies  irrespective of the sign of $A_t$.
The LHC, on the other hand, does guarantee a $5\sigma$ discovery
in the range  $\mH<800\GeV$ for $100\fb^{-1}$
 of integrated luminosity; and a
$3\sigma$ evidence of a charged Higgs boson with a mass up to $1 \TeV$
for any SUSY spectrum. With only $30\fb^{-1}$ the LHC can not
discover a  superheavy charged Higgs boson
$\mH>1.1\TeV$, but the discovery is possible for $\mH<1.1\TeV$
and $\mu<0$. Once we will have data from the LHC this picture
 could change significantly. If SUSY particles are
found, and we know (roughly) their mass scale, the band in
Fig.~\ref{fig:xsecmhc} will get significantly
shrunk, providing clean predictions. If,  on the
contrary, a charged Higgs boson is found, but the scale of the
SUSY particles is still not  known with sufficient
accuracy, one can use Fig.~\ref{fig:xsecmhc} to extract
information on the SUSY particles properties.}
{Clearly the potential impact of the SUSY corrections is not at all
negligible and should not be ignored in any serious analysis of this subject.%
}

\FIGURE[t]{
\begin{tabular}{cc}
\resizebox{!}{6cm}{\includegraphics{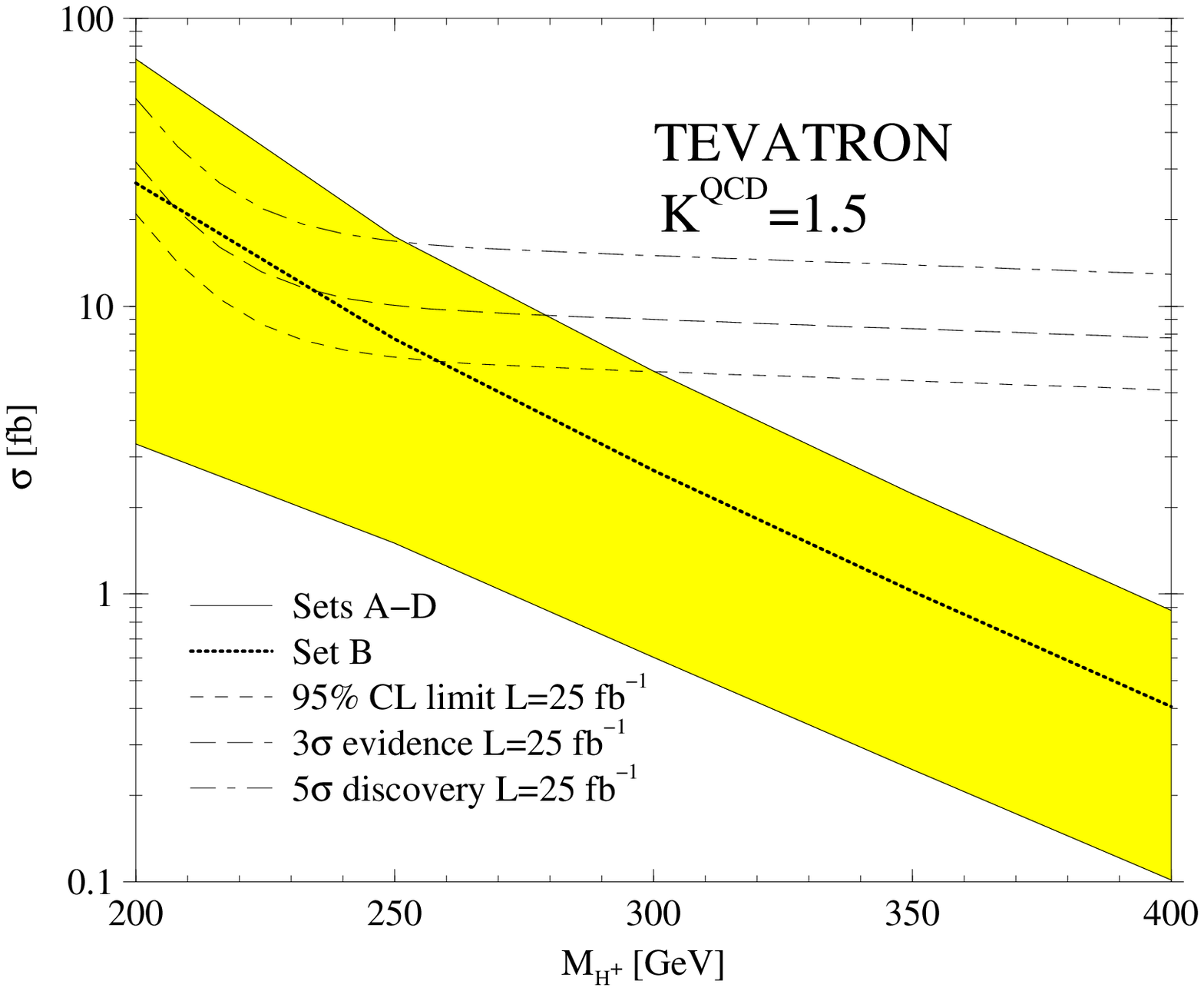}} & %
\resizebox{!}{6cm}{\includegraphics{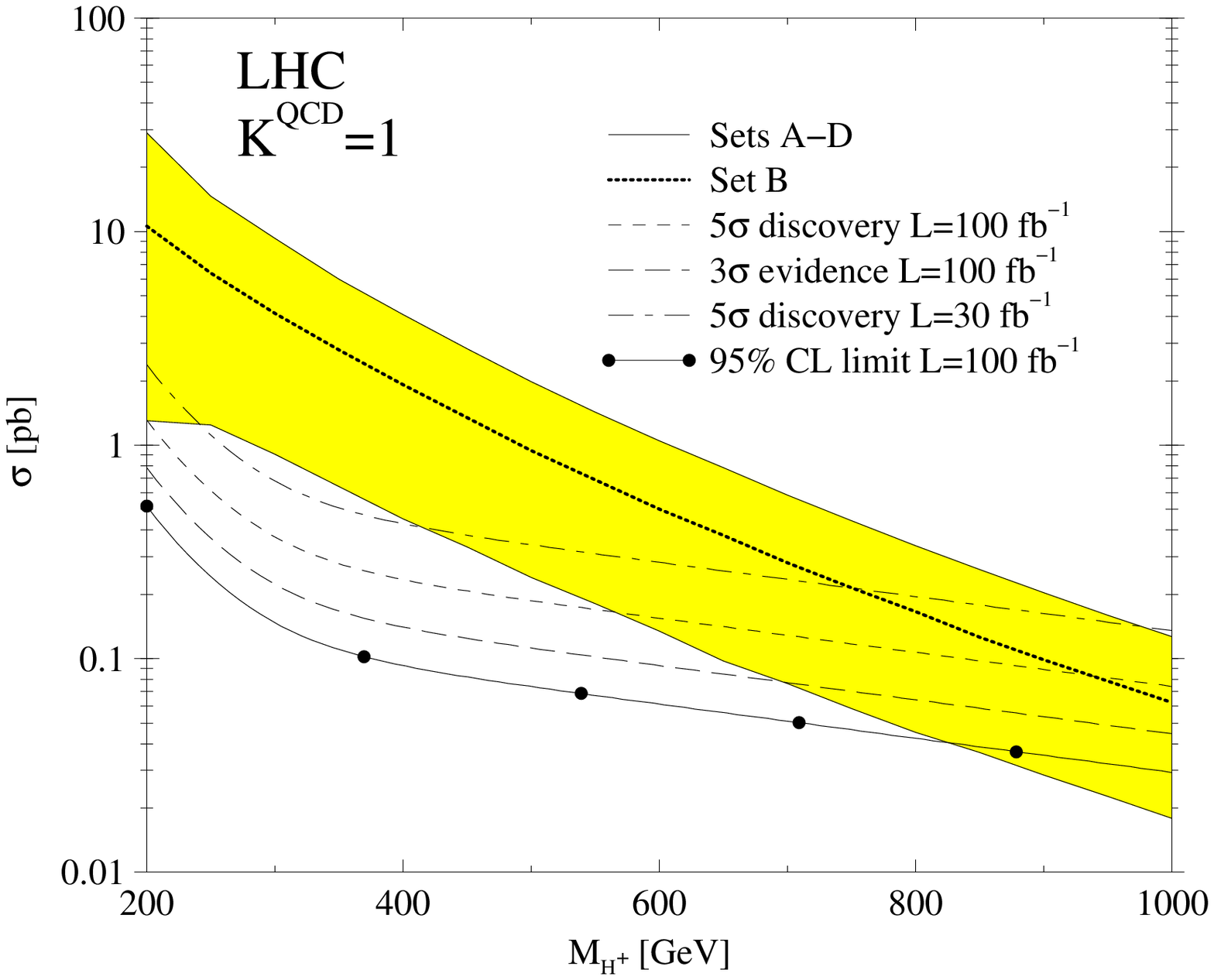}} \\
(a) & (b)
\end{tabular}
\caption{%
{Total effective cross-section for the
process~(\ref{tbh}) as a function of the charged Higgs boson mass
\textbf{a)} at the Tevatron {with a QCD correction factor $K^{\rm
QCD}=1.5$} and \textbf{b)} the LHC (with $K^{\rm
QCD}=1$).  In both cases
$\tan\beta=50$. Shown is the total cross-section 
 interpolating between parameters sets
 A and D (shaded region) and for the parameter Set B.
Also shown are the cross-sections necessary for
discovery and exclusion from Fig.~\ref{fig:lhc-eff}.}}
\label{fig:xsecmhc}
}

{By iterating this procedure we find the regions of the
  $\tb-\mH$ plane in which each experiment can find (or exclude) the
  existence of the charged Higgs boson. We present the result of this
  analysis in Figs.~\ref{fig:discovery}a and b. 
 These figures make transparent what
 was stated
  above: the presence of the SUSY corrections alters significantly
  the  Higgs boson
  discovery potential of the  hadron colliders.  From the figures it is also patent  the
  Tevatron could have a non-negligible chance to find a charged Higgs boson  in the
  intermediate range $M_{H^{\pm}}<280\GeV$.}

\FIGURE[t]{
\begin{tabular}{cc}
\resizebox{!}{6.8cm}{\includegraphics{TEV_discovery_mHc_ma_new.eps}} & %
\resizebox{!}{6.8cm}{\includegraphics*{LHC_discovery_mHc_ma_new.eps}} \\
(a) & (b)
\end{tabular}
\caption{%
{Discovery/Exclusion regions for the Tevatron and the LHC,
in the $\tb-\mH$ plane. 
{Also shown is the value of $M_{A^0}$ corresponding to each
$\mH$ according to the MSSM.}}} \label{fig:discovery}
}

{We have also shown  in Fig.~\ref{fig:discovery}
  the value of the pseudoscalar Higgs boson mass $M_{A^0}$ corresponding  to
  each value of $\mH$. We
  have correlated $M_{A^0}$ with $\mH$ using the one-loop
  corrections to the Higgs bosons masses as provided by the program
  \textit{FeynHiggsFast}~\cite{FeynHiggsFast}. Note that in this case
  $\mH$ changes with $\tb$, but is also different for each set of SUSY
  parameters. However, the change of $M_{A^0}$ among  the different sets
  is very small (below 5\%), and the one-to-one correspondence
  $\mH\leftrightarrow M_{A^0}$ is a very good approximation.}

Since the performed analysis of the SUSY corrections is meaningful only
for large \tb\  ($\tb\gsim20$), the shifts in the discovery regions in
Fig.~\ref{fig:discovery}b are not sufficiently reliable for
$\tb\lsim20$. Unfortunately this means we can not compare with the
results of Ref.~\cite{tev3-t-hb}, given for $M_{A^0}<500\GeV$ (which
correspond to \tb\  values $\lesssim 20$ in Fig.~\ref{fig:discovery}b).  
Sticking to the large \tb\  scenario, we see that for $\tb=60$ a charged
Higgs 
boson discovery is guaranteed for $\mH<1\TeV$. For $\mu<0$ a
charged Higgs boson could be discovered up to a mass of $1\TeV$ for
any $\tb>30$.

{
We should also recall here that our results show a certain degree of
uncertainty due to the dependence of the cross-sections on the particular PDF
parametrization and the variation with the renormalization and factorization
scales, which we have already illustrated in section~\ref{sect:svb}. These 
issues are particularly relevant for the Tevatron, where a charged Higgs boson
of order $250\GeV$ could be discovered only if the QCD and SUSY effects add up
with the same positive sign. 
In the worst possible scenario, the $K^{\rm QCD}=1.5$ factor that is
needed in combination with the typical SUSY effects from the
parameter Set B in Table~\ref{tab:SUSYpar} could be effectively reduced.
In this adverse circumstance the
charged Higgs discovery at 
the Tevatron would only be possible if the SUSY effects would be still larger,
namely at least as large as those from the parameter Set A 
in that table. In any case the lesson is once again the same: without the
collaboration of SUSY (or some other form of physics beyond the SM), the
charged Higgs discovery is impossible at the Tevatron.  On the other hand, the
aforementioned uncertainties are much more tamed for the LHC, and our results
shown in Fig.~\ref{fig:xsecmhc}b and \ref{fig:discovery}b should not be too much affected by them.
}
\section{Summary and conclusions}

\label{sec:conclu}

 We have presented a thorough study of the process
\pptbH\ in hadron colliders in order to assess the possibility to
see a SUSY charged Higgs boson at the Tevatron II and at the LHC.
 A necessary starting condition for this is that
$\tan\beta$ is sufficiently large  (viz.
$\tan\beta>20$)  so as to insure that the
cross-section of the process \pptbH\ is sizeable.
Beyond this tree-level requirement, our study of the quantum
corrections to that process within the MSSM has shown that they
are dominated by exceptionally important effects that can be
absorbed into an effective $tbH^{+}$ vertex, and therefore in
practice they can be treated at the level of an ``improved Born
approximation''. After the
inclusion of the MSSM effects we have found that the  ``effective $K$%
-factor''  (\ref{KMSSM}) for the process \pptbH\
could be either smaller than expected, perhaps barely $K^{\rm
MSSM}\lesssim 1$, or on the contrary substantially higher, say
 $K^{\rm MSSM}=2-3$. In both cases it should be
possible to exploit this process to unravel the nature of the new
physics in which the purportedly discovered charged Higgs boson
is integrated. This could be performed by detecting the existence
of large radiative corrections beyond the SM, namely sizeable
quantum effects that remain after subtracting the full set of
conventional QCD corrections at the NLO.  This will
 typically be the case if the squark masses are
heavy at an intermediate level of order of a few hundred $\GeV$.
 The gluino, notwithstanding, can be as heavy as
$1\TeV$. Remarkably enough, in certain circumstances explained
in the text {\it all} the sparticle masses, and not only the
gluino, could be of order of $1\TeV$, and still
the MSSM quantum effects would remain large ($\sim 50\%$).
  This feature shows that sound hints of SUSY can be derived from
 the sole study of Higgs boson production even if the whole sparticle spectrum is
 barely accessible to the future colliders.

{If $\tan\beta\gtrsim m_{t}/m_{b}> 35$, then the upgraded
$2\TeV$-Tevatron can exclude charged Higgs boson masses below $300\GeV$
at the $95\%$ C.L.
On the other hand, it is not likely that the Tevatron will be able
             to produce any signal at the $5\,\sigma$ significance level
             to ``officially'' discover a SUSY charged Higgs boson. 
 The only
             chance would be in the narrow window {$M_{H^+}=(180-250)
             \GeV$} and assuming the most favorable region in the MSSM
             parameter space (typically, Set A in Table
             \ref{tab:SUSYpar}).
In any case, the Tevatron can provide reasonable
evidence (at the $3\,\sigma$ level) of a $M_{H^+}=(200-300)%
\GeV$ Higgs boson, and also it can do a good complementary job at
high $\tan\beta$ using the
neutral Higgs bosons channels $p\overline{p}(pp)\rightarrow h\,\overline{b}b+X$ ($%
h=h^{0},H^{0},A^{0}$) already studied in the literature
 \,\cite{HiggsRunII,tev3-t-hb,CMS2001}, where the
absence of a top quark in the final state increases the signal in
some of them provided the Higgs boson is not too heavy. 
{In this respect we have seen
that in the light charged Higgs boson mass scenario, the partial decay
channel $H^+\to\tau^+\nu_\tau$ has a large branching ratio, and in some
favorable cases it could be successfully used at the Tevatron II to
explore this region, as it was the case for the Run
I\,\cite{TeVHpm,CDF,DPR1,tbmh}.}
Fortunately, the situation with the LHC
is much more rewarding under similar circumstances. In fact, the LHC
will be able to discover a 
charged Higgs boson up to $800\GeV$ or else to exclude it up to
$1\TeV$ at least. Remarkably, these limits can be significantly
improved up to $1.1\TeV$ and $1.5\TeV$ respectively in the
presence of SUSY corrections.}

{At the end of the day the strategy should be clear and can be
put in a nutshell. If the signal \pptbH\ is found it will
automatically imply new physics. Moreover, under the assumption
that the discovered charged Higgs boson is a member of a
 Type II two-Higgs-doublet model it will
strongly suggest a high value of 
$\tan\beta$,  certainly not below $20$, and most
likely above $30$.
Then an accurate measurement of the $%
tbH^{+}$ vertex should be possible at a level better than $10\%$, together
with its correlation with the neutral Higgs bosons vertices $\overline{b}bh$ ($%
h=h^{0},H^{0},A^{0}$) obtained both at the LHC and at the Tevatron from the
measurement of the companion processes $p\overline{p}(pp)\rightarrow h\,%
\overline{b}b+X$. After subtraction of the conventional QCD
corrections the genuine non-standard quantum effects could
possibly be disentangled after accurate comparison of all these
signals~{\cite{GHP,tau}}. As we have shown, the
``remnant'' of this subtraction could be quite large in the MSSM
case, viz. of the order of the full QCD effects themselves! Then
a pattern of sparticle masses (or at least of favoured regions in
the MSSM parameter
space) responsible for these virtual effects could emerge (see e.g. figures~%
\ref{fig:corrtb1}-\ref{fig:LHCcorr}), and of course it should be
cross-checked with the results obtained from other independent
experiments, e.g. at the LC~{\cite{Carenanew}}. Altogether the
bottom line of this strategy could definitely point towards the
supersymmetric nature of the found Higgs bosons until the
sparticles themselves  will eventually be
produced and correctly identified.}

\acknowledgments  

We thank M. Spira for helpful conversations on QCD effects. J.G. is
thankful to Germ{\'a}n Rodrigo for discussions.  A.B. acknowledges the
support of the U.S. Department of Energy under contract number
DE-FG02-97ER41022.  The work of J.S. has been supported in part by MECYT
and FEDER under project FPA2001-3598. J.S. also acknowledges the financial support provided through the European
Community's Human Potential Programme under contract HPRN-CT-2000-00149
Physics at Colliders. The work of J.G. and D.G. has been 
supported by the European Union under contracts Nr.  HPMF-CT-1999-00150
and ERBFMBICT-983539 respectively. Part of the calculations has been
done using the QCM cluster of the DFG Forschergruppe
``Quantenfeldtheorie, Computeralgebra und Monte-Carlo Simulation''.

\input{biblio_new.tex}

\end{document}

%% file: biblio_new.tex


%% file: bggs_new.bbl
\begin{thebibliography}{10}


\bibitem{LEP115}
R.~Barate \textit{et al.} {[}ALEPH Collaboration{]}, \plb{495}{2000}{1}, \hepex{0011045}; \\
M.~Acciarri \textit{et al.} {[}L3 Collaboration{]}, \plb{495}{2000}{18}, \hepex{0011043}; \\
P. Igo-Kemenes for the LEP Working Group on Higgs boson searches, talk
given at the LEPC on Nov. 3rd, 2000.


\bibitem{HiggsRunII}
M.~Carena \textit{et al.},
``Report of the Tevatron Higgs working group of the Tevatron Run 2
SUSY/Higgs Workshop'',
\hepph{0010338}.

\bibitem{tev3-t-hb} ATLAS Technical proposal, CERN-LHCC-94-38, 254pp. (1994); CMS Technical proposal,
CERN-LHCC-94-43, 272pp (1994).

\bibitem{CMS2001} See also the recent review: ``Summary of the CMS
Discovery Potential for the MSSM SUSY Higgses'', D. Denegri
\textit{et al.}, CMS NOTE 2001/032, \hepph{0112045}.


\bibitem{Ellisnew}
{J.~Ellis, S.~Heinemeyer, K.A.~Olive, G.~Weiglein,
\plb{515}{2001}{348},
\hepph{0105061}.}

\bibitem{LHCtop}
{M. Beneke, I. Efthymiopoulos, M.L. Mangano, J. Womersley
   (\textit{conveners}), ``Top quark physics'', report of the
   ``1999 CERN Workshop on SM physics (and more) at the LHC'', CERN
   2000-004, G. Altarelli, M.L. Mangano eds., \hepph{0003033}.}


\bibitem{Hunter}  J.F. Gunion, H.E. Haber, G.L. Kane, S. Dawson, \textit{The
Higgs Hunters' Guide} (Addison-Wesley, Menlo-Park, 1990).

\bibitem{MSSMreps}
H.P.~Nilles, 
\prep{110}{1984}{1}; \\
H.E.~Haber, G.L.~Kane,
\prep{117}{1985}{75}; \\
A.B.~Lahanas, D.V.~Nanopoulos, 
\prep{145}{1987}{1}.



\bibitem{CHHHWW}
M. Carena, H.E. Haber, S. Heinemeyer, W. Hollik, C.E.M. Wagner, G. Weiglein,
\npb{580}{2000}{29},
\hepph{0001002}; \\
J.R.~Espinosa, R.~Zhang,
\npb{586}{2000}{3},
\hepph{0003246},  and references therein.

\bibitem{SUSYtbH}
J.A.~Coarasa, D.~Garcia, J.~Guasch, R.A.~Jim\'{e}nez, J.~Sol\`{a},
\epjc{2}{1998}{373}, \hepph{9607485}; \\
J. Guasch, R.A. Jim\'enez, J. Sol\`a,\, \plb{360}{1995}{47}, \hepph{9507461}.

\bibitem{CJS1}
J.A.~Coarasa, R.A.~Jim\'{e}nez, J.~Sol\`{a}, \plb{389}{1996}{312}, \hepph{9511402}.

\bibitem{SUSYHtoTB}
{R.A.~Jim\'{e}nez, J.~Sol\`{a}, \plb{389}{1996}{53}, \hepph{9511292};} \\
{A.~Bartl \textit{et al.},
\plb{378}{1996}{167}, \hepph{9511385};\\
J.~A.~Coarasa, D.~Garcia, J.~Guasch, R.~A.~Jim\'enez, J.~Sol\`a,
\plb{425}{1998}{329}, \hepph{9711472};
\textit{ibid.} in: ``Proc. of the 4th international symposium on
Radiative Corrections (RADCOR98)'', p 498, World Scientific 1999,
ed. J. Sol{\`a}, \hepph{9903213}. }


\bibitem{LEPHpm}
M.~Acciarri \textit{et al.} {[}L3 Collaboration{]}, \plb{496}{2000}{34}, \hepex{0009010}; \\
R.~Barate \textit{et al.} {[}ALEPH Collaboration{]}, \plb{487}{2000}{253}, \hepex{0008005}.

\bibitem{TeVHpm}
Talk given by D.~Chakraborty at the SUSY~99 conference, Fermilab, June
1999; \\
B.~Abbott \textit{et al.} {[}D\O\  Collaboration{]}, \prl{82}{1999}{4975}, \hepex{9902028}.

\bibitem{CDF} F. Abe \textit{et al.} (CDF Collab.), \prl{79}{1997}{357}, \hepex{9704003}; 
\textit{ibid.} \prd{54}{1996}{735}, \hepex{9601003}.

\bibitem{DPR1} M. Guchait, D.P. Roy, \prd{55}{1997}{7263}, \hepph{9610514}.

\bibitem{tbmh}
J.~Guasch, J.~Sol\`{a}, \plb{416}{1998}{353},
\hepph{9707535}.

\bibitem{BSG-HPM}
T.~Goto, Y.~Okada, \ptp{94}{1995}{407},
\hepph{9412225}.

\bibitem{BSG-HPM-R}
M.A.~Diaz, E.~Torrente-Lujan, J.W.~Valle, \npb{551}{1999}{78}, \hepph{9808412}.


\bibitem{Bsg2l}
{M.~Ciuchini, G.~Degrassi, P.~Gambino, G.F.~Giudice,
\npb{534}{1998}{3},
\hepph{9806308}v2;\\
   G.~Degrassi, P.~Gambino, G.~F.~Giudice,
   \jhep{0012}{2000}{009}, 
   \hepph{0009337}; \\
M.~Carena, D.~Garcia, U.~Nierste, C.E.~Wagner,
\plb{499}{2001}{141},
\hepph{0010003}.}


\bibitem{eeHH} N.G.~Deshpande, X.~Tata, D.A.~Dicus, \prd{29}{1984}{1527}; \\
J.F.~Gunion, J.~Kelly, \prd{56}{1997}{1730},
\hepph{9610495}; \\
J.~Guasch, W.~Hollik, A.~Kraft, \npb{596}{2001}{66};
  \hepph{9911452}; \\
A.~Kiiskinen, P.~Poyhonen, M.~Battaglia,
\hepph{0101239}.

\bibitem{ppHH}
S.S.~Willenbrock, \prd{35}{1987}{173};\\
A.~Krause, T.~Plehn, M.~Spira, P.M.~Zerwas, \npb{519}{1998}{85}, \hepph{9707430};\\
A.A.~Barrientos Bendez\'{u}, B.A.~Kniehl, \npb{568}{2000}{305}, \hepph{9908385};\\
 O.~Brein, W.~Hollik, \epjc{13}{2000}{175}, \hepph{9908529}.


\bibitem{eeHW}
S.~Moretti, K.~Odagiri, \epjc{1}{1998}{633},
\hepph{9705389}; \\
A.~Arhrib, M.~Capdequi Peyranere, W.~Hollik, G.~Moultaka,
\npb{581}{2000}{34},
\hepph{9912527};\\
S.~Kanemura, S.~Moretti, K.~Odagiri,
\jhep{0102}{2001}{011},
\hepph{0012030}.

\bibitem{KZ}
Z.~Kunszt, F.~Zwirner, \npb{385}{1992}{3},
\hepph{9203223}.

\bibitem{ppHW}
A.A.~Barrientos Bendez\'{u}, B.A.~Kniehl, \prd{59}{1999}{015009}, \hepph{9807480};\\
S.~Moretti, K.~Odagiri, \prd{59}{1999}{055008}, \hepph{9809244}; \\
O.~Brein, W.~Hollik, S.~Kanemura, \prd{63}{2001}{095001},
\hepph{0008308}.

\bibitem{BH}
H.E.~Logan, U.~Nierste, \npb{586}{2000}{39},
\hepph{0004139}; \\
Y.~Grossman, H.E.~Haber, Y.~Nir, \plb{357}{1995}{630}, \hepph{9507213};\\
Y.~Grossman, Z.~Ligeti, \plb{332}{1994}{373},
\hepph{9403376}.

\bibitem{TauK}
{S.~Towers,
\hepex{0004022}.}

\bibitem{Dattaetal} A.~Datta, A.~Djouadi, M.~Guchait and Y.~Mambrini,
\prd{65}{2002}{015007}, \hepph{0107271}; \\
D.~Cavalli {\it et al.},
``The Higgs working group: Summary report,'' proceedings of the Workshop \textit{Physics at TeV Colliders}, Les Houches, France, 21 May - 1 June 2001,
\hepph{0203056}, and references therein.


\bibitem{prep} A. Belyaev, J. Guasch, J. Sol\`a, in preparation.

\bibitem{CGHS}
J.A. Coarasa, J. Guasch, W. Hollik, J. Sol{\`a}, \plb{442}{1998}{326}, \hepph{9808278}.


\bibitem{GHP} {See e.g. J.~Guasch, W.~Hollik, S.~Pe\~naranda,
\plb{515}{2001}{367},
\hepph{0106027}.}


\bibitem{Carenanew} {See e.g. M.~Carena, H.E.~Haber, H.E.~Logan, S.~Mrenna,
\prd{65}{2002}{055005},
\hepph{0106116}, and references therein.}


\bibitem{Dmb}
L.J.~Hall, R.~Rattazzi, U.~Sarid, \prd{50}{1994}{7048}, \hepph{9306309}; \\
M.~Carena, M.~Olechowski, S.~Pokorski, C.E.M.~Wagner, \npb{426}{1994}{269}, \hepph{9402253}.

\bibitem{Haberetal}
{H.E.~Haber \textit{et al.},
\prd{63}{2001}{055004},
\hepph{0007006}.}

\bibitem{eff}
M.~Carena, D.~Garcia, U.~Nierste, C.E.M.~Wagner, \npb{577}{2000}{88},
\hepph{9912516}.


\bibitem{tau}
J.A.~Coarasa, R.A.~Jim\'{e}nez, J.~Sol\`{a}, \plb{406}{1997}{337}, \hepph{9701392}.

\bibitem{CMW}
M. Carena, S. Mrenna, C.E.M. Wagner, \prd{60}{1999}{075010}, \hepph{9808312}.

\bibitem{Gunion}
J.F.~Gunion, \plb{322}{1994}{125}, \hepph{9312201}.

\bibitem{Barger}
V.~Barger, R.J.~Phillips, D.P.~Roy, \plb{324}{1994}{236}, \hepph{9311372}.

\bibitem{treepptbH}
S.~Moretti, D.P.~Roy, \plb{470}{1999}{209},
\hepph{9909435}.

\bibitem{Borzumati}
F.~Borzumati, J.~Kneur, N.~Polonsky, \prd{60}{1999}{115011}, \hepph{9905443}.

\bibitem{morepptbH}
S.~Moretti, K.~Odagiri, \prd{55}{1997}{5627}, \hepph{9611374}; \\
C.S.~Huang, S.~Zhu, \prd{60}{1999}{075012},
\hepph{9812201};\\
L.G.~Jin, C.S.~Li, R.J.~Oakes, S.H.~Zhu, \epjc{14}{2000}{91}, \hepph{9907482}.

\bibitem{DP-4b}
D.J.~Miller, S.~Moretti, D.P.~Roy, W.J.~Stirling, \prd{61}{2000}{055011}, \hepph{9906230}.

\bibitem{Guchait} M.~Guchait, S.~Moretti,  \jhep{0201}{2002}{001},
\hepph{0110020}.

\bibitem{Coarasa}
J.A.~Coarasa, J.~Guasch, J.~Sol\`{a}, \hepph{9909397}. Contributed to
Physics at Run II: Workshop on Supersymmetry/Higgs: Summary Meeting, Batavia,
Ill, 19--21 Nov. 1998.

\bibitem{BGGS1} {A.~Belyaev, D.~Garcia, J.~Guasch, J.~Sol\`a,
\prd{65}{2002}{031701}(R),
\hepph{0105053}.}

\bibitem{SpiraNLO}
{W.~Beenakker \textit{et al.},
\prl{87}{2001}{201805},
\hepph{0107081}.
}

\bibitem{DRNLO}
{L.~Reina, S.~Dawson,
\prl{87}{2001}{201804},
\hepph{0107101}; \\
L.~Reina, S.~Dawson, D.~Wackeroth,
Report: FSU-HEP-2001-0602, \hepph{0109066}; \textit{ibid.} Report:
FSU-HEP-2001-1019,
\hepph{0110299}.
}

\bibitem{Olness}
F.I.~Olness, W.~Tung, \npb{308}{1988}{813}; \\
R.M.~Barnett, H.E.~Haber, D.E.~Soper, \npb{306}{1988}{697};\\
D.~Dicus, T.~Stelzer, Z.~Sullivan, S.~Willenbrock,
  \prd{59}{1999}{094016}, \hepph{9811492}.

\bibitem{Dawson}
S.~Dawson, L.~Reina, \prd{57}{1998}{5851}, \hepph{9712400}.

\bibitem{barnett}
        R.M. Barnett, G. Senjanovic, D. Wyler,
        \prd{30}{1984}{1529}; \\
        P.N. Pandita, \plb{151}{1985}{51}; \\
        P.Q. Hung, S. Pokorski, Report FERMILAB-PUB-87/211-T, 1987; \\
        J.Lorenzo Diaz-Cruz, Hong-Jian He,Tim
        Tait, C.P. Yuan, \prl{80}{1998}{4641},
        \hepph{9802294}; \\
        C. Balazs, J.L. Diaz-Cruz, H.J. He, T. Tait, C.P. Yuan,
        \prd{59}{1999}{055016}, \hepph{9807349}.


\bibitem{Nason} P. Nason, S. Dawson, R.K. Ellis, \npb{303}{1988}{607}; \\
W. Beenakker, H. Kuijf, W.L. van Neerven, J. Smith, \prd{40}{1989}{54}.


\bibitem{spira_privat} M.Spira, private communication.

\bibitem{shouhua}
{S.-H. Zhu, \hepph{0112109}.
}



\bibitem{SpiraHouches}
{M. Spira, talk at the workshop ``Physics at TeV Colliders''.
Higgs Working Group. Les Houches, France, May 2001.}


\bibitem{QCDtbH}
C.S.~Li, T.C.~Yuan, \prd{42}{1990}{3088}, Erratum
\textit{ibid.} \textbf{D47} (1993) 2156; \\
M.~Drees, D.P.~Roy, \plb{269}{1991}{155}; \\
C.~Li, Y.~Wei, J.~Yang, \plb{285}{1992}{137}; \\
A.~Czarnecki, S.~Davidson, \prd{48}{1993}{4183},
\hepph{9301237}.



\bibitem{AP}
G.~Altarelli, G.~Parisi, \npb{126}{1977}{298};\\
G.~Altarelli,  \prep{81}{1982}{1}.

\bibitem{TW}
A.S.~Belyaev, E.E.~Boos, L.V.~Dudko, \prd{59}{1999}{075001}, \hepph{9806332}.


\bibitem{gut}
B.~Ananthanarayan, G.~Lazarides, Q.~Shafi, \prd{44}{1991}{1613}; \\
T.~Banks, \npb{303}{1988}{172}; \\
M.~Olechowski, S.~Pokorski, \plb{214}{1988}{393};
\\
S.~Dimopoulos, L.J.~Hall, S.~Raby, \prl{68}{1992}{1984};
\prd{45}{1992}{4192}; \\
G.W.~Anderson, S.~Raby, S.~Dimopoulos, L.J.~Hall, \prd{47}{1993}{3702}, \hepph{9209250}; \\
R. Hempfling, \prd{49}{1994}{6168}.

\bibitem{LEPtb}
  [LEP Higgs Working Group Collaboration],
\hepex{0107030}.



\bibitem{CompHEP}
A.~Pukhov \textit{et al.}, {}``CompHEP - a package for evaluation of Feynman
diagrams and integration over multi-particle phase space. User's manual for
version 33,{}'' preprint INP MSU 98-41/542, \hepph{9908288}.

\bibitem{tev1-t-hb}
B.K.~Bullock, K.~Hagiwara, A.D.~Martin, \npb{395}{1993}{499}.

\bibitem{tev2-t-hb}
S.~Raychaudhuri, D.P.~Roy,
\prd{52}{1995}{1556}, \hepph{9503251};
\textit{ibid.} \prd{53}{1996}{4902}, 
\hepph{9507388}.


\bibitem{DP-3b}
S.~Moretti, D.P.~Roy, \plb{470}{1999}{209},
\hepph{9909435}.

\bibitem{CTEQ4}
H.L.~Lai \textit{et al.}, \prd{51}{1995}{4763}, \hepph{9410404}.

\bibitem{MRST}
A.~Martin, R.G. Roberts, W.J. Stirling, R.S. Thorne, \epjc{4}{1998}{463}, \hepph{9803445}.

\bibitem{MoriondHtaunu}
{A.~Tricomi, contribution to the ``36th Rencontres de
    Moriond on QCD and Hadronic Interactions'', Les Arcs, France, 17-24
  Mar 2001,
\hepph{0105199}.}


\bibitem{d0tag}
Proc. of Supersymmetry/Higgs RUN II workshop, February - November, 1998,
\texttt{http://fnth37.fnal.gov/susy.html}.

\bibitem{cmstag}
CMS collab., S.~Abdullin \textit{et al.}, CMS-NOTE-1998-006, \hepph{9806366}.


\bibitem{pythia}
T.~Sjostrand, 
\cpc{82}{1994}{74}; \\
S.~Mrenna, 
\cpc{101}{1997}{232},
\hepph{9609360}.

\bibitem{comp-pyth}
A.S.~Belyaev {\it et al.}, \hepph{0101232}.

\bibitem{colorbreak}
J.M. Fr\`ere, D.R.T. Jones, S. Raby, \npb{222}{1983}{11}; \\
M. Claudson, L. Hall, I. Hinchliffe, \npb{228}{1983}{501}; \\
C. Kounnas, A.B. Lahanas, D.V.
Nanopoulos, M. Quir\'os,\ \npb{236}{1984}{438}; \\
J.F. Gunion, H.E. Haber, M. Sher,\ \npb{306}{1988}{1}.

\bibitem{LoopTools}
{T.~Hahn, M.~P{\'e}rez-Victoria,
\cpc{118}{1999}{153}
\hepph{9807565};\\
T.~Hahn, \textit{LoopTools user's guide},
\texttt{http://www.feynarts.de/looptools};\\
G.~J.~van Oldenborgh,
\cpc{66}{1991}{1}.
}

\bibitem{bsgamm}  R. Barbieri, G.F. Giudice, \plb{309}{1993}{86}, \hepph{9303270}; \\
J.N. Ng, \plb{315}{1993}{372}, \hepph{9307301};\\
M. Carena, C.E.M. Wagner \npb{452}{1995}{45}, \hepph{9408253}.


\bibitem{gminus2ex}
{H.~N.~Brown {\it et al.}  [Muon g-2 Collaboration],
\prl{86}{2001}{2227},
\hepex{0102017}.}


\bibitem{gminus2SUSY}
{See e.g. A.~Czarnecki, W.~J.~Marciano,
\prd{64}{2001}{013014},
\hepph{0102122}, and references therein.}




\bibitem{HtbQCD} {C.~Li, R.~J.~Oakes,
\prd{43}{1991}{855}.}


\bibitem{FeynHiggsFast}
{S.~Heinemeyer, W.~Hollik, G.~Weiglein,
\cpc{124}{2000}{76}, \hepph{9812320};
\hepph{0002213}.}






\end{thebibliography}
